\documentclass[letterpaper,11pt]{article}

\usepackage[utf8]{inputenc}
\usepackage[T1]{fontenc}

\usepackage{jcapmod}

\usepackage{amsmath,amsfonts,amssymb}
\usepackage{mathtools,physics}
\usepackage{textalpha}

\usepackage{hyperref}
\usepackage{graphicx}

\usepackage[letterpaper,margin=2.55cm]{geometry}

\numberwithin{equation}{section}

\def\sqrtb{\mathpalette\DHLhksqrt}
\def\DHLhksqrt#1#2{%
\setbox0=\hbox{$#1\sqrt{#2\,}$}\dimen0=\ht0
\advance\dimen0-0.2\ht0
\setbox2=\hbox{\vrule height\ht0 depth -\dimen0}%
{\box0\lower0.4pt\box2}}

\DeclareRobustCommand{\loooongrightarrow}{%
  \DOTSB\relbar\joinrel\relbar\joinrel\relbar\joinrel\relbar\joinrel\rightarrow
}

\makeatletter
\DeclareFontFamily{OMX}{MnSymbolE}{}
\DeclareSymbolFont{MnLargeSymbols}{OMX}{MnSymbolE}{m}{n}
\SetSymbolFont{MnLargeSymbols}{bold}{OMX}{MnSymbolE}{b}{n}
\DeclareFontShape{OMX}{MnSymbolE}{m}{n}{
    <-6>  MnSymbolE5
   <6-7>  MnSymbolE6
   <7-8>  MnSymbolE7
   <8-9>  MnSymbolE8
   <9-10> MnSymbolE9
  <10-12> MnSymbolE10
  <12->   MnSymbolE12
}{}
\DeclareFontShape{OMX}{MnSymbolE}{b}{n}{
    <-6>  MnSymbolE-Bold5
   <6-7>  MnSymbolE-Bold6
   <7-8>  MnSymbolE-Bold7
   <8-9>  MnSymbolE-Bold8
   <9-10> MnSymbolE-Bold9
  <10-12> MnSymbolE-Bold10
  <12->   MnSymbolE-Bold12
}{}

\let\llangle\@undefined
\let\rrangle\@undefined
\DeclareMathDelimiter{\llangle}{\mathopen}%
                     {MnLargeSymbols}{'164}{MnLargeSymbols}{'164}
\DeclareMathDelimiter{\rrangle}{\mathclose}%
                     {MnLargeSymbols}{'171}{MnLargeSymbols}{'171}
\makeatother

\begin{document}


\begin{titlepage}

\baselineskip=15.5pt \thispagestyle{empty}

\begin{center}
    {\fontsize{20.74}{24}\selectfont \bfseries Fingerprints of a Non-Inflationary Universe\\ \vspace*{5pt} from Massive Fields} 
\end{center}

\vspace{0.1cm}

\begin{center}
    {\fontsize{12}{18}\selectfont Jerome Quintin,$^{1,2}$ Xingang Chen,$^{3}$ and Reza Ebadi$^{4,5}$}
\end{center}

\begin{center}
    \vskip8pt
    \textsl{$^1$ Department of Applied Mathematics and Waterloo Centre for Astrophysics,\\ University of Waterloo, Waterloo, ON N2L 3G1, Canada}\\
    \vskip4pt
    \textsl{$^2$ Perimeter Institute for Theoretical Physics, Waterloo, ON N2L 2Y5, Canada}\\
    \vskip4pt
    \textsl{$^3$ Institute for Theory and Computation,\\ Harvard-Smithsonian Center for Astrophysics, Cambridge, MA 02138, USA}\\
	\vskip4pt
    \textsl{$^4$ Department of Physics, University of Maryland, College Park, MD 20742, USA}\\
    \vskip4pt
    \textsl{$^5$ Quantum Technology Center, University of Maryland, College Park, MD 20742, USA}
\end{center}

\vspace{1.2cm}

\hrule
\vspace{0.3cm}
\noindent {\bf Abstract}\\[0.1cm]
We construct explicit models of classical primordial standard clocks in an alternative to inflation, namely the slowly contracting ekpyrotic scenario. We study the phenomenology of massive spectator fields added to a state-of-the-art ekpyrotic model, with coupling functions that allow for these heavy fields to be classically excited while the background is slowly contracting. We perform numerical computations of the corrections to the scalar primordial power spectrum and compare with analytical estimates. Our full numerical results reveal so-called clock signals, sharp feature signals, as well as signals that link the two together. The models are found to predict oscillatory features that are resolutely different from what is calculated in inflation, and thus, such features represent unique fingerprints of a slowly contracting universe. This confirms the capability of primordial standard clocks to model-independently discriminate among very early universe scenarios.
\vskip10pt
\hrule
\vskip10pt

\end{titlepage}


\thispagestyle{empty}
\tableofcontents
\newpage
\pagenumbering{arabic}
\setcounter{page}{1}



\section{Introduction}

Despite the tremendous success of the `hot big bang' model, the enigmatic nature of its initial conditions requires a primordial epoch that is distinct from the current standard cosmological model. Several qualitatively different scenarios (rapidly/slowly expanding/contracting; e.g., \cite{Guth:1980zm,Linde:1981mu,Albrecht:1982wi,Starobinsky:1980te,Sato:1980yn,Khoury:2001wf,Lehners:2007ac,Gasperini:1992em,Gasperini:2002bn,Wands:1998yp,Finelli:2001sr,Brandenberger:1988aj,Nayeri:2005ck,Brandenberger:2006pr,Boyle:2004gv,Tolley:2007nq,Khoury:2010gw,Khoury:2008wj,Khoury:2009my,Khoury:2011ii,Creminelli:2010ba,Joyce:2011ta,Hinterbichler:2011qk,Geshnizjani:2011dk,Hinterbichler:2014tka,Carrillo-Gonzalez:2020ejs,Afshordi:2016guo,Mylova:2021eld,Bramberger:2017tid,Boyle:2018tzc,Agrawal:2020xek, Turok:2022fgq,Turok:2023amx}) have been proposed to account for this primordial universe, with the inflationary scenario \cite{Guth:1980zm,Linde:1981mu,Albrecht:1982wi,Starobinsky:1980te,Sato:1980yn} being the most studied and perhaps the simplest and most unproblematic candidate. However, the key to distinguishing these candidates lies in observational evidence. Because each scenario may be realized by different models, among many properties of physics beyond the standard model that might be measured in future observations, the ones that could be used to directly pin down the type of scenario --- instead of selecting models across different scenarios --- are those distinctively and model-independently predicted by each scenario.

So far, there are two promising candidates for such observables. One is primordial gravitational waves \cite{Grishchuk:1974ny,Starobinsky:1979ty,Rubakov:1982df}, which are quantum fluctuations of the spacetime metric sourced by the background evolution of the primordial universe. These could be detected through the B-mode polarization in the cosmic microwave background (CMB) \cite{Seljak:1996ti,Seljak:1996gy,Kamionkowski:1996zd} and can be used to model-independently distinguish certain scenarios that have slowly evolving backgrounds from others with rapidly evolving backgrounds. Another candidate is primordial standard clock signals \cite{Chen:2011zf,Chen:2011tu,Chen:2015lza,Chen:2018cgg}, induced by classical or quantum oscillations of massive fields in the primordial universe. These so-called `clock signals' could be observed as scale- or shape-dependent oscillatory components in correlation functions of the primordial density perturbations \cite{Mukhanov:1981xt,Mukhanov:1990me}. The pattern of these oscillations directly measures the time dependence of the scale factor (see \cite{Chen:2016cbe,Chen:2016qce} for short reviews) --- arguably the defining property of the scenario of the primordial universe. 

In this paper, we will be interested in models of primordial standard clocks in which the signals are induced by the classical oscillations of massive fields, namely classical primordial standard clocks \cite{Chen:2011zf,Chen:2011tu,Chen:2012ja, Saito:2012pd,Battefeld:2013xka,Saito:2013aqa, Gao:2013ota,Noumi:2013cfa, Chen:2014joa,Chen:2014cwa,Huang:2016quc,Domenech:2018bnf,Braglia:2021ckn,Braglia:2021sun,Braglia:2021rej, Chen:2022vzh}. Full models of classical primordial standard clocks have so far only been explored within the context of the inflationary scenario \cite{Chen:2014joa,Chen:2014cwa,Braglia:2021ckn,Braglia:2021sun,Braglia:2021rej}. While it is possible to derive some model-independent properties of primordial standard clocks without delving into the details of model constructions \cite{Chen:2011zf,Chen:2011tu}, the examination of complete models has proven to be valuable. Such studies not only provide examples with comprehensive details that supplement the model-independent properties, but also increase the efficiency of data comparison.
Moreover, to compare with data, it would be important to not just use the inflationary predictions but also predictions from different scenarios.
To address this gap, our paper aims to construct and analyze full models of classical primordial standard clocks for an alternative scenario, namely the ekpyrotic scenario \cite{Khoury:2001wf,Lehners:2007ac}. This scenario posits a primordial universe that undergoes a slowly contracting phase before experiencing a big bounce and transitioning into standard big bang cosmology.

Classical primordial standard clock signals naturally fall within a broader category of beyond-the-standard-model signals known as primordial features (see \cite{Chen:2010xka,Chluba:2015bqa,Slosar:2019gvt,Achucarro:2022qrl} for reviews). Primordial features are components of the density perturbations that exhibit significant scale dependence. Furthermore, full models of classical primordial standard clocks generate different types of primordial features\footnote{For examples of different scale-dependent features in ekpyrotic and other non-inflationary scenarios, see \cite{Chen:2011zf,Chen:2011tu,Chen:2012ja,Chen:2014cwa,Raveendran:2018yyh,Chen:2018cgg,Domenech:2020qay}.} in addition to the \emph{clock signal}, namely the \emph{sharp feature signal} induced directly by the feature that excites the oscillation of the massive field, as well as its connection with the clock signal in terms of the signal profile. These include some highly model-dependent aspects, or possibly other unforeseen model-independent aspects, that can only be learned through full models.

\paragraph*{Outline} This paper is organized as follows. We start by reviewing the `vanilla' ekpyrotic scenario in section \ref{sec:vanillaEk}, namely ekpyrotic cosmology without massive spectator fields. We focus the presentation on the model that has become the canonical example in terms of generating a nearly scale-invariant power spectrum of scalar perturbations. In section \ref{sec:models}, we introduce a massive field and couple it to `vanilla' ekpyrosis. We assume that, during the course of the background evolution, a sharp feature excites classical oscillation of the massive field. We give several examples of such sharp features, and we gain some analytical understanding on the background evolution of those fields and their effects on the other background quantities. We then solve the background equations numerically for these examples with different coupling functions and different model parameters, thus exploring the phenomenology of massive spectator fields in ekpyrotic cosmology as thoroughly as possible. We introduce linear cosmological perturbations in section \ref{sec:signals} and derive the perturbed action and equations of motion. Those equations are then solved analytically using the in-in formalism, as well as numerically. In doing so, we compute the predicted features in the scalar power spectrum for different standard clock implementations. We end with our conclusions and a discussion in section \ref{sec:conclusions}.

\paragraph*{Summary of the main results} We provide a condensed summary of our results here for the busy reader:
\begin{itemize}
    \item A period of slow contraction (ekpyrosis) can support a massive spectator field, i.e., it is possible to add a massive field to the theory and excite it such that the background evolution is not disrupted.
    \item To classically excite a massive field in ekpyrosis, the direct coupling between the ekpyotic scalar and the massive scalar has to be field dependent and go to zero in the past. Furthermore, the more stable background solutions are found when the coupling goes to zero in the future as well since otherwise it can change the character of the ekpyrotic scalar such that the massive field eventually dominates the background.
    \item If the coupling is quickly turned on and off again, the resulting corrections to the scalar two-point function match the analytical templates of a classical primordial standard clock, in particular the clock signal \cite{Chen:2011zf,Chen:2011tu}, which clearly separates from the sharp feature signal and which is very distinct from what one obtains in inflation. This becomes a clear model-independent fingerprint of ekpyrotic cosmology.
    \item If the coupling remains for an extended period of time or if it is turned on and off rather slowly, the clock signal gets entangled in a more complicated way with the sharp feature signal, the latter of which is not a model-independent signature of ekpyrosis.
\end{itemize}

\section{Review of `vanilla' ekpyrotic cosmology}\label{sec:vanillaEk}

Let us begin by reviewing the basics of ekpyrotic cosmology (more can be found in reviews such as \cite{Lehners:2008vx,Lehners:2011kr}). Originally motivated as a modulus in a higher-dimensional brane theory \cite{Khoury:2001wf,Khoury:2001bz}, the ekpyrotic scalar field $\phi$ has a steep, negative potential, which we parametrize\footnote{One would like to derive the effective theory from a more fundamental, ultraviolet-complete one, such as string theory. As for inflation, this might be challenging, though a priori not impossible --- see, e.g., \cite{Lehners:2018vgi,Bernardo:2021wnv,Shiu:2023yzt} and references therein. We leave this issue aside in this work.} as
\begin{equation}
	V(\phi)=-V_0e^{-\sqrtb{2/p}\phi}\,,\label{eq:potential}
\end{equation}
with $V_0>0$ and $0<p<1/3$. The field has a canonical kinetic term and is minimally coupled to Einstein gravity, and at the background level it is responsible for driving a smoothing phase of slowly contracting cosmology. At the perturbation level, the ekpyrotic field generates unenhanced blue spectra (and thus unobservable) of adiabatic scalar fluctuations and tensor fluctuations.

The theory possesses a second scalar field, a massless field kinetically coupled to the ekpyrotic field, which is solely responsible for the generation of the structures of our universe. This spectator field $\chi$ is non-dynamical at the background level, but it generates a nearly scale-invariant power spectrum of entropy (a.k.a.~isocurvature) fluctuations on super-horizon scales, which are later converted into curvature perturbations. The action of the full theory reads \cite{Qiu:2013eoa,Li:2013hga,Fertig:2013kwa,Ijjas:2014fja}
\begin{equation}
	S=\int\dd^4x\,\sqrtb{-g}\left(\frac{R}{2}-\frac{1}{2}(\partial\phi)^2-V(\phi)-\frac{1}{2}\Omega^2(\phi)(\partial\chi)^2\right)\label{eq:actionvanilla}
\end{equation}
in units of $c=\hbar=8\pi G_\mathrm{N}=1$, where $R$ is the Ricci scalar of the metric tensor $g_{\mu\nu}$ with determinant $g$. Also, $(\partial\phi)^2$ is shorthand notation for $g^{\mu\nu}\partial_\mu\phi\partial_\nu\phi$. The kinetic coupling function $\Omega^2(\phi)$ is taken to have a similar functional form to the potential,
\begin{equation}
	\Omega^2(\phi)=e^{-\sqrtb{2/b}\phi}\,,\label{eq:OmegaCoupling}
\end{equation}
with some parameter $0<b<1/3$, $b\approx p$.

On a flat, homogeneous and isotropic cosmological background,
\begin{equation}
	g_{\mu\nu}\dd x^\mu\dd x^\nu=-\dd t^2+a(t)^2\delta_{ij}\dd x^i\dd x^j\,,
\end{equation}
the field equations read
\begin{subequations}
\begin{align}
	&3H^2=\frac{1}{2}\dot\phi^2+V(\phi)+\frac{1}{2}\Omega^2(\phi)\dot\chi^2\,,\\
	&2\dot H=-\dot\phi^2-\Omega^2(\phi)\dot\chi^2\,,\\
	&\ddot\phi+3H\dot\phi+V_{,\phi}=\Omega\Omega_{,\phi}\dot\chi^2\\
	&\ddot\chi+3H\dot\chi=-2\frac{\Omega_{,\phi}}{\Omega}\dot\phi\dot\chi\,,\label{eq:chibackEOM}
\end{align}
\end{subequations}
where $H\equiv\dot a/a$ defines the Hubble parameter, a dot is a derivative with respect to $t$, and a comma denotes a partial derivative. At this point, $\phi=\phi(t)$ and $\chi=\chi(t)$, i.e., the scalar fields are taken to be homogeneous. An immediate observation from \eqref{eq:chibackEOM} is that if $\chi$ is initially unexcited, the spectator field will stably remain so with $\dot\chi=0$ throughout \cite{Ijjas:2014fja,Levy:2015awa}. Then, the scale factor and ekpyrotic field follow the following scaling solution,
\begin{equation}
	a(t)\propto(-t)^p\,,\qquad\phi(t)=\sqrtb{2p}\ln\left((-t)\sqrtb{\frac{V_0}{p(1-3p)}}\right)\,,\label{eq:backScaling}
\end{equation}
for $t<0$. The energy density of the ekpyrotic field scales as $\rho_\phi\equiv\dot\phi^2/2+V(\phi)\propto a^{-2/p}$, hence it dominates over matter components (e.g., the energy density of the vacuum, of spatial curvature, dust, radiation, and anisotropies are proportional to $a^0$, $a^{-2}$, $a^{-3}$, $a^{-4}$, and $a^{-6}$, respectively). This suggests the ekpyrotic field smooths the universe through slow contraction \cite{Erickson:2003zm,Lidsey:2005wr}, even if the universe was initially highly anisotropic and inhomogeneous. This has been carefully studied using full numerical relativity techniques \cite{Garfinkle:2008ei,Cook:2020oaj,Ijjas:2020dws,Ijjas:2021gkf,Ijjas:2021wml,Ijjas:2021zyf,Ijjas:2022qsv,Kist:2022mew,Ijjas:2023dnb,Ijjas:2024oqn}, confirming the smoothing power and attractor nature of the theory.
Often, it shall be useful to characterize the background evolution by its equation of state\footnote{An equation of state is usually a property of a fluid, such as the ratio of a perfect fluid's pressure to its energy density, often denoted by $w$. In cosmology, though, if the universe is dominated by a single perfect fluid with equation of state parameter $w$, then one can show through the Friedmann equations that $\epsilon=3(1+w)/2$ --- a simple linear relation. Therefore, while $\epsilon$ characterizes the background evolution of the universe, it is naturally related to the matter's equation of state through general relativity. When the matter sector consists of several components, we shall say that $\epsilon$ characterizes the `effective equation of state' of the system.}, which we define as $\epsilon\equiv -\dot H/H^2$. For the scaling solution \eqref{eq:backScaling}, this means $\epsilon=1/p$, and ekpyrosis corresponds to $\epsilon>3$ since $0<p<1/3$, though smoothing is more easily achieved with $\epsilon\gg 3$ (i.e., $0<p\ll 1$). This is analogous to inflation, where $\epsilon$ is the slow-roll parameter, satisfying $0<\epsilon<1$, though again typically one needs $\epsilon\ll 1$ (i.e., $p\gg 1$).

Let us review the properties of the spectator field $\chi$ since it is the one solely responsible for the generation of a nearly scale-invariant power spectrum of scalar perturbations. We introduce a linear perturbation $\delta\chi(t,\mathbf{x})\equiv\chi(t,\mathbf{x})-\chi(t)$, but recalling the background $\chi(t)$ remains at some fixed irrelevant constant value, we may well take $\chi(t)=0$, in which case the scalar field is purely a perturbation, i.e., $\delta\chi(t,\mathbf{x})=\chi(t,\mathbf{x})$. Thus, in what follows we shall simply denote the entropy perturbation $\delta\chi$ by $\chi$. Then working in the spatially flat gauge of the metric, the $\chi$ part of the action \eqref{eq:actionvanilla} expanded to second order in cosmological perturbations reads (after solving for the Hamiltonian and momentum constraints to eliminate the metric perturbations of the lapse and shift, integrating by parts, and using the background equations of motion to simplify)
\begin{equation}
	S^{(2)}_s=\frac{1}{2}\int\dd^3x\,\dd t\,a^3\Omega^2(\phi)\left(\dot\chi^2-a^{-2}(\partial_i\chi)^2\right)\,.\label{eq:S2chivanilla}
\end{equation}
Defining the Mukhanov-Sasaki variable $v\equiv z\chi$ with $z\equiv a\Omega(\phi)$, transforming to Fourier space, going to conformal time with $\dd\tau\equiv a^{-1}\dd t$, and integrating by parts, the action can be equivalently written as
\begin{equation}
	S^{(2)}_s=\frac{1}{2}\int\dd^3k\,\dd\tau\left(v_k'^2+\frac{z''}{z}v_k^2-k^2v_k^2\right)\,,\label{eq:Ss2v}
\end{equation}
where a prime denotes a derivative with respect to $\tau$, and where $k$ denotes the Fourier wavenumber (the subscript $k$ on a perturbation variable indicates its Fourier transform). From this, we obtain the standard equation of motion
\begin{equation}
	v_k''+\left(k^2-\frac{z''}{z}\right)v_k=0\,.\label{eq:vsEOM}
\end{equation}
In conformal time, the scaling solution \eqref{eq:backScaling} is
\begin{equation}
	a(\tau)\propto(-(1-p)\tau)^{\frac{p}{1-p}}\,,\qquad\phi(\tau)=\sqrtb{2p}\ln\left((-(1-p)\tau)^{\frac{1}{1-p}}\sqrt{\frac{V_0}{p(1-3p)}}\right)\,,\qquad \tau<0\,,
\end{equation}
from which follows
\begin{equation}
    z=a\Omega(\phi)\propto\frac{1}{(-\tau)^n}\,,\qquad n=\frac{\sqrtb{p/b}-p}{1-p}\stackrel{b\approx p}{\approx}1\,,\label{eq:zbar}
\end{equation}
and thus,
\begin{equation}
	\frac{z''}{z}=\frac{n(n+1)}{(-\tau)^2}=\frac{\nu^2-1/4}{(-\tau)^2}\,,\qquad\nu=n+\frac{1}{2}\stackrel{n\approx 1}{\approx}\frac{3}{2}\,.\label{eq:zppozbar}
\end{equation}
The solution to the resulting ordinary differential equation, $v_k''+(k^2-(\nu^2-1/4)/\tau^2)v_k=0$, which asymptotes the usual Bunch-Davies initial quantum vacuum,
\begin{equation}\label{eq:BD}
    v_k(\tau)\stackrel{k\tau\to-\infty}{\simeq}\frac{1}{\sqrtb{2k}}e^{-ik\tau}\,,
\end{equation}
is expressed in terms of the Hankel function of the first kind as
\begin{equation}
	v_k(\tau)=-\frac{\sqrtb{-\pi\tau}}{2}H_\nu^{(1)}(-k\tau)\,,\label{eq:mode_func_canonical_v}
\end{equation}
up to an irrelevant phase. The resulting spectrum in the limit $k\tau\to 0^-$ can be checked to be
\begin{equation}\label{eq:powerspectrum_ekpyrotic}
    \mathcal{P}_s(k)\equiv\frac{k^3}{2\pi^2}|\chi_k|^2=\frac{k^3}{2\pi^2}\frac{|v_k|^2}{z^2}\sim k^{3-2\nu}\,,
\end{equation}
where one has to be careful to pick the growing mode $(-k\tau)^{-\nu}$ in the super-horizon expansion ($k\tau\to 0^-$) of the Hankel function. Therefore, scale invariance is obtained if $\nu=3/2$, so if $n=1$, hence if $b=p$. The spectral index is, in fact,
\begin{equation}
    n_s-1=3-2\nu=2(1-n)=2\frac{1-\sqrtb{p/b}}{1-p}\stackrel{p\ll 1}{\approx}2\left(1-\sqrtb{\frac{p}{b}}\right)\,,\label{eq:ns}
\end{equation}
hence a red tilt is obtained if $b<p$ (ever so slightly). In all numerical computation below, we shall set $b$ as a function of the numerical value for $p$ such that $n_s\approx 0.965$ is obtained according to \eqref{eq:ns}, namely $b/p\approx 0.966$.

It is important to stress at this point that the scalar power spectrum \eqref{eq:powerspectrum_ekpyrotic}, which has the desired amplitude and red tilt, represents the two-point correlation function of the entropy perturbation $\chi$. In other words, this is the power spectrum of an isocurvature perturbation, not of the curvature perturbation $\mathcal{R}$. What typically happens in ekpyrosis, though, is that there is some kind of potential barrier or reheating surface at an angle in $\chi$-$\phi$ field space, which has the effect of converting the isocurvature perturbations into curvature perturbations (see, e.g., \cite{Battefeld:2007st,Lehners:2007wc,Lehners:2008my,Lehners:2009qu,Lehners:2009ja,Lehners:2010fy,Fertig:2013kwa,Ijjas:2014fja,Fertig:2015ola,Fertig:2016czu,Raveendran:2018yyh,Ijjas:2020cyh,Ijjas:2021ewd} for detailed analyses of this process). There are different ways to model this conversion, but the generic outcome is that we may simply treat the entropy spectrum $\mathcal{P}_s(k)$ as the power spectrum of curvature perturbation $\mathcal{P}_\mathcal{R}(k)$ at the onset of standard big bang cosmology. This further rests on the assumption that it is possible to have a stable non-singular transition from contraction to expansion --- the bounce --- and that this transition does not affect the perturbations on large scales. Once again, a bounce can be modelled in different ways, but successful realizations confirm the tendency of the power spectra not to be affected by the bouncing physics (see, e.g., \cite{Xue:2013bva,Battarra:2014tga,Quintin:2015rta,Koehn:2015vvy,Fertig:2016czu,Quintin:2019orx,Kim:2020iwq}). We will mention this again in the discussion section, but in this work we do not model and analyze the conversion of isocurvature perturbations into curvature perturbations, nor the evolution of the perturbations through a non-singular bouncing phase. We shall essentially assume that the late-time, large-scale $\mathcal{P}_s(k)$ obtained at the end of ekpyrosis can be mapped onto $\mathcal{P}_\mathcal{R}(k)$ at the onset of radiation-dominated expansion.

\section{Standard clock models in ekpyrotic cosmology}\label{sec:models}

\subsection{Setup and analytical background solutions}\label{sec:backAnalytical}

Heavy fields should exist in ultraviolet completions of any primordial universe models. Similar to inflation, we assume that the physics that gives rise to the low-energy effective theory of the ekpyrotic cosmology is complicated enough so that there may be sharp features in the potential or in field space that could excite classical oscillations of some of these heavy fields.
It is well known that, during inflation, the amplitude of heavy field oscillations redshifts (at the same rate as pressureless matter), so inflationary cosmology can easily support such small disturbances. However, the consequences in a contracting cosmology are much less known and have not been well studied.
So, the very first question we shall address in this work is whether a phase of a slowly contracting cosmology (i.e., ekpyrosis) can support the existence of oscillating heavy fields. In other words, we shall explore different field theory models, which incorporate a massive spectator field that will be classically excited in the contracting phase, and analyze the background evolution of all the fields at play. Cosmological perturbations will be addressed in the next section.

The general setup consists of general relativity and a multidimensional curved field space,
\begin{equation}
	S=\int\dd^4x\,\sqrtb{-g}\left(\frac{R}{2}-\frac{1}{2}\mathcal{G}_{IJ}(\Phi^K)g^{\mu\nu}\partial_\mu\Phi^I\partial_\nu\Phi^J-V(\Phi^I)\right) \,,
\end{equation}
where our field space consists of $\Phi^I=\{\phi,\chi,\sigma\}$, i.e., the ekpyrotic background field $\phi$, the entropy field $\chi$, and a new massive spectator scalar field $\sigma$ (there could be several massive fields, but we consider a single one for simplicity; this can be easily generalized). The potential $V$ and field-space metric $\mathcal{G}_{IJ}$ can a priori depend on all the fields at play. The action \eqref{eq:actionvanilla} is of this form, except it does not include any massive field $\sigma$. Taking \eqref{eq:actionvanilla} as our starting point, we shall consider our full potential and field-space metric such that the action has the form
\begin{align}
	S=\int\dd^4x\,\sqrtb{-g}\bigg(\frac{R}{2}&-\frac{1}{2}\big(1+\Xi(\phi)\sigma\big)(\partial\phi)^2-\frac{1}{2}\big(1+\Upsilon(\phi)\sigma\big)\Omega^2(\phi)(\partial\chi)^2-\frac{1}{2}(\partial\sigma)^2\nonumber\\
	&-V(\phi)-\frac{1}{2}m_\sigma^2\sigma^2\bigg)\,,\label{eq:actionGenModels}
\end{align}
where $\Omega^2(\phi)$ and $V(\phi)$ are the same ekpyrotic kinetic coupling \eqref{eq:OmegaCoupling} and potential \eqref{eq:potential} as before, and $m_\sigma$ denotes the mass of the spectator field $\sigma$. Two new coupling functions $\Xi(\phi)$ and $\Upsilon(\phi)$ are introduced, which shall control the strength of the direct couplings in the kinetic terms for $\phi$ (background) and $\chi$ (perturbations). The motivation for writing the couplings as $1+\Xi(\phi)\sigma$ and $1+\Upsilon(\phi)\sigma$ is that $\sigma$ should be a subdominant spectator field, hence its coupling to other fields should be at the level of a perturbative correction. It should thus be the case that $\Xi(\phi)\sigma$ and $\Upsilon(\phi)\sigma$ remain smaller than unity for the theory to remain weakly coupled.

Note that the above does not exhaust all possible ways of adding a massive field $\sigma$ to the theory, but it is a fairly simple approach, hence it is our focus to start. Additional possibilities shall be reserved for follow-up works. Likewise, beyond $\sigma$ being a scalar, the coupling to massive fields with non-zero spin could be the focus of future work.

Noting that the coupling $\Upsilon(\phi)$ does not affect the background scaling for $\chi$ (i.e., we can still set $\chi(t)=0$ and this remains stable throughout, though this always needs to be checked after the fact), we shall specify $\Upsilon(\phi)$ only in the next section that addresses perturbations and ignore $\chi$ altogether for the rest of this section. The background equations of motion become
\begin{subequations}\label{eq:backEOMsFull}
\begin{align}
	&3H^2=\frac{1}{2}\big(1+\Xi(\phi)\sigma\big)\dot\phi^2+\frac{1}{2}\dot\sigma^2+V(\phi)+\frac{1}{2}m_\sigma^2\sigma^2\,,\label{eq:FriedConstraint}\\
 	&2\dot H=-\big(1+\Xi(\phi)\sigma\big)\dot\phi^2-\dot\sigma^2\,,\\
 	&\big(1+\Xi(\phi)\sigma\big)(\ddot\phi+3H\dot\phi)+V_{,\phi}=-\frac{1}{2}\Xi_{,\phi}\sigma\dot\phi^2-\Xi(\phi)\dot\sigma\dot\phi\,,\\
 	&\ddot\sigma+3H\dot\sigma+m_\sigma^2\sigma=\frac{1}{2}\Xi(\phi)\dot\phi^2\,.\label{eq:backEOMsF1s}
\end{align}
\end{subequations}

At this point, we need to specify the form of $\Xi(\phi)$, and different couplings can be considered. 
Each case is an example of a sharp feature that excites the classical oscillation of the massive field $\sigma$.
The idea is that we wish $\sigma$ to be initially unexcited, hence the coupling should be zero at first and then acquire a constant value at some critical value in $\phi$-space (call it $\phi_0$) to trigger $\sigma$ dynamics. The constant coupling may or may not remain forever afterwards (it could stop after some other field value, $\phi_\mathrm{e}$), so we consider three cases,
\begin{subequations}\label{eq:Xi3}
\begin{align}
	\Xi_\mathrm{step}(\phi)&\approx\frac{1}{\varrho}\Theta(\phi_0-\phi)=\begin{cases}
                 1/\varrho & \phi\leq\phi_0\\
                 0 & \phi>\phi_0
                \end{cases}\,,\label{eq:Xistep}\\
	\Xi_\mathrm{plateau}(\phi)&\approx\frac{1}{\varrho}\Theta(\phi_0-\phi)\Theta(\phi-\phi_\mathrm{e})=\begin{cases}
                 1/\varrho & \phi_\mathrm{e}\leq\phi\leq\phi_0\\
                 0 & \mathrm{otherwise}
                \end{cases}\,,\label{eq:Xiplateau}\\
	\Xi_\mathrm{bump}(\phi)&\approx\frac{1}{\varrho}\Theta(\phi_0-\phi)\Theta(\phi-(1-\varepsilon)\phi_0)\approx\begin{cases}
                 1/\varrho & \phi\approx\phi_0\\
                 0 & \mathrm{otherwise}
                \end{cases}\,,\qquad 0<\varepsilon\ll 1\,,\label{eq:Xibump}
\end{align}
\end{subequations}
written in terms of the Heaviside step function $\Theta$, whose convention is clear from the above.
Note that $\phi$ evolves from large to small field values according to the scaling solution \eqref{eq:backScaling} [recall the early- and late-time limits correspond to $t\to-\infty$ and $t\to 0^-$, respectively].
In every case, when the coupling is $\Xi\approx 1/\varrho$, one can imagine the trajectory in curved field space as being bent with curvature radius $\varrho$: the smaller the radius, the tighter the bending, the bigger the coupling; and vice versa.

In all three cases, we set $\sigma=\dot\sigma=0$ initially (i.e., at some field value $\phi_\mathrm{ini}>\phi_0$). Thus for this first interval, $\phi_\mathrm{ini}\geq\phi>\phi_0$, we recover the scaling solution \eqref{eq:backScaling}. From here on, we denote by a bar the `unperturbed' scaling solution \eqref{eq:backScaling} [i.e., $\bar a$, $\bar\phi$, $\bar H$, etc., are given according to \eqref{eq:backScaling}], and they indicate the background solutions when there is no massive field contribution to the dynamics. Once $\phi$ is near $\phi_0$, the coupling becomes $\Xi\approx 1/\varrho$, and we see from \eqref{eq:backEOMsF1s} that the massive field will receive an initial (centrifugal) force, thus triggering $\sigma$ oscillations at the bottom of the $m_\sigma^2\sigma^2$ potential (perpendicular to the adiabatic direction given by the $\phi$ trajectory). Thinking of $\sigma(t)$ as being a subdominant contribution to $\phi(t)\equiv\bar\phi(t)+\Delta\phi(t)$ and $H(t)\equiv\bar H(t)+\Delta H(t)$, we can make a Born approximation to evaluate the $\sigma$ equation of motion when $\Xi\approx 1/\varrho$ as
\begin{equation}
	\ddot\sigma+3\bar H\dot\sigma+m_\sigma^2\sigma\approx\frac{1}{2\varrho}\dot{\bar\phi}^2\Theta(\phi_0-\phi)\qquad\implies\qquad\ddot\sigma+\frac{3p}{t}\dot\sigma+m_\sigma^2\sigma\approx\frac{p}{\varrho t^2}\Theta(t-t_0)\,,\label{eq:sigmanonhomo}
\end{equation}
where $\bar\phi(t_0)=\phi_0$ defines $t_0$. The above assumes the step coupling function \eqref{eq:Xistep}, but it is useful to first analyze this case and then describe the plateau and bump cases \eqref{eq:Xiplateau}--\eqref{eq:Xibump} later.
The solution to the above equation can be expressed as the sum of a homogeneous solution $\bar{\sigma}(t)$ and a particular solution $\sigma_\mathrm{p}(t)$, i.e., $\sigma(t)=\bar{\sigma}(t)+\sigma_\mathrm{p}(t)$.

The two linearly independent solutions to the homogeneous equation
\begin{equation}
	\ddot{\bar{\sigma}}+\frac{3p}{t}\dot{\bar{\sigma}}+m_\sigma^2\bar{\sigma}=0
\end{equation}
are expressed in terms of the Bessel functions of the first and second kind as $\bar\sigma_{1}(t)=(-t)^\alpha J_\alpha(-m_\sigma t)$ and $\bar\sigma_{2}(t)=(-t)^\alpha Y_\alpha(-m_\sigma t)$, hence
\begin{equation}
	\bar{\sigma}(t)=c_1\bar\sigma_1(t)+c_2\bar\sigma_2(t)=(-t)^\alpha\left(c_1J_\alpha(-m_\sigma t)+c_2Y_\alpha(-m_\sigma t)\right)\,,\qquad\alpha\equiv\frac{1-3p}{2}\stackrel{p\ll 1}{\approx}\frac{1}{2}\,,\label{eq:Besselapproxsigma}
\end{equation}
with integration constants $c_1$ and $c_2$. At early times, when the massive field is effectively heavy, it oscillates according to
\begin{equation}
	\bar{\sigma}(t)\stackrel{m_\sigma|t|\gg 1}{\simeq}\frac{1}{(-m_\sigma t)^{3p/2}}\left(c_1\sin\left(-m_\sigma t+\frac{3\pi p}{4}\right)-c_2\cos\left(-m_\sigma t+\frac{3\pi p}{4}\right)\right)\,,\label{eq:sigmabarearlytime}
\end{equation}
where a constant factor has been absorbed in the integration constants. (This is the prototypical background evolution of heavy fields used for primordial standard clocks \cite{Chen:2011zf,Chen:2011tu,Chen:2014joa,Chen:2014cwa}.) Oppositely at late times, when the massive field is effectively light, it scales according to
\begin{equation}
	\bar{\sigma}(t)\stackrel{m_\sigma|t|\ll 1}{\simeq}c_1+c_2(-m_\sigma t)^{1-3p}\,,\label{eq:sigmabarlatetime}
\end{equation}
where again the integration constants have been rescaled. Importantly, the second term above is a decaying mode, hence the homogeneous solution is asymptotically constant. The contribution of the homogeneous solution to the massive field energy density $\rho_\sigma\equiv\dot\sigma^2/2+m_\sigma^2\sigma^2/2$ in the different regimes can be expressed as
\begin{subequations}
\begin{align}
	\rho_{\bar{\sigma}}&\stackrel{m_\sigma|t|\gg 1}{\sim}1/\bar{a}^{3}+\textrm{(subdominant~oscillations)}\,,\\
	\rho_{\bar{\sigma}}&\stackrel{m_\sigma|t|\ll 1}{\sim}1/\bar{a}^{6}\,.
\end{align}
\end{subequations}
This confirms that the homogeneous solution to the massive field remains a subdominant contribution to the total energy density at all times, which is rather dominated by the ekpyrotic field with $\rho_\phi\sim 1/\bar{a}^{2/p}$, $0<p<1/3$.

When there is a constant coupling $\Xi\approx 1/\varrho$ present in the model, there is an additional solution to the non-homogeneous equation \eqref{eq:sigmanonhomo},
\begin{equation}
    \sigma_\mathrm{p}(t)=\bar\sigma_2(t)\Sigma_1(t)-\bar\sigma_1(t)\Sigma_2(t)\,,
\end{equation}
where
\begin{equation}
    \Sigma_1(t)=\frac{p}{\varrho}\int^t\dd\tilde t\,\frac{\bar\sigma_1(\tilde t)}{\tilde t^2W(\tilde t)}\,,\qquad\Sigma_2(t)=\frac{p}{\varrho}\int^t\dd\tilde t\,\frac{\bar\sigma_2(\tilde t)}{\tilde t^2W(\tilde t)}\,,
\end{equation}
together with the Wronksian
\begin{equation}
    W(t)\equiv\bar\sigma_1(t)\dot{\bar\sigma}_2(t)-\bar\sigma_2(t)\dot{\bar\sigma}_1(t)=-\frac{2}{\pi}(-t)^{-3p}\,,
\end{equation}
hence
\begin{equation}
    \Sigma_1(t)=-\frac{\pi p}{2\varrho}\int^t\dd\tilde t\,\frac{J_\alpha(-m_\sigma\tilde t)}{(-\tilde t)^{1+\alpha}}=-\frac{\pi pm_\sigma^\alpha}{2^{2+\alpha}\varrho}G_{1,3}^{2,0}\Big(y^2\Big|\begin{array}{c}
        1 \\
        0,0,-\alpha
    \end{array}\Big)\,,\qquad y\equiv-\frac{1}{2}m_\sigma t\,,
\end{equation}
and
\begin{align}
    \Sigma_2(t)&=\frac{\pi p}{2\varrho}\int^t\dd\tilde t\,\frac{Y_\alpha(-m_\sigma\tilde t)}{(-\tilde t)^{1+\alpha}}\nonumber\\
    &=-\frac{\pi pm_\sigma^\alpha}{2^{2+\alpha}\varrho\sin(\pi\alpha)}\bigg(\frac{{}_1\!F_2(-\alpha;1-\alpha,1-\alpha;-y^2)}{\alpha^2\Gamma(-\alpha)y^{2\alpha}}+\cos(\pi\alpha)\,G_{1,3}^{2,0}\Big(y^2\Big|\begin{array}{c}
        1 \\
        0,0,-\alpha
    \end{array}\Big)\bigg)\,.
\end{align}
In the above, the integrals of the Bessel functions involve special functions such as the Meijer $G$-function and the generalized hypergeometric function ${}_pF_q$.
Putting everything together, we find
\begin{align}
    \sigma_\mathrm{p}(t)=\frac{\pi p}{4\varrho\sin(\pi\alpha)}\bigg( & \frac{J_\alpha(-m_\sigma t){}_1\!F_2(-\alpha;1-\alpha,1-\alpha;-y^2)}{\alpha^2\Gamma(-\alpha)y^\alpha}\nonumber\\
    &+e^{-i\pi\alpha}y^\alpha J_{-\alpha}(+m_\sigma t)\,G_{1,3}^{2,0}\Big(y^2\Big|\begin{array}{c}
        1 \\
        0,0,-\alpha
    \end{array}\Big)\bigg)\,,\label{eq:sigmapfull}
\end{align}
where the simplification $\cos(\pi\alpha)J_\alpha(-m_\sigma t)-\sin(\pi\alpha)Y_\alpha(-m_\sigma t)=e^{-i\pi\alpha}J_{-\alpha}(+m_\sigma t)$ was used.

It is more enlightening to directly express the particular solution in the different limits of interest:
\begin{equation}
	\sigma_\mathrm{p}(t)\stackrel{m_\sigma|t|\gg 1}{\simeq}\frac{p\sqrt{\pi}}{2^{\alpha+3/2}\varrho}\Gamma(-\alpha)\sec(\frac{3\pi p}{2})\frac{1}{(-m_\sigma t)^{3p/2}}\sin(-m_\sigma t+\frac{3\pi p}{4})\,,\label{eq:sigmapearlytime}
\end{equation}
which is of the same form as the homogeneous solution in the same early-time limit [cf.~\eqref{eq:sigmabarearlytime}], i.e., oscillatory in $t$ with frequency $m_\sigma$ and amplitude growing as $1/(-m_\sigma t)^{3p/2}$; and in the late-time limit,
\begin{equation}
	\sigma_\mathrm{p}(t)\stackrel{m_\sigma|t|\ll 1}{\simeq}-\frac{p}{\varrho(1-3p)}\ln(-m_\sigma t)+C\,,\label{eq:sigmaplatetime}
\end{equation}
which is slowly blowing up as $m_\sigma t\to 0^-$.
The constant $C$ is found to be approximately equal to $-\gamma_\mathrm{E}p/\varrho$ in the small-$p$ limit, where $\gamma_\mathrm{E}$ is the Euler-Mascheroni constant.
This latter scaling solution is comparable to that of the ekpyrotic field since its energy density is $\rho_{\sigma_\mathrm{p}}\sim t^{-2}\sim 1/\bar{a}^{2/p}\sim\rho_\phi$; however, it would never appear to dominate over $\phi$ within the regime of validity of the approximations made here.

Imposing the initial conditions that $\sigma=\dot\sigma=0$ right up to the critical time $t_0$ determines the integration constants $c_1$ and $c_2$ in \eqref{eq:Besselapproxsigma}, and the full solution is then found to be
\begin{equation}
    \sigma(t)=\Big(\big(\Sigma_2(t_0)-\Sigma_2(t)\big)\bar\sigma_1(t)+\big(\Sigma_1(t)-\Sigma_1(t_0)\big)\bar\sigma_2(t)\Big)\Theta(t-t_0)\,.\label{eq:sigmaAnalyticalFull}
\end{equation}
Interestingly, in the limits where both $m_\sigma t\to-\infty$ and $m_\sigma t_0\to-\infty$, we find that $\sigma(t)\simeq 0$ at all times. This is because \eqref{eq:sigmapearlytime} depicts the same oscillatory behavior as \eqref{eq:sigmabarearlytime} at early times, so the two solutions must be precisely out of phase at critical time $t_0$ in order to match the initial conditions --- in terms of \eqref{eq:sigmabarearlytime} and \eqref{eq:sigmapearlytime}, the initial conditions set $c_2=0$ and $c_1=-\sigma_\mathrm{p}(t_0)/\bar\sigma_1(t_0)$, but then this means $\sigma(t)=c_1\bar\sigma_1(t)+c_2\bar\sigma_2(t)+\sigma_\mathrm{p}(t)=0$ for all $t\geq t_0$. As such, even after the critical time the massive field remains at rest. This exact cancellation disappears, however, as soon as one goes away from the limit $m_\sigma t\to-\infty$, though it is hard to find a good approximation to the sub-leading contribution. Numerical examples are shown later (see Fig.~\ref{fig:2Comp}).

With the knowledge of the general $\sigma(t)$ solutions (from an approximate, analytical perspective), let us describe the resulting expected dynamics of the different scenarios defined by the coupling $\Xi(\phi)$ in \eqref{eq:Xi3}. In the case of a `step' [eq.~\eqref{eq:Xistep}], $\sigma$ is initially at rest at the bottom of its potential since $\Xi(\phi)=0$ when $\phi>\phi_0$. Once $\phi$ drops below $\phi_0$, then $\Xi(\phi)\approx 1/\varrho$, and the $\sigma$ equation of motion becomes \eqref{eq:sigmanonhomo}, hence the resulting full solution is the sum of the respective homogeneous and particular solutions [cf.~\eqref{eq:sigmaAnalyticalFull}]. Assuming that near $\phi\approx\phi_0$ we are in a regime where $m_\sigma|t|\gg 1$, oscillations in $\sigma$ may be suppressed, and the exact behavior is given by a combination of Bessel, generalized hypergeometric, and Meijer functions, which is more easily understood numerically in the next subsection. At late times, the homogeneous solution $\bar\sigma$ asymptotes a constant and effectively becomes massless (with stiff equation of state), but the particular solution $\sigma_\mathrm{p}$ slowly grows without bound, with an energy density growing at the same rate as that of the dominant background ekpyrotic field. In itself, this implies that the massive field should not disrupt the background driven by the ekpyrotic field $\phi$, but rather it should effectively become an ekpyrotic field of its own (with same effective equation of state as $\phi$); $\sigma$ would thus contribute to the total `ekpyrotic' energy density, yet it would still be a subdominant component. The caveat, though, is that this assumes no backreaction of the other fields, and as we will see numerically, the growth of the massive field induces corrections to the background, which may in turn disrupt the ekpyrotic domination and rather lead to $\epsilon \to 3$.

We can understand the above behavior from a qualitative perspective by recalling the curved field space picture, where a constant coupling $\Xi\approx 1/\varrho$ creates a constant-radius bend to the adiabatic $\phi$ trajectory as it rolls down its potential, while perpendicular to this trajectory is the massive field potential direction. Because of the bending in $\phi$, the perpendicular direction for $\sigma$ receives a centrifugal force, and at late times, together with the Hubble anti-friction, the massive field is pushed always higher up its potential and has an energy density scaling similar to the adiabatic ekpyrotic field. Interestingly, this is analogous to what would happen in a similar situation in inflation. Indeed, we could imagine $\phi$ being a slowly rolling inflaton, with sudden constant-radius bend in its trajectory, perpendicular to which one has an $m_\sigma^2\sigma^2$ massive field potential. In such a case, after the sharp transition, the massive field would be excited and start oscillating, though at some shifted height on its potential --- this averaged height would remain constant, though, due to a balance between the centrifugal force and the massive field potential. Due to Hubble friction in such a case, the oscillations would damp down at late times, and one would be left with the massive field staying at some non-zero constant value on its potential (i.e., away from the minimum), effectively acting as a cosmological constant, thus contributing to the inflationary background. Note in this case the massive field would never severely backreact onto the inflaton and disrupt the inflationary background.

Back to our ekpyrotic scenario, we can then consider the case of a plateau coupling [given by eq.~\eqref{eq:Xiplateau}]. We recover the same dynamics as for the step coupling, except that $\Xi$ returns to $0$ at a later field value $\phi_\mathrm{e}$. Thus, once $\phi<\phi_\mathrm{e}$, there is no external force term left in the $\sigma$ equation of motion, and we only recover the homogeneous solution \eqref{eq:Besselapproxsigma}, which is oscillating (effectively pressureless) at early times [$m_\sigma|t|\gg 1$; eq.~\eqref{eq:sigmabarearlytime}] and frozen and asymptotically constant (effectively massless) at late times [$m_\sigma|t|\ll 1$; eq.~\eqref{eq:sigmabarlatetime}]. The early-time regime for the massive field in such a case (though after $\phi<\phi_\mathrm{e})$ recovers the prototypical massive field background oscillations of a primordial standard clock \cite{Chen:2011zf,Chen:2011tu,Chen:2014joa,Chen:2014cwa}.

In the case of a bump [eq.~\eqref{eq:Xibump}], we essentially only have the homogeneous solution throughout. Indeed, the coupling is non-zero in a very small region in field space, so it only acts as a trigger for the $\sigma$ dynamics --- $\sigma$ then acts as a massive primordial standard clock field as long as $m_\sigma|t|\gg 1$. In the case of a step function for $\Xi$ that is nearly instantaneously `turned on and off,' the particular solution \eqref{eq:sigmapfull} would simply determine the initial conditions for the homogeneous solution \eqref{eq:Besselapproxsigma}. In a more realistic situation, the coupling is not instantaneous, and we need to resort to numerical techniques to evaluate the full solution. Qualitatively, viewed as a curved field space, the adiabatic trajectory effectively undergoes a sharp turn in such a situation, with a straight trajectory before and after the turn. In contrast, in the plateau scenario, the trajectory is straight, then it curves with a constant radius for some field range $[\phi_\mathrm{e},\phi_0]$, before straightening again; and in the case of a step, the trajectory is first straight and curved with constant radius forever afterwards. When the trajectory is curved, the massive field effectively feels a centrifugal force.

In the evolution of the massive field $\sigma$, the freezing time that separates the regimes where $\sigma$ oscillates (when $m_\sigma |t|\gg 1$) and when it does not (when $m_\sigma |t|\ll 1$) can be quantified by the horizon-exit time $t_{\textrm{h-e}}$, defined by $m_\mathrm{h}(t_{\textrm{h-e}})=m_\sigma$, where the notions of horizons and the definition of $m_\mathrm{h}(t)$ can be found in Appendix \ref{app:mass-horizon}.
As shown there,
we find a horizon-exit time
\begin{equation}
	t_{\textrm{h-e}}=-\frac{1-p}{m_\sigma}\stackrel{p\ll 1}{\simeq}-\frac{1}{m_\sigma}\,,\label{eq:th}
\end{equation}
where the last approximation holds for slowly contracting backgrounds. This confirms $m_\sigma|t|=1$ is a good divider for the early and late-time limits.

To support the above descriptions and analytical estimates, we turn to solving the full set of coupled background equations numerically.

\subsection{Numerical background solutions}

\subsubsection{Methodology}\label{sec:backNumMethod}

In order to solve the full background dynamics numerically, we first reexpress the coupling function for the three scenarios of \eqref{eq:Xi3} as smooth functions. Indeed, while \eqref{eq:Xi3} are expressed using the Heaviside function, this is unrealistic beyond an approximation, and in order to use numerical techniques, we shall replace the Heaviside function by smooth variants. In fact, we consider the following parametrizations,
\begin{subequations}
\begin{align}
	\Xi_\mathrm{step}(\phi)&=\frac{1}{2\varrho}\left(1-\tanh\left(\frac{\phi-\phi_0}{\delta}\right)\right)\,,\label{eq:Xisteptanh}\\
	\Xi_\mathrm{plateau}(\phi)&=\frac{1}{2\varrho}\left(\tanh\left(\frac{\phi-\phi_\mathrm{e}}{\delta}\right)-\tanh\left(\frac{\phi-\phi_0}{\delta}\right)\right)\,,\label{eq:Xiplateautanh}\\
	\Xi_\mathrm{bump}(\phi)&=\frac{1}{\varrho}\exp\left(-\frac{(\phi-\phi_0)^2}{\delta^2}\right)\,,\label{eq:Xibumpexp}
\end{align}
\end{subequations}
where in each case $\delta$ is a real positive constant controlling the sharpness of the transitions at $\phi_0$ and $\phi_\mathrm{e}$.

Since the time scales over which we wish to solve the background equations can span a large number of orders of magnitude, it is useful to perform a change of time variable. A common `logarithmic' time function is the $e$-folding number, whose definition we take to be
\begin{equation}
 \dd\mathcal{N}\equiv\dd\ln(a|H|)\,.
\end{equation}
This is a more appropriate generalization of the inflationary $e$-folding time $\dd N\equiv\dd\ln a$, especially for contracting alternatives to inflation. Indeed, in a slowly contracting universe, the scale factor barely changes, hence $N\sim\ln a$ is not representative. Rather, the comoving wavenumber $k$ of fluctuating modes that later correspond to the CMB spans a range in $a|H|$, and here it is $H$ that evolves over many orders of magnitude (as opposed to inflation where it is the scale factor that spans many orders of magnitude, while $H$ is approximately constant).
We further make use of the effective equation of state parameter $\epsilon\equiv -\dot H/H^2$. In inflation, this is the usual slow-roll parameter, which is small and approximately constant. In vanilla ekpyrosis, it is also approximately constant, but large. Indeed, the expected scaling solution from \eqref{eq:backScaling} is $\bar\epsilon=-\dot{\bar H}/\bar H^2=1/p>3$.

Note that we can relate derivatives with respect to $t$ to derivatives with respect to $\mathcal{N}$ through the chain rule and the identity $\dot{\mathcal{N}}=H(1-\epsilon)$.
The equations of motion \eqref{eq:backEOMsFull} can thus be reexpressed as
\begin{align}
	&(3-\epsilon)H^2=V(\phi)+\frac{1}{2}m_\sigma^2\sigma^2\,,\nonumber\\
 	&\frac{\epsilon}{(1-\epsilon)^2}=\frac{1}{2}\big(1+\Xi(\phi)\sigma\big)(\phi_{,\mathcal{N}})^2+\frac{1}{2}(\sigma_{,\mathcal{N}})^2\,,\nonumber\\
 	&\big(1+\Xi(\phi)\sigma\big)\left(\phi_{,\mathcal{N}\mathcal{N}}+\left(\frac{3-\epsilon-\epsilon_{,\mathcal{N}}}{1-\epsilon}\right)\phi_{,\mathcal{N}}\right)+\frac{V_{,\phi}}{(1-\epsilon)^2H^2}=-\frac{1}{2}\Xi_{,\phi}\sigma(\phi_{,\mathcal{N}})^2-\Xi(\phi)\sigma_{,\mathcal{N}}\phi_{,\mathcal{N}}\,,\nonumber\\
 	&\sigma_{,\mathcal{N}\mathcal{N}}+\left(\frac{3-\epsilon-\epsilon_{,\mathcal{N}}}{1-\epsilon}\right)\sigma_{,\mathcal{N}}+\frac{m_\sigma^2\sigma}{(1-\epsilon)^2H^2}=\frac{1}{2}\Xi(\phi)(\phi_{,\mathcal{N}})^2\,.\label{eq:backEOMsM1efolds}
\end{align}
Setting our initial conditions at
$\mathcal{N}=0$ with the massive field at rest at the bottom of its potential, $\sigma(\mathcal{N}=0)=\sigma_{,\mathcal{N}}(\mathcal{N}=0)=0$, we can set the background according to the scaling solution,
\begin{equation}
	\bar H(\mathcal{N})=H_\mathrm{ini}\exp(\frac{\mathcal{N}}{1-p})\,,\qquad \bar\phi(\mathcal{N})=\sqrtb{\frac{p}{2}}\left(-\frac{2\mathcal{N}}{1-p}+\ln(\frac{p}{1-3p}\frac{V_0}{H_\mathrm{ini}^2})\right)\,,\label{eq:HandphibarofcalN}
\end{equation}
so in particular $\epsilon_\mathrm{ini}\equiv\epsilon(\mathcal{N}=0)=\bar\epsilon=1/p$. We thus infer $\phi_{,\mathcal{N}}(\mathcal{N}=0)=-\sqrtb{2p}/(1-p)$, together with
\begin{equation}
    \phi_\mathrm{ini}\equiv\phi(\mathcal{N}=0)=\sqrtb{\frac{p}{2}}\ln(\frac{p}{1-3p}\frac{V_0}{H_\mathrm{ini}^2})\,,
\end{equation}
or equivalently,
\begin{equation}
    H_\mathrm{ini}\equiv H(\mathcal{N}=0)=-\sqrtb{\frac{pV_0}{1-3p}}\exp(-\frac{\phi_\mathrm{ini}}{\sqrtb{2p}})\,.
\end{equation}
In this section, unless otherwise stated we always set $\phi_\mathrm{ini}=50$, $V_0=10^{-8}$, and the transition sharpness parameter is fixed to $\delta=10^{-3}$.

\subsubsection{Bump-like coupling function}

\begin{figure}
	\centering
	\includegraphics[height=12.2cm]{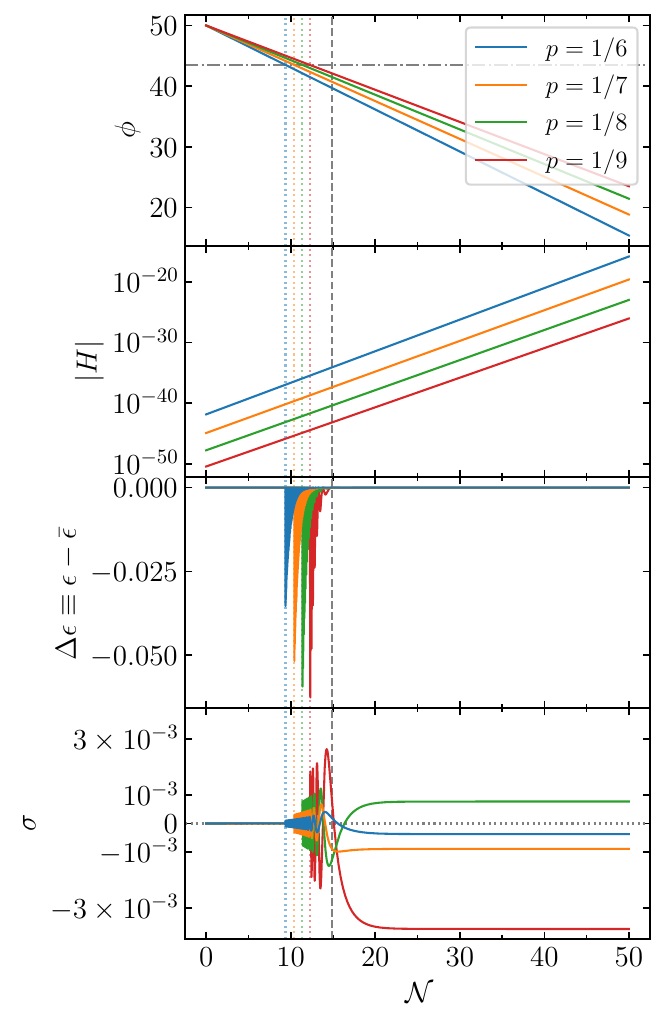}
	\hspace*{-0.02\textwidth}
	\includegraphics[height=12.2cm]{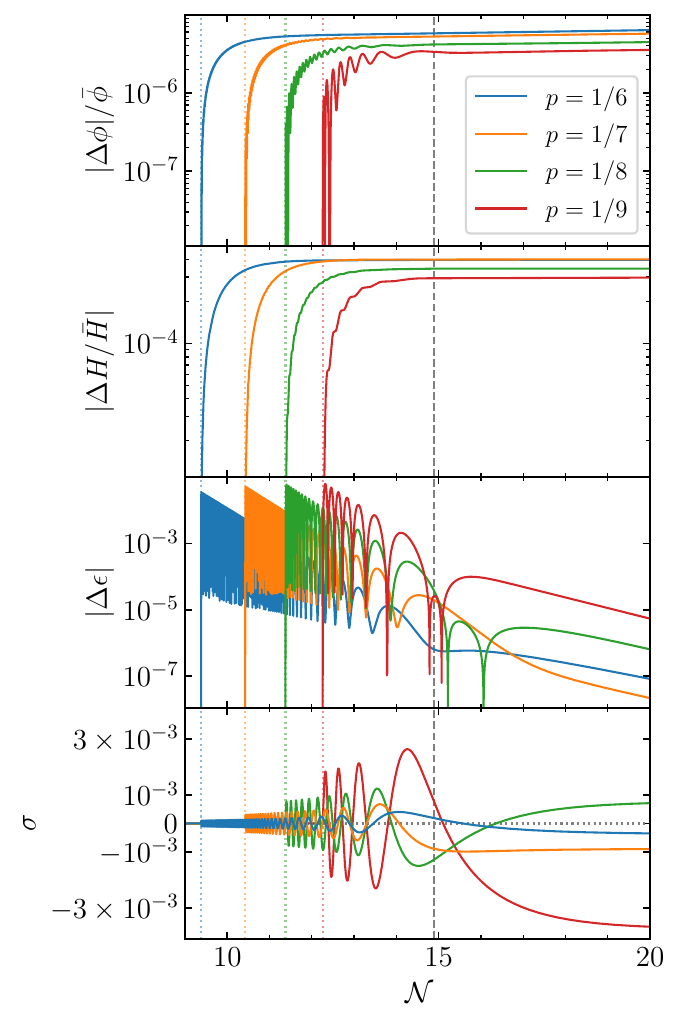}
	\caption{{\footnotesize{Bump with $\phi_0=43.5$, $\varrho=0.01$, and $m_\sigma=2\times 10^8|H_\mathrm{ini}|$. Descriptions in the text.}}}
	\label{fig:1p}
\end{figure}

Let us start by solving the equations for the case of a bump in $\Xi$ [eq.~\eqref{eq:Xibumpexp}]. This is first depicted in Fig.~\ref{fig:1p}, where we set $\phi_0=43.5$, $\varrho=0.01$, and $m_\sigma=2\times 10^8|H_\mathrm{ini}|$ and vary $p$ from $1/6$ to $1/9$. Thus, curves of different color represent different scaling equations of state from $\bar\epsilon=6$ to $\bar\epsilon=9$. The panel on the left-hand side shows the evolution of the ekpyrotic field $\phi$, the Hubble parameter in absolute value $|H|$, the change in the equation of state compared to the scaling solution $\Delta\epsilon\equiv\epsilon-\bar\epsilon$, and the massive field $\sigma$, all as functions of the $e$-folding number $\mathcal{N}$. The value of $\phi_0$ is depicted by the horizontal dash-dotted gray line in the top plot, hence when the solid colored curves first cross that line, the coupling kicks in and triggers massive field oscillations, as can be seen in the bottom plot. The time when this happens for the different values of $p$ is depicted by the vertical dotted colored lines.
From the scaling solution \eqref{eq:HandphibarofcalN}, $\bar\phi(\mathcal{N}_0)\approx\phi_0$ gives an approximation for the critical $e$-folding time $\mathcal{N}_0$ as
\begin{equation}
    \mathcal{N}_0\approx (1-p)\left(-\frac{\phi_0}{\sqrtb{2p}}+\frac{1}{2}\ln(\frac{p}{1-3p}\frac{V_0}{H_\mathrm{ini}^2})\right)\,.\label{eq:Nc}
\end{equation}
The horizon-exit time is indicated by the vertical dashed gray line, and as we can see, the massive field oscillations stop at that time, and $\sigma$ simply approaches a constant to the future.
Combining \eqref{eq:th} and the scaling solution \eqref{eq:HandphibarofcalN},
\begin{equation}
    \mathcal{N}_{\textrm{h-e}}\approx (1-p)\ln(\frac{m_\sigma}{(1-p)|H_\mathrm{ini}|})\label{eq:Nh}
\end{equation}
is a good approximation for the $e$-folding horizon-exit time.

The panel on the right-hand side of Fig.~\ref{fig:1p} presents the same solutions as on the left, but zooming in the time range during which the massive field is excited. Also, the change in the equation of state is plotted on a logarithmic scale, and similarly, the top two plots show the relative change in the ekpyrotic field $|\Delta\phi/\bar\phi|=|(\phi-\bar\phi)/\bar\phi|$ and in the Hubble parameter $|\Delta H/\bar H|=|(H-\bar H)/\bar H|$. Of those background quantities, the equation of state appears to be the one that is the most sensitive to the massive field oscillations, i.e., it is the one that receives the most oscillations due to backreaction from $\sigma$. The Hubble parameter and the ekpyrotic field also receive corrections (again of oscillatory type), though not as significant.

\begin{figure}
	\centering
	\includegraphics[width=0.9\textwidth]{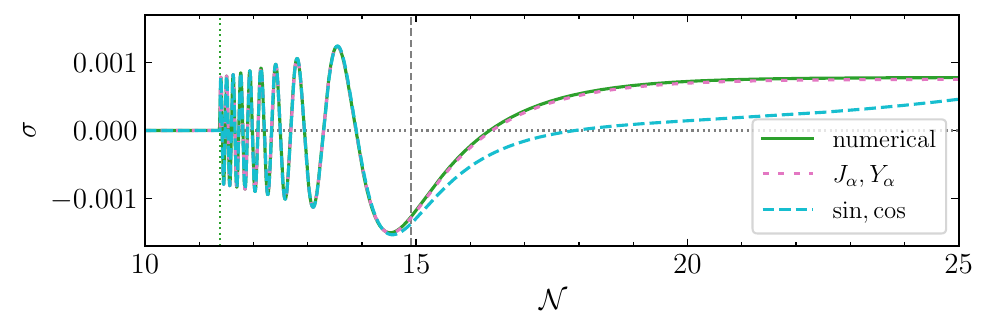}
	\caption{{\footnotesize{Massive field evolution for a bump with $p=1/8$, $\phi_0=43.5$, $\varrho=0.01$, and $m_\sigma=2\times 10^8|H_\mathrm{ini}|$. The full numerical solution is compared to analytical estimates in terms of the Bessel functions and in terms of the sine and cosine functions.}}}
	\label{fig:1Comp}
\end{figure}

From an analytical point of view, the fact that the scaling solutions for $H$ and $\phi$ are not affected much by the massive field oscillations justifies the Born approximation previously done to estimate the $\sigma$ solutions.\footnote{This is true as long as the theory remains weakly coupled, which is to say that $|\sigma|\ll\varrho$, hence $\varrho$ should not be taken to be too small. The effect of changing $\varrho$ will be discussed shortly.} This can be seen in Fig.~\ref{fig:1Comp}, which shows the comparison between the numerical solution for $\sigma$ and the analytical estimates, both in terms of the full Bessel functions $J_\alpha,Y_\alpha$ [eq.~\eqref{eq:Besselapproxsigma}] and in terms of the trigonometric $\sin,\cos$ functions that are an early-time approximation to the Bessel functions [eq.~\eqref{eq:sigmabarearlytime}]. For this plot, we use $p=1/8$ and the same parameter values for $\phi_0$, $\varrho$, and $m_\sigma$ as in Fig.~\ref{fig:1p}. To plot the analytical expressions \eqref{eq:Besselapproxsigma} and \eqref{eq:sigmabarearlytime}, we used the numerical solution to determine the value of the integration constants $c_1,c_2$, by matching the numerical and analytical solutions (and their first derivatives) at the onset of $\sigma$ oscillations (depicted by the vertical dotted green line). We also resorted to the scaling approximation $\bar H=p/t$ and its expression \eqref{eq:HandphibarofcalN} in terms of the $e$-folding number to write $t$ in terms of $\mathcal{N}$. At early times (before the horizon-exit time depicted by the vertical dashed gray line), we see that the oscillations are well described by either the Bessel functions or the trigonometric functions. After horizon exit, the trigonometric functions are no longer particularly good approximations, but the analytical solution in terms of Bessel functions applies and still matches very well the numerical solution.

\begin{figure}
	\centering
	\includegraphics[height=10.04cm]{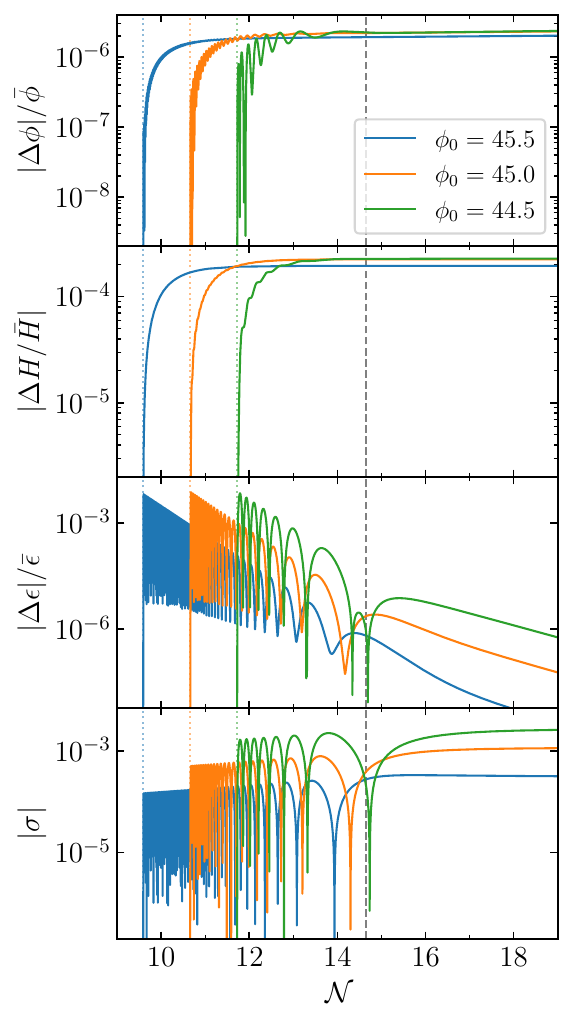}
    \hspace*{-0.023\textwidth}
	\includegraphics[height=10.04cm]{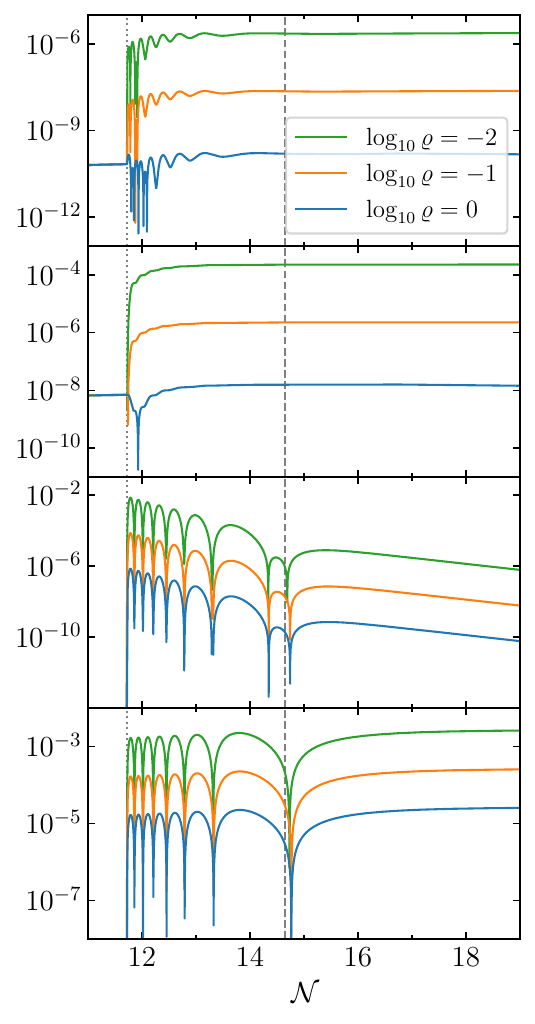}
    \hspace*{-0.015\textwidth}
	\includegraphics[height=10.04cm]{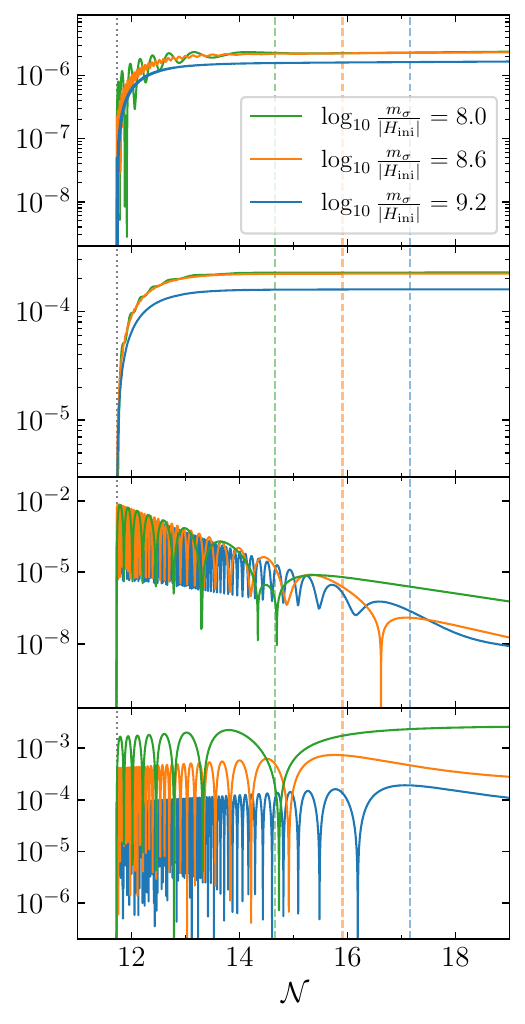}
	\caption{{\footnotesize{Bump with $p=1/11$ in all these cases. \textit{Left:} $\varrho=0.01$ and $m_\sigma=10^8|H_\mathrm{ini}|$ are kept fix, but $\phi_0$ is varied. \textit{Center:} $\phi_0=44.5$ and $m_\sigma=10^8|H_\mathrm{ini}|$ are kept fix, but $\varrho$ is varied. \textit{Right:} $\varrho=0.01$ and $\phi_0=44.5$ are kept fix, but $m_\sigma$ is varied. More descriptions in the text.}}}
	\label{fig:1phirhom}
\end{figure}

Turning to Fig.~\ref{fig:1phirhom}, we explore the effect of changing the other model parameters. The value of $\phi_0$ is varied in the left panel (keeping $p=1/11$, $\varrho=0.01$, and $m_\sigma=10^8|H_\mathrm{ini}|$ fixed), the value of $\varrho$ is varied in the middle panel (keeping $p=1/11$, $\phi_0=44.5$, and $m_\sigma=10^8|H_\mathrm{ini}|$ fixed), and the value of $m_\sigma$ is varied in the right panel (keeping $p=1/11$, $\varrho=0.01$, and $\phi_0=44.5$ fixed). In each panel, the top three plots depict the relative difference in $\phi$, $H$, and $\epsilon$ between their full numerical solution and their scaling analytical approximation. The bottom plot shows the evolution of the massive field (in absolute value to use a logarithmic scale).

Looking at the left planel of Fig.~\ref{fig:1phirhom}, we see that changing $\phi_0$ has the effect of changing the critical time at which the massive field is excited (depicted by the vertical dotted lines of color). Since $\dot\phi^2$ approximately grows as $1/t^2$, the later the coupling kicks in, the larger the trigger force term on the right-hand side of the massive field equation \eqref{eq:backEOMsF1s}, which is proportional to $\dot\phi^2$. Therefore, while the massive field has the same frequency of oscillation at any given time regardless of $\phi_0$, the amplitude of the oscillations is larger the smaller $\phi_0$ is (i.e., the later the massive field is excited). Interestingly, a larger amplitude in $|\sigma|$ leads to marginally larger backreaction in the other background quantities. Indeed, the induced oscillations in $\phi$ and $H$ are more pronounced the smaller $\phi_0$ is, but the overall relative correction remains of the same order. In all cases, the oscillations stop at the same horizon-exit time, depicted by the vertical dashed gray line, and $\sigma$ approaches a constant at late times.

Looking at the middle panel of Fig.~\ref{fig:1phirhom}, it is straightforward to understand that the coupling parameter $\varrho$ controls the amplitude of the massive field oscillations and of the induced oscillations in the other background quantities. Indeed, the smaller $\varrho$ is (recall, a smaller `turning radius' means a larger coupling), the larger the relative changes in $\phi$, $H$, and $\epsilon$, and the larger $|\sigma|$ becomes. We can see here the limitation of increasing the coupling constant (decreasing $\varrho$) since the backreaction on $\epsilon$ could become important. The case $\varrho=0.01$ indicates a $\sim 1\%$ relative change in $\epsilon$ right after the trigger time (depicted by the vertical dotted gray line). This is reasonable, but a bigger coupling (by a couple more orders of magnitude) could start disrupting the ekpyrotic nature of the background --- this can also be the case with a sustained coupling (the cases of plateau and step functions for the coupling) as we will see shortly.

In the right panel of Fig.~\ref{fig:1phirhom}, we see that changing the mass of the massive field has three effects: (1) at any given time, the larger $m_\sigma$ is, the larger the frequency of oscillations in $\sigma$ is; (2) the larger $m_\sigma$ is, the smaller the amplitude in oscillations is; and (3) the larger $m_\sigma$ is, the longer the oscillations last (i.e., the later the horizon-exit time becomes, as depicted by the vertical dashed colored lines). There is some analytic understanding to these three effects since: (1) the $\sigma$ oscillation frequency is well approximated by $m_\sigma$ [recall \eqref{eq:sigmabarearlytime}]; (2) the amplitude in $\sigma$ oscillations is approximately controlled by $1/m_\sigma^{3p/2}$ [again according to \eqref{eq:sigmabarearlytime}]; (3) the horizon-exit time is roughly $\mathcal{N}_{\textrm{h-e}}\sim\ln(m_\sigma/|H_\mathrm{ini}|)$ [recall \eqref{eq:Nh}].
Those three effects on $\sigma$ directly translate into effects in the other background quantities, i.e., changing $m_\sigma$ changes the frequency, size, and duration of the oscillations in $\phi$, $H$, and $\epsilon$ (to varying degrees).

\subsubsection{Plateau- and step-like coupling function}\label{sec:plateauStepBack}

\begin{figure}
	\centering
	\includegraphics[height=13.2cm]{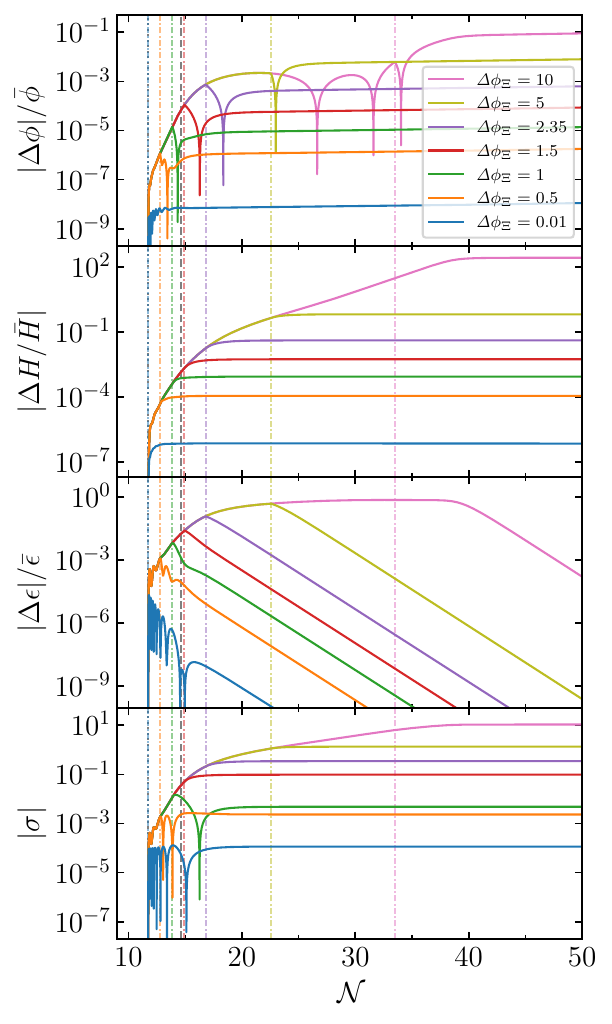}
	\includegraphics[height=13.2cm]{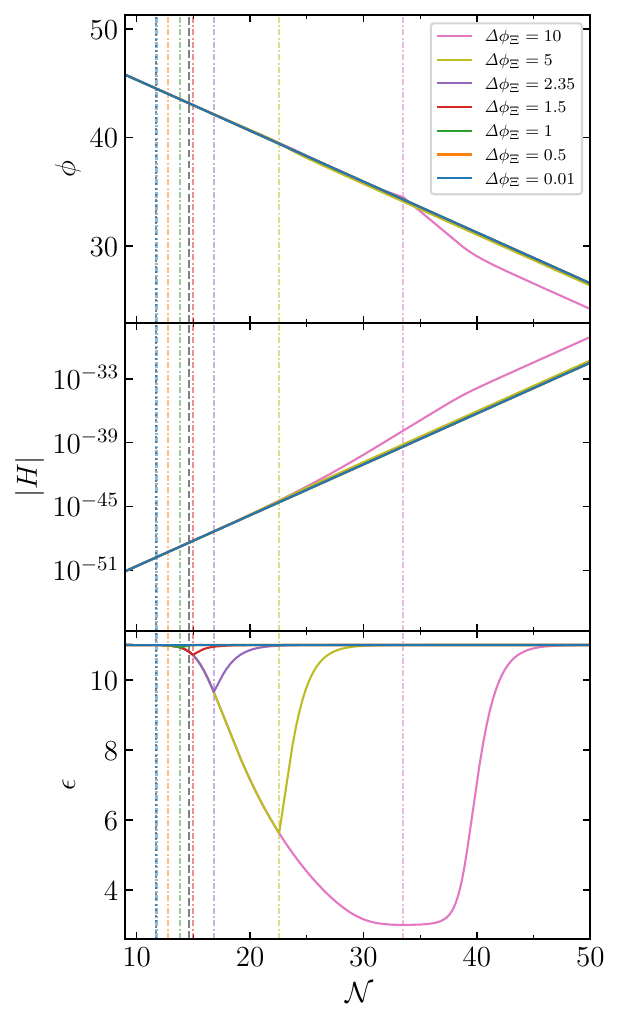}
	\caption{{\footnotesize{Plateau with $\phi_0=44.5$, $p=1/11$, $\varrho=1$, and $m_\sigma=10^8|H_\mathrm{ini}|$ kept fix. The value of $\phi_\mathrm{e}$ is varied, and the legend shows the corresponding values for $\mathit{\Delta}\phi_\Xi\equiv\phi_0-\phi_\mathrm{e}$. More descriptions in the text.}}}
	\label{fig:2}
\end{figure}

\begin{figure}
	\centering
	\includegraphics[width=0.9\textwidth]{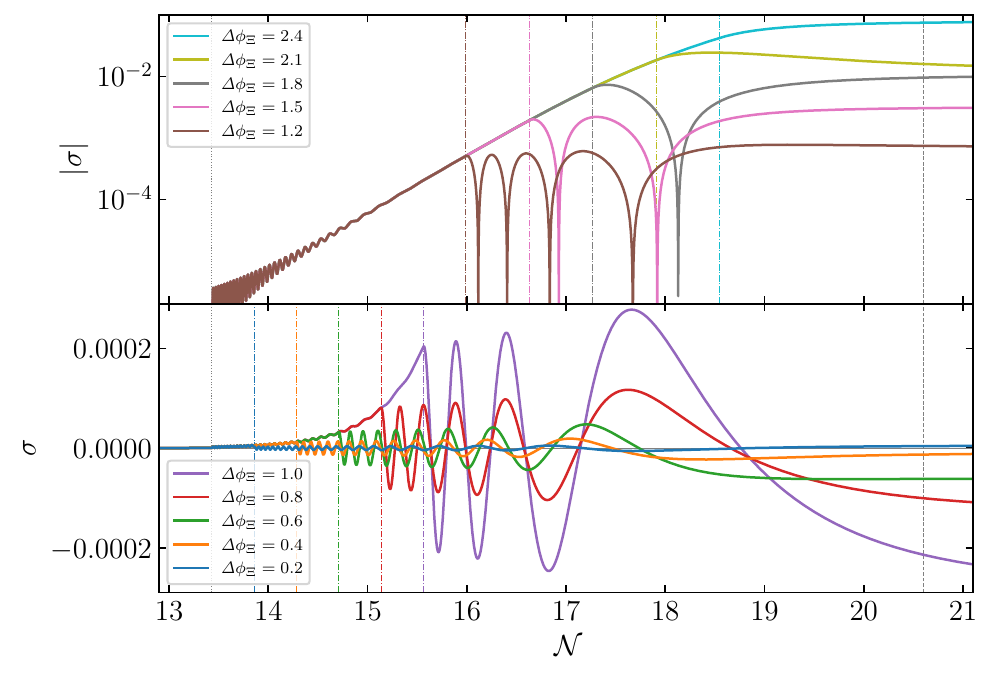}
	\caption{{\footnotesize{Plateau with $\phi_\mathrm{ini}=19.5$, $\phi_0=13.2$, $p=1/11$, $\varrho=1$, and $m_\sigma=10^{9.8}|H_\mathrm{ini}|$ kept fix. The value of $\phi_\mathrm{e}$ is varied, and the legend shows the corresponding values for $\mathit{\Delta}\phi_\Xi\equiv\phi_0-\phi_\mathrm{e}$. The top plot shows the logarithmic value of the absolute value of the massive field to have a better appreciation of the oscillations on a large range of scales. In contrast, a linear scale is used in the bottom plot when $\mathit{\Delta}\phi_\Xi$ is smaller.}}}
	\label{fig:plateau_more}
\end{figure}

Figures \ref{fig:1p} to \ref{fig:1phirhom} cover the phenomenology of the bump coupling function $\Xi$, so we now turn to Figs.~\ref{fig:2} and \ref{fig:plateau_more}, which present the results in the case of a plateau function for $\Xi$ [eq.~\eqref{eq:Xiplateautanh}]. In Fig.~\ref{fig:2}, we fix $\phi_0=44.5$, $p=1/11$, $\varrho=0.01$, and $m_\sigma=10^8|H_\mathrm{ini}|$, but vary the value of $\phi_\mathrm{e}$, i.e., the width of the plateau coupling function, which controls how long the coupling remains before shutting off. We denote the field-space width when the coupling is `on' by $\mathit{\Delta}\phi_\Xi\equiv\phi_0-\phi_\mathrm{e}$. The left panel of Fig.~\ref{fig:2} presents the same information as in the panels of Fig.~\ref{fig:1phirhom}. In particular, the vertical dashed gray line depicts the horizon-exit time. An important addition, though, is the vertical dash-dotted colored lines, which depict the times at which $\phi_\mathrm{e}$ is crossed, i.e., when it is that the coupling $\Xi$ is shut off. Different values of $\mathit{\Delta}\phi_\Xi$ are shown, which show cases where the coupling is turned off both before and after the horizon-exit time.

Looking at the case $\mathit{\Delta}\phi_\Xi\equiv\phi_0-\phi_\mathrm{e}=0.01$ in the left panel of Fig.~\ref{fig:2} (the blue curve), we see that a very narrow plateau produces more or less the same evolution as a bump coupling function does (the blue curve here resembles, both qualitatively and quantitatively, the blue curve in the middle panel of Fig.~\ref{fig:1phirhom}, which has the same parameter values). As the plateau coupling is made wider, say the orange curve with $\mathit{\Delta}\phi_\Xi=0.5$, we already start seeing important differences. Indeed, the coupling has the effect of pushing $\sigma$ to greater values (though still with some oscillations), and correspondingly, it leads to greater corrections in $\epsilon$, $H$, and $\phi$. This lasts until $\phi_\mathrm{e}$ is crossed, after which we recover the same evolution as in the bump case (when the coupling is switched off). The later the coupling is switched off (the wider the plateau), the greater the effect is --- one effectively looses the oscillatory behavior in $\sigma$ and instead the massive field eventually gets to disrupt the ekpyrotic background. Indeed, we can see for the very wide plateaus that the relative change in the equation of state can reach order unity. This can be explicitly seen in the right panel of Fig.~\ref{fig:2}, where we plot the actual values of $\phi$, $H$, and $\epsilon$. It is clear that the longer the coupling is sustained, the smaller the equation of state can become (asymptotically approaching $\epsilon=3$, i.e., the equation of state of a massless scalar field). However, as long as the coupling switches off eventually (e.g., near $\mathcal{N}\approx 33.5$ for the pink curve), the background will be driven back to its ekpyrotic scaling solution (this is the strength of the attractor nature of the ekpyrotic field). That being said, if we have a step coupling function for $\Xi$ [eq.~\eqref{eq:Xisteptanh}], where $\phi_\mathrm{e}$ is effectively pushed to $-\infty$, then the ekpyrotic background will be resolutely disrupted.

Figure \ref{fig:plateau_more} presents more examples of a plateau coupling. We exceptionally use a different value for $\phi_\mathrm{ini}$ as stated in the caption (this will be reused in the next section), but we note that this does not qualitatively affect the results. As in Fig.~\ref{fig:2}, we make $\mathit{\Delta}\phi_\Xi$ incrementally larger (starting in the bottom plot and then going to the upper plot), and we see that when the coupling is `on' the massive field has small oscillations at the same time as being pushed up its potential. 
These oscillations have the characteristic frequencies and envelops of the `standard clock' oscillations, but the evolution of the average values is very different and they keep growing.
When the coupling is turned off (depicted by the dash-dotted vertical lines of color), the field falls at the bottom of its potential and undergoes standard clock oscillations with increasingly larger amplitude the longer the coupling is `on' (since the field is `released' from higher). When the coupling is turned off too close to the freezing time (depicted by the vertical dashed gray line), so when $\mathit{\Delta}\phi_\Xi\gtrsim 2$, the massive field does not get to have any standard clock oscillations after the coupling is turned off.

\begin{figure}
	\centering
	\includegraphics[height=7.2cm]{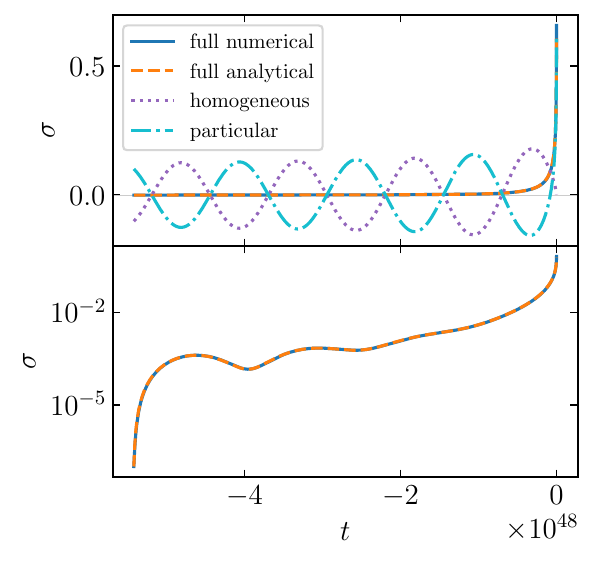}
	\includegraphics[height=7.2cm]{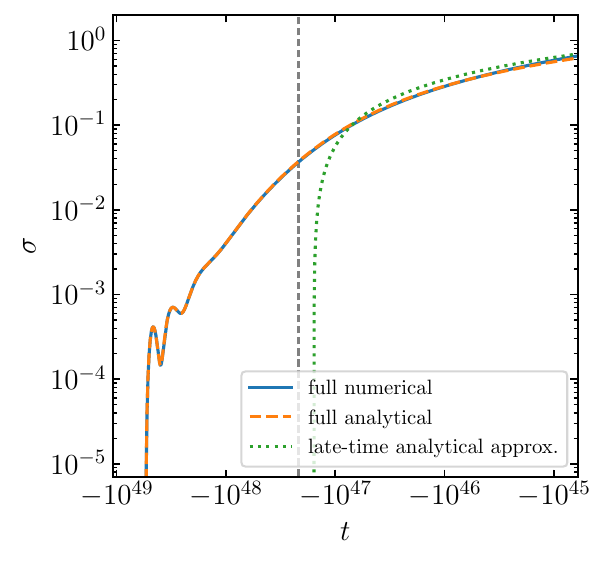}
	\caption{{\footnotesize{Step with $\phi_\mathrm{ini}=50$, $\phi_0=44.5$, $p=1/11$, $\varrho=1$, and $m_\sigma=10^8|H_\mathrm{ini}|$. The full numerical solution for the massive field $\sigma$ is compared to its analytical solution, which is the sum of a homogeneous solution and a particular solution.
    More descriptions in the text.}}}
	\label{fig:2Comp}
\end{figure}

Let us show a first example of step coupling function in Fig.~\ref{fig:2Comp}, though this is really the same thing as a plateau before $\phi_\mathrm{e}$ is reached. Let us focus our attention on the time interval before the backreaction becomes too large, e.g., looking at the pink curve in Fig.~\ref{fig:2} we are only looking at the interval $10\lesssim\mathcal{N}\lesssim 30$. Figure \ref{fig:2Comp} presents the solution as a function of the physical $t$, instead of $\mathcal{N}$, in order to ease the comparison with analytical approximations. The solid blue curve shows the full numerical solution, while the dashed orange curve shows the full analytical estimate, which is the sum [eq.~\eqref{eq:sigmaAnalyticalFull}] of the homogeneous solution \eqref{eq:Besselapproxsigma} and the particular solution \eqref{eq:sigmapfull} (which were derived under the assumption of no backreaction on $H$ and $\phi$). To better understand the solution, we plot the homogeneous (dotted purple) and particular (dash-dotted cyan) solutions separately in the top-left panel of Fig.~\ref{fig:2Comp}. This shows the behavior we alluded to before: the homogeneous and particular solutions, which are essentially both oscillating as $\sin(-m_\sigma t)$ [recall eqs.~\eqref{eq:sigmabarearlytime} and \eqref{eq:sigmapearlytime}], are nearly exactly out of phase, hence the oscillations are mostly cancelled out. Recall this is to be expected in the limit where both $m_\sigma|t|$ and $m_\sigma|t_0|$ are very large, and generally in such a case, the `initial conditions' at $t_0$ are fully set by the particular solution due to the coupling term. Recall also that this is specific to the fact that we take the massive field to be classically at rest ($\sigma=\dot\sigma=0$) before being classically excited. What is left after `cancellation', i.e., the full solution that is the sum of the homogeneous and particular solutions, is a growing function for $\sigma$ with small wiggles. This is better seen on a logarithmic scale for $\sigma$ --- see the bottom-left plot of Fig.~\ref{fig:2Comp}. On the right panel (now on a log-log scale), we contrast the full solution to the
late-time analytical approximation \eqref{eq:sigmaplatetime} (dotted green).
As expected, it is a good approximation at very late times.
The early/late-time divide is the horizon-exit time depicted by the vertical dashed gray line as before.

\begin{figure}
	\centering
	\includegraphics[height=13.2cm]{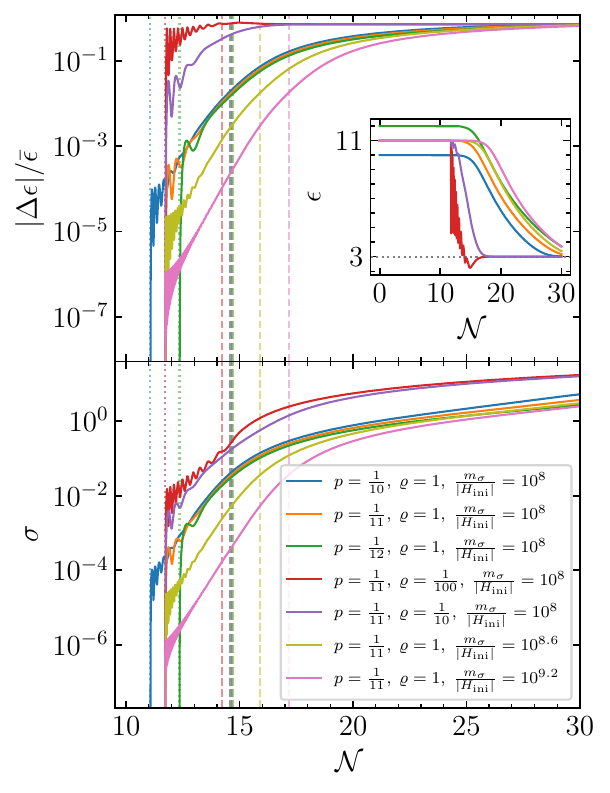}
	\caption{{\footnotesize{Step with $\phi_0=44.5$, but varying $p$, $\varrho$, and $m_\sigma$ as shown by the legend. As before, the dotted vertical lines indicate the critical times $\mathcal{N}_0$, while the dashed ones indicate the horizon-exit times $\mathcal{N}_{\textrm{h-e}}$. The top plot shows the relative difference in the equation of state, while the inset shows the actual value of $\epsilon$.}}}
	\label{fig:3}
\end{figure}

Given the understanding of the competition between the homogeneous and particular solutions when the coupling $\Xi\approx 1/\varrho$ remains present, we present in Fig.~\ref{fig:3} many more examples of a step coupling function. We set $\phi_0=44.5$ in every case, but vary $p$, $\varrho$, and $m_\sigma$. The orange curve of Fig.~\ref{fig:3} corresponds to the solution shown in Fig.~\ref{fig:2Comp}, which exhibits very few oscillations, but we see that different values of $p$, $\varrho$, and $m_\sigma$ generally lead to more oscillations, though again with the pattern that the force term pushes the massive field up its potential in average (and monotonically after horizon exit --- the horizon-exit times are depicted by the vertical dashed lines). 
As in the previous examples (Figs.~\ref{fig:plateau_more} and \ref{fig:2Comp}), these oscillations have the same frequencies and envelops as those of the standard clocks, but their average values keep growing due to the constant coupling to the ekpyrotic field.
In all cases, we can see that the backreaction on the background is quite significant, as the initially large equation of state is disrupted and forced down to $\epsilon=3$, asymptotically --- this is most explicit in the inset plot, which shows the value of $\epsilon$ instead of the relative difference. In the late-time limit, as long as the coupling term remains present, the displacement of the massive field can eventually become super-Planckian, which may signal the breakdown of the effective field theory \cite{Ooguri:2006in,Klaewer:2016kiy,Grimm:2018ohb,Ooguri:2018wrx,Rudelius:2023mjy}. From the late-time analytical estimate of the particular solution, eq.~\eqref{eq:sigmaplatetime}, we estimate the field displacement to be given by $\mathit{\Delta}\sigma\sim (p/\varrho)\mathit{\Delta}\mathcal{N}$ in the small-$p$ limit. The smaller $p$ is, the longer the regime of validity of the effective field theory; the same thing can be said (as expected) the smaller the coupling is (so the larger $\varrho$ is).

\begin{figure}
	\centering
	\includegraphics[width=\textwidth]{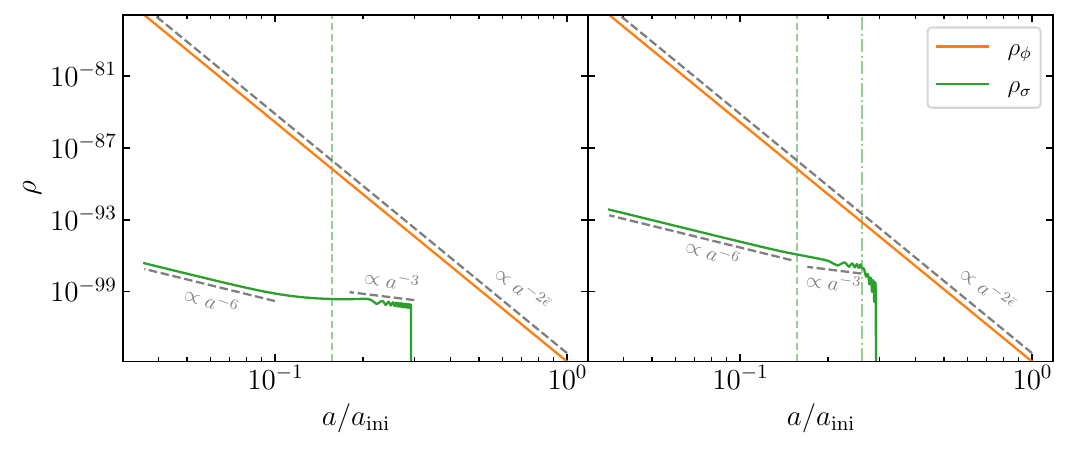}\\
    \includegraphics[width=\textwidth]{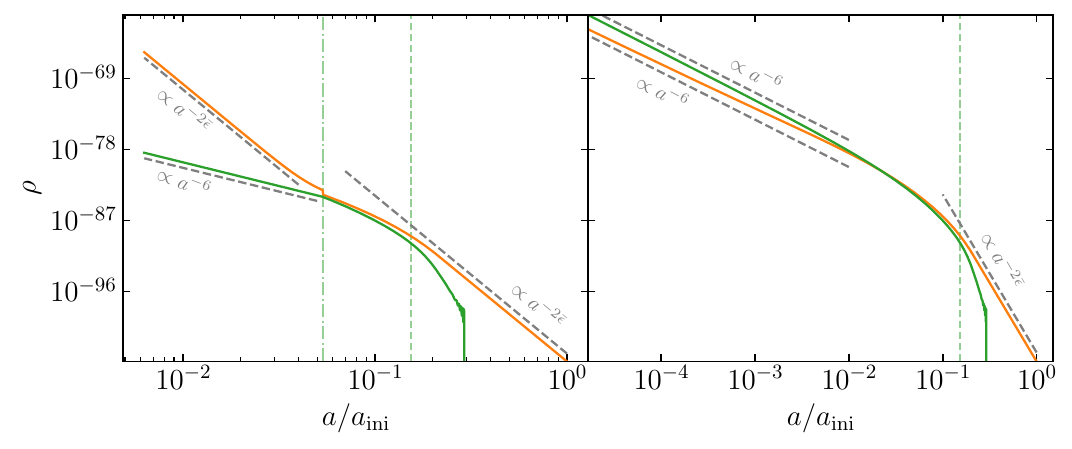}
	\caption{{\footnotesize{Evolution of the ekpyrotic field's energy density $\rho_\phi$ in orange and of the massive field's energy density $\rho_\sigma$ in green as functions of the scale factor $a$ (normalized by an arbitrary initial scale, $a_\mathrm{ini}$). Since the universe contracts, forward evolution is read from right to left. The model parameters are: $\phi_\mathrm{ini}=50$, $\phi_0=44.5$, $p=1/10$ (so $\bar\epsilon=10$), $m_\sigma/|H_\mathrm{ini}|=10^8$, and $\varrho=1$. A bump coupling function is used for the top left plot; a plateau coupling function is used for the top right and bottom left plots, with $\mathit{\Delta}\phi_\Xi=0.5$ and $\mathit{\Delta}\phi_\Xi=5.5$, respectively; and a step coupling function is used in the bottom right plot. We add a vertical dashed line to denote the freezing time, and in the plateau cases, we add a vertical dash-dotted line to denote when the plateau coupling shuts off. We also add dashed gray lines to represent the standard scaling of an ekpyrotic field ($\rho\propto a^{-2\bar\epsilon}$), massless field ($\rho\propto a^{-6}$), and massive field ($\rho\propto a^{-3}$).}}}
	\label{fig:rhos}
\end{figure}

To complement the above results, let us present the evolution of the respective energy densities in Fig.~\ref{fig:rhos}. We define the energy densities by writing the Friedmann constraint equation \eqref{eq:FriedConstraint} as $3H^2=\rho_\phi+\rho_\sigma$, so $\rho_\phi=(1+\Xi(\phi)\sigma)\dot\phi^2/2+V(\phi)$ and $\rho_\sigma=\dot\sigma^2/2+m_\sigma^2\sigma^2/2$. Upon solving the full set of background equations numerically, Fig.~\ref{fig:rhos} depicts how the coupling between $\phi$ and $\sigma$ leads to energy exchange between the two fields. Keeping $p$, $\phi_0$, $\varrho$, and $m_\sigma$ fixed, we show the cases of a bump (top left), small plateau (top right), large plateau (bottom left), and step (bottom right). In the bump case, it is quite clear that $\sigma$ always remains sub-dominant compared to $\phi$, i.e., it is a spectator field, whose energy density grows more slowly than the background field. When $-m_\sigma t\gg 1$, $\sigma$ acts as a heavy field, whose equation of state averages that of pressureless matter ($\rho_\sigma\sim a^{-3}$), while after freezing ($-m_\sigma t\ll 1$; $-m_\sigma t\sim 1$ is the vertical dashed line), $\sigma$ acts as a massless field with stiff equation of state ($\rho_\sigma\sim a^{-6}$). This is in agreement with \eqref{eq:sigmabarlatetime}. For a small plateau, the same conclusions are reached, except when the plateau coupling is present --- reading the plot from right to left, this corresponds to the interval from the time $\sigma$ is excited until the vertical dash-dotted line. In that interval, $\rho_\sigma$ grows in average at least as fast as $\rho_\phi$, if not faster. This is even more explicit for the larger plateau, where one can see that, as long as the coupling is `on', $\phi$ effectively `decays' and `sources' $\sigma$. Indeed, $\rho_\sigma$ grows faster than $\rho_\phi$, and $\rho_\phi$ eventually deviates from its standard background scaling (its effective equation of state decreases). In this example, this happens until the two fields have nearly the same energy density, but then, as soon as the coupling is turned off, $\rho_\phi$ sharply goes back to its ekpyrotic scaling, while $\sigma$ goes back to being a spectator (effectively massless by that point). In the case of a step coupling, as the coupling is always present after exciting $\sigma$, $\rho_\sigma$ eventually dominates over $\rho_\phi$, and both eventually scale as $a^{-6}$. This is why $\epsilon$ approaches $3$ in Fig.~\ref{fig:3}.

\paragraph*{}
Now that we presented successful realizations of a massive spectator field classically excited in a phase of ekpyrotic contraction and described the possible background phenomenology, we turn to the exploration of the cosmological perturbations and what corrections a massive field induces on the usual correlation functions.

\section{Standard clock signals in ekpyrotic cosmology}\label{sec:signals}

\subsection{Perturbation equations}\label{sec:pertEqns}

Let us begin with action \eqref{eq:actionGenModels} again, and going beyond the FLRW background, let us introduce cosmological perturbations.
In addition to $\chi(t,\mathbf{x})$, which we now reintroduce since it is solely a perturbation [recall the discussion above \eqref{eq:S2chivanilla}], we perturb the other two scalar fields and the metric as
\begin{align}
    \phi(t,\mathbf{x})&=\phi(t)+\delta\phi(t,\mathbf{x})\,,\qquad
    \sigma(t,\mathbf{x})=\sigma(t)+\delta\sigma(t,\mathbf{x})\,,\nonumber\\
    g_{\mu\nu}\dd x^\mu\dd x^\nu&=-\big(1+2\Phi(t,\mathbf{x})\big)\dd t^2+2a(t)B_{,i}(t,\mathbf{x})\dd x^i\dd t+a(t)^2\big(1-2\Psi(t,\mathbf{x})\big)\delta_{ij}\dd x^i\dd x^j\,.\label{eq:perturbedFields}
\end{align}
Note that the focus here is on scalar perturbations of the metric, i.e., we are ignoring vector and tensor perturbations in this current work.\footnote{This is certainly justified since we recall that in ekpyrotic cosmology vector perturbations are negligible and tensor perturbations are unobservable (they have a deeply blue spectrum).}
Let us work in the spatially flat gauge, where $\Psi\equiv 0$, so that all propagating perturbation degrees of freedom are characterized by the field fluctuations $\delta\phi$ (the non-amplified blue adiabatic perturbation), $\delta\sigma$ (perturbations of the massive spectator `clock' field), and $\chi$ (the amplified nearly scale-invariant entropy perturbation). In this case, the perturbed action to second order reads\footnote{The derivation of the second-order action may be found in Appendix \ref{app:2ndaction}. In short, we transformed to Fourier space, eliminated the perturbations of the lapse $\Phi$ and shift $B$ using the Hamiltonian and momentum constraints, integrated by parts, and simplified using the background equations of motion. Note that the subscript $k$ on a perturbation variable indicates its Fourier transform, and $k$ represents the wavenumber.}
\begin{align}
    S^{(2)}=&~\frac{1}{2}\int\dd^3k\,\dd t\,a^3\bigg(\big(1+\Upsilon(\phi)\sigma\big)\Omega^2(\phi)\Big(\dot{\chi}_k^2-\frac{k^2}{a^2}\chi_k^2\Big)\nonumber\\
    &+\dot{\delta\sigma}_k^2-\mathcal{F}_\sigma\,\delta\sigma_k^2+\big(1+\Xi(\phi)\sigma\big)\Big(\dot{\delta\phi}_k^2-\mathcal{F}_\phi\,\delta\phi_k^2\Big)+2\Xi(\phi)\dot\phi\,\delta\sigma_k\,\dot{\delta\phi}_k-\mathcal{F}_{\sigma\phi}\,\delta\sigma_k\,\delta\phi_k\bigg)\,,\label{eq:S21}
\end{align}
where $\phi$ and $\sigma$ are understood to denote only the background quantities $\phi(t)$ and $\sigma(t)$. In the above, we defined
\begin{subequations}
\begin{align}
    \mathcal{F}_\sigma\equiv&~\frac{k^2}{a^2}+m_\sigma^2+2\frac{m_\sigma^2}{H}\sigma\dot\sigma-\frac{\dot\sigma^4}{2H^2}+\dot\sigma^2\left(3-\frac{\big(1+\Xi(\phi)\sigma\big)\dot\phi^2}{2H^2}\right)\,,\\
    \mathcal{F}_\phi\equiv&~\frac{k^2}{a^2}+2\frac{\dot\phi}{H}V_{,\phi}+\big(1+\Xi(\phi)\sigma\big)\left(3\dot\phi^2-\frac{\dot\sigma^2\dot\phi^2}{2H^2}\right)-\big(1+\Xi(\phi)\sigma\big)^2\frac{\dot\phi^4}{2H^2}\nonumber\\
    &+\frac{V_{,\phi\phi}+\frac{1}{2}\Xi_{,\phi\phi}\sigma\dot\phi^2}{1+\Xi(\phi)\sigma}+\frac{\Xi_{,\phi}\dot\sigma\dot\phi-\Xi_{,\phi}V_{,\phi}\sigma-\frac{1}{2}\Xi_{,\phi}^2\sigma^2\dot\phi^2}{(1+\Xi(\phi)\sigma)^2}\,,\\
    \mathcal{F}_{\sigma\phi}\equiv&~2\frac{\dot\sigma}{H}V_{,\phi}-\Xi_{,\phi}\dot\phi^2+\big(1+\Xi(\phi)\sigma\big)\left(2\frac{m_\sigma^2}{H}\sigma+6\dot\sigma-\frac{\dot\sigma^3}{H^2}\right)\dot\phi-\big(1+\Xi(\phi)\sigma\big)^2\frac{\dot\sigma\dot\phi^3}{H^2}\,.
\end{align}
\end{subequations}

Recall $\chi_k$ is the entropy fluctuation that acquires near scale invariance (it is later converted to curvature perturbations), and hence it is the perturbation for which we wish to find the correction due to the (often oscillatory) massive field $\sigma$. We can define $s\equiv\Omega(\phi)\sqrtb{1+\Upsilon(\phi)\sigma}\,\chi$ and $v\equiv as\equiv\tilde z\chi$ --- so $\tilde z=a\Omega(\phi)\sqrtb{1+\Upsilon(\phi)\sigma}=z\sqrtb{1+\Upsilon(\phi)\sigma}$, recalling from below \eqref{eq:S2chivanilla} the definition $z\equiv a\Omega(\phi)$ --- such that we can write the $\chi_k$ part of the above perturbed action exactly as in \eqref{eq:Ss2v}, simply with $z$ replaced by $\tilde z$. The equation of motion for the Mukhanov-Sasaki variable $v_k$ is then as before [eq.~\eqref{eq:vsEOM}], with $z$ replaced by $\tilde z$. The time dependence of the effective frequency, controlled by $\tilde z''/\tilde z$, depends on up to the second derivative of $a$, $\phi$, and now $\sigma$ as well if the coupling $\Upsilon(\phi)$ is non-zero. We thus see that the induced oscillations at the background level [in $a$, $\phi$, and explicitly $\sigma$ (and derivatives thereof)] may induce oscillations in the perturbations (in $v$, so equivalently in $\chi$). Concretely, if we investigate the $\chi_k$ equation of motion, which we may express as
\begin{equation}
    \chi_k''+\frac{(\tilde z^2)'}{\tilde z^2}\chi_k'+k^2\chi_k=0\,,
\end{equation}
we note that the coefficient of the first-derivative term, $\chi_k'$, is separable as
\begin{equation}
    \frac{(\tilde z^2)'}{\tilde z^2}=\frac{(z^2)'}{z^2}+\frac{\Upsilon\sigma}{1+\Upsilon\sigma}\left(\frac{\Upsilon'}{\Upsilon}+\frac{\sigma'}{\sigma}\right)\,,\qquad \frac{(z^2)'}{z^2}=2\mathcal{H}-\sqrtb{\frac{2}{b}}\phi'\,,
\end{equation}
where $\mathcal{H}\equiv a'/a$ is the conformal Hubble parameter.
Further separating the background quantities into their scaling behavior plus perturbations due to the massive field as $\mathcal{H}=\mathcal{\bar H}+\Delta\mathcal{H}$ and $\phi=\bar\phi+\Delta\phi$ (so also $z=\bar z+\Delta z$), we find
\begin{equation}
    \frac{(\tilde z^2)'}{\tilde z^2}=\frac{(\bar z^2)'}{\bar z^2}+2\Delta\mathcal{H}-\sqrtb{\frac{2}{b}}\Delta\phi'+\frac{\Upsilon\sigma}{1+\Upsilon\sigma}\left(\frac{\Upsilon'}{\Upsilon}+\frac{\sigma'}{\sigma}\right)\,,\label{eq:zt2po2zt2}
\end{equation}
where
\begin{equation}
    \frac{(\bar z^2)'}{\bar z^2}=2\mathcal{\bar H}-\sqrtb{\frac{2}{b}}\bar\phi'=\frac{p-\sqrtb{p/b}}{1-p}\left(\frac{2}{\tau}\right)
\end{equation}
is the contribution that yields the unperturbed near scale-invariant power spectrum derived in section \ref{sec:vanillaEk}, which we now call $\mathcal{\bar P}_s(k)$. The other terms in \eqref{eq:zt2po2zt2} are now present due to the massive field oscillations, either explicitly through $\sigma$ and $\sigma'$ or implicitly through $\Delta\mathcal{H}$ and $\Delta\phi'$. Those will generate corrections to the power spectrum, $\Delta\mathcal{P}_s(k)$. Of course, $\Delta\mathcal{H}$ and $\Delta\phi'$ are related to $\sigma$ and $\sigma'$ by the background equations of motion \eqref{eq:backEOMsFull}, but it is difficult to simplify \eqref{eq:zt2po2zt2} analytically any further. Interestingly, even if we did not include a direct $\sigma$-$\chi$ coupling in the Lagrangian (so imagining taking $\Upsilon\equiv 0$), the above shows that there would still be a correction to the power spectrum due to the induced oscillations in $\mathcal{H}$ and $\phi'$. Whether it is the implicit oscillations (due to the gravitational coupling) or the explicit oscillations in the massive field (due to a direct $\sigma$-$\chi$ coupling) that contribute the most to the power spectrum depends in a non-trivial way on the relative size of the couplings $\Xi$ and $\Upsilon$. In other words, whether for instance $\Upsilon\sigma'/(1+\Upsilon\sigma)$ dominates over $\Delta\mathcal{H}$ and $\Delta\phi'$ is hard to determine other than numerically at this point.

The challenge is thus to solve for the perturbations to find what are the induced oscillations due to the presence of the massive field. As for the background, we shall explore the analytical and numerical avenues separately; this constitutes the next two subsections.

\subsection{Analytical results}\label{sec:analytical}

\subsubsection{General considerations}\label{sec:pertGen}

Let us begin with a short comment about the (generally coupled) $\{\delta\phi_k,\delta\sigma_k\}$ perturbations in \eqref{eq:S21}. When the massive field $\sigma$ is left unexcited at the bottom of its potential [when $\Xi(\phi)=0$, so $\sigma(t)=0$], the action reduces to
\begin{align}
    S^{(2)}_{\{\delta\phi,\delta\sigma\}}=\frac{1}{2}\int\dd^3k\,\dd t\,a^3\bigg( & \dot{\delta\phi}_k^2-\left(\frac{k^2}{a^2}+V_{,\phi\phi}-2H^2(3\epsilon-\epsilon^2+\epsilon\eta)\right)\delta\phi_k^2\nonumber\\
    &+\dot{\delta\sigma}_k^2-\left(\frac{k^2}{a^2}+m_\sigma^2\right)\delta\sigma_k^2\bigg)\,,
\end{align}
where we defined $\eta\equiv\dot\epsilon/(H\epsilon)$.
Note that this is the usual second-order perturbed action for a general adiabatic field with an uncoupled massive spectator field (see, e.g., \cite{Wang:2013zva}). Using the background equations in such a case, we note $V_{,\phi\phi}-2H^2(3\epsilon-\epsilon^2+\epsilon\eta)=-(1/4)H^2\eta(6-2\epsilon+\eta+2\eta^{(2)})$, where $\eta^{(2)}\equiv\dot\eta/(H\eta)$. Thus, for the scaling background solution where the equation of state $\epsilon$ is constant, we have $\eta=0$, hence this term vanishes.
The resulting equations of motion are
\begin{equation}
    \ddot{\delta\phi}_k+3\bar H\,\dot{\delta\phi}_k+\frac{k^2}{\bar a^2}\delta\phi_k=0\,,\qquad\ddot{\delta\sigma}_k+3\bar H\,\dot{\delta\sigma}_k+\left(\frac{k^2}{\bar a^2}+m_\sigma^2\right)\delta\sigma_k=0\,.
\end{equation}
Imposing a Bunch-Davies vacuum in the limit $-k\tau\to\infty$ and assuming a slowly contracting background (in fact, let us take the $p\to 0^+$ limit to simplify the argument), the mode functions remain in their Bunch-Davies states throughout time (as for tensor modes).
The solutions can be expressed as $|\delta\phi_k|^2\simeq|\delta\sigma_k|^2\simeq(2k)^{-1}$, hence $\mathcal{P}_{\delta\phi}(k)\simeq\mathcal{P}_{\delta\sigma}(k)\simeq k^2/(2\pi)^2$. 
Such vacuum blue spectra are not observable on cosmological scales, so one would normally simply neglect the $\phi$ and $\sigma$ perturbations altogether.

Then, assuming the addition of a non-zero background coupling $\Xi$ excites the massive field $\sigma$ without disrupting the ekpyrotic scaling solution, we may expect the $\delta\phi$ and $\delta\sigma$ blue spectra to only receive small oscillatory corrections, which would still be unobservable. Even if the background coupling were large and the $\sigma$ oscillations had a significant amplitude, inducing say $\mathcal{O}(1)$ corrections to the ekpyrotic scaling background, we could expect $\mathcal{O}(1)$ oscillatory corrections to the power spectra, but those would not affect the previous conclusion. Indeed, corrections that could lead to observable signals in the power spectrum would have to be, explicitly reinserting the units, $\mathcal{O}(M_\mathrm{Pl}^2/k_\mathrm{cmb}^2)$, where the CMB scales are $10^{-4}\lesssim k_\mathrm{cmb}/\mathrm{Mpc}^{-1}\lesssim 1$, i.e., more than 50 orders of magnitude away from the Planck scale.

Accordingly, we shall not investigate further the evolution and spectrum of $\delta\phi_k$ and $\delta\sigma_k$ and simply ignore the perturbations to the $\phi$ and $\sigma$ fields from here on.
In other words, the focus shall be on the observable $\chi_k$ perturbations.
We will comment some more on $\delta\sigma$ fluctuations and quantum standard clocks in the discussion section, though.
For the rest of this section, we derive analytical results for the classical clock signal as well as an example of sharp feature signal.

According to \eqref{eq:actionGenModels}, the massive field and entropy field essentially interact via the action
\begin{equation}
	S_{\sigma\chi} = \int\dd^4x\,\sqrtb{-g}\left(-\frac{1}{2}\big(1+\Upsilon(\phi)\sigma\big)\Omega^2(\phi)(\partial\chi)^2\right)\,.
\end{equation}
The coupling function $\Upsilon(\phi)$ can in general be time dependent (i.e., $\phi$ dependent).
This would introduce feature signals in addition to the ones induced by the field $\sigma$. Therefore, let us consider a simple case with a constant coupling $\Upsilon(\phi) = 1/\Lambda>0$. 
This leads to the following simplified second-order action [essentially rewriting the $\sigma$-$\chi$ part of \eqref{eq:S21}, now in real space]
\begin{equation}
	S_{\sigma\chi}^{(2)} = \frac{1}{2}\int\dd^3x\,\dd \tau\,z^2\left(1+\dfrac{\sigma}{\Lambda}\right)\bigg(\chi^{\prime 2}-(\partial_i\chi)^2\bigg)\equiv\int\dd^3x\,\dd\tau\,\mathcal{L}^{(2)}\,,
\end{equation}
where we recall $z=a\Omega(\phi)$, and where $\phi$ and $\sigma$ indicate the background solutions $\phi(\tau)$ and $\sigma(\tau)$ as before. The energy scale $\Lambda$ can be interpreted as a scale proportional to the radius of the turning trajectory in $\chi$-$\sigma$ field space. Alternatively, it can be interpreted as a generic cutoff scale in an effective field theory perspective (see, e.g., \cite{Chen:2022vzh}), where the dimension-$5$ operator $\sigma(\partial\chi)^2$ is suppressed by $1/\Lambda$.

Let us write the quadratic Lagrangian as
\begin{equation}
    \mathcal{L}^{(2)} = \mathcal{L}^{(2)}_{0} + \delta\mathcal{L}^{(2)}\,,
\end{equation}
with
\begin{subequations}
\begin{align}
    \mathcal{L}^{(2)}_{0} &= \frac{1}{2}z^2\left(\chi^{\prime 2}-(\partial_i\chi)^2\right)\,,\\
    \delta\mathcal{L}^{(2)} &= \frac{1}{2}\dfrac{\sigma}{\Lambda}z^2\left(\chi^{\prime 2}-(\partial_i\chi)^2\right)\,.
\end{align}
\end{subequations}
The effect of $\sigma$ on the primordial power spectrum is perturbative as long as $|\delta\mathcal{L}^{(2)}|\ll|\mathcal{L}^{(2)}_{0}|$, which is guaranteed if $|\sigma|/\Lambda\ll1$.  The associated Hamiltonian densities\footnote{It shall be clear from context whether $\mathcal{H}$ is the conformal Hubble parameter or a Hamiltonian density. Note, also, that the sub-/superscript `$I$' indicates the interaction picture and is not a running index.} can be calculated as follows:
\begin{subequations}\label{eq:HamDens}
\begin{align}
    \mathcal{H}_{0}^{(2)} &= \frac{1}{2}z^2\left(\chi_I^{\prime 2}+(\partial_i\chi_I)^2\right)\,;\label{eq:quadratic_hamiltonian_kin}\\
    \mathcal{H}_I^{(2)} &=-\frac{1}{2}\dfrac{\sigma}{\Lambda}z^2\left(\chi_I^{\prime 2}-\left(\partial_i\chi_I\right)^2\right)\,.\label{eq:quadratic_hamiltonian_interaction}
\end{align}
\end{subequations}
Indeed, the Hamiltonian density is defined in terms of $\chi$ and its conjugate momentum density $\pi_\chi$, where the relation between $\chi'$ and $\pi_\chi$ is as follows:
\begin{equation}
    \pi_\chi\equiv\dfrac{\partial\mathcal{L}^{(2)}}{\partial\chi'}=\left(1+\dfrac{\sigma}{\Lambda}\right)z^2\chi'\qquad\implies\qquad \chi'\simeq \left(1-\dfrac{\sigma}{\Lambda}\right)\dfrac{1}{z^2}\pi_\chi\,.
\end{equation}
Therefore, the Hamiltonian density is
\begin{align}
    \mathcal{H}^{(2)}=\chi'\pi_\chi-\mathcal{L}^{(2)}&=\frac{1}{2}\left(1+\dfrac{\sigma}{\Lambda}\right)z^2\left(\chi^{\prime 2}+(\partial_i\chi)^2\right)\nonumber\\
    &=\frac{1}{2}\left(1+\dfrac{\sigma}{\Lambda}\right)z^2\left(\left(1+\dfrac{\sigma}{\Lambda}\right)^{-2}\dfrac{1}{z^4}\pi_\chi^2+(\partial_i\chi)^2\right)\,.
\end{align}
This yields the kinematic ($\mathcal{H}_0^{(2)}$) and interaction ($\mathcal{H}_I^{(2)}$) Hamiltonian densities,
\begin{subequations}
\begin{align}
    \mathcal{H}_{0}^{(2)} &= \frac{1}{2}z^2\left(\dfrac{1}{z^4}(\pi_\chi^I)^{2}+(\partial_i\chi_I)^2\right)\,,\\
    \mathcal{H}_I^{(2)} &= -\frac{1}{2}\dfrac{\sigma}{\Lambda}z^2\left(\dfrac{1}{z^4}(\pi_\chi^I)^{2}-(\partial_i\chi_I)^2\right)\,,
\end{align}
\end{subequations}
where $\pi_\chi$ and $\chi$ are replaced by the variables in the interaction picture, $\pi_\chi^I$ and $\chi_I$, respectively, which evolve with $\mathcal{H}_0^{(2)}$.
Finally, we use one of Hamilton's equations to replace $\pi_\chi^I$ with $\chi_I'$,
\begin{equation}
    \chi_I'=\dfrac{\partial\mathcal{H}_{0}^{(2)}}{\partial\pi_\chi^I}=\dfrac{1}{z^2}\pi_\chi^I\,,
\end{equation}
which results in \eqref{eq:HamDens}.

Now we formulate the expected signals in terms of corrections to the power spectrum using the in-in formalism. The $\chi$ two-point correlation function is
\begin{equation}\label{eq:inin_main}
    \langle{\hat{\chi}{}^2}\rangle = \bra{0} \left(\bar{T}  e^{i \int_{-\infty}^{\tau_\mathrm{e}} \dd\tau\, \mathcal{H}_I^{(2)}}\right) \, \hat{\chi}(\tau_\mathrm{e})^2 \, \left(T e^{ -i \int_{-\infty}^{\tau_\mathrm{e}} \dd\tau \,\mathcal{H}_I^{(2)}}\right) \ket{0}\,,
\end{equation}
where $(\bar{T})~{T}$ stands for the (anti-)time-ordering operator and where $\tau_\mathrm{e}$ corresponds to the conformal time at the end of the ekpyrotic phase. To extract and isolate the correction due to $\sigma$ oscillations in $\mathcal{H}_I^{(2)}$, we shall treat $\mathcal{H}_0^{(2)}$ as unperturbed by the presence of massive field oscillations. Specifically, this shall mean replacing $z=a\Omega(\phi)$ by $\bar z=\bar a\Omega(\bar\phi)$. Then,
we assume a mode function $\bar{\chi}_k$ for the Fourier-space quantized $\hat\chi_\mathbf{k}$ as
\begin{equation}
    \hat{\chi}_\mathbf{k}(\tau) = \bar{\chi}_k(\tau) \, \hat{b}_\mathbf{k} + \bar{\chi}_k^*(\tau) \, \hat{b}^\dagger_{-\mathbf{k}}\,,
\end{equation}
where $\hat{b}^\dagger_{-\mathbf{k}}$ and $\hat{b}_\mathbf{k}$ are the usual creation and annihilation operators;
$\bar{\chi}_k$ is related to the unperturbed canonically normalized mode function $\bar v_k$ [given by \eqref{eq:mode_func_canonical_v}] as
\begin{equation}
    \bar{\chi}_k(\tau) = \dfrac{\bar v_k(\tau)}{\bar z(\tau)}\,.
\end{equation}
The mode function $\bar v_k$ is consistent with an initial Bunch-Davies quantum vacuum \eqref{eq:BD}
and in the super-horizon limit follows
\begin{align}
    \bar v_k(\tau) \stackrel{k\tau\to 0^-}{\simeq} &~\dfrac{-\sqrtb{-\pi\tau}}{2}\left(\dfrac{1+i\cot(\pi\nu)}{\Gamma(1 + \nu)} \left( \dfrac{-k\tau} {2} \right)^{\nu} - i \dfrac{\Gamma(\nu)}{\pi} \left( \dfrac{-k\tau} {2} \right)^{-\nu}\right)\nonumber\\
    \simeq &~ \dfrac{i\sqrtb{-\pi\tau}}{2} \dfrac{\Gamma(\nu)}{\pi} \left( \dfrac{-k\tau} {2} \right)^{-\nu}\,,\label{eq:superHubble_chi}
\end{align}
where we recall from \eqref{eq:zbar}--\eqref{eq:zppozbar} that $\nu=\frac{1}{2}(1-3p+2\sqrtb{p/b})/(1-p)$, which follows from writing $\bar z''/\bar z=(\nu^2-1/4)/\tau^2$.
In the last equality of \eqref{eq:superHubble_chi}, we used the fact that for $\nu>0$ only the term $\propto(-k\tau)^{-\nu}$ is growing. Note from this that $\bar v_k^*(\tau)\simeq-\bar v_k(\tau)$ [and hence $\bar{\chi}_k(\tau)\simeq-\bar{\chi}_k^*(\tau)$] in the super-horizon limit.

Let us go back to \eqref{eq:inin_main} and expand the time-evolution operators to first order. We have
\begin{equation}\label{eq:inin_first_order}
    \langle{\hat{\chi}{}^2}\rangle=\bra{0}\hat{\chi}(\tau_\mathrm{e})^2\ket{0}+2\,\mathrm{Im}\bra{0}\hat{\chi}(\tau_\mathrm{e})^2\int_{-\infty}^{\tau_\mathrm{e}} \dd\tau\,\mathcal{H}_I^{(2)}(\tau)\ket{0}+\cdots,
\end{equation}
where $\cdots$ stands for higher-order contributions of the interaction Hamiltonian. In Fourier space and using \eqref{eq:quadratic_hamiltonian_interaction}, we have
\begin{align}
    \llangle\hat{\chi}_\mathbf{k}\hat{\chi}_{-\mathbf{k}}\rrangle&\simeq|\bar{\chi}_k(\tau_\mathrm{e})|^2-\mathrm{Im}\left(\bar{\chi}_k(\tau_\mathrm{e})^2\int_{-\infty}^{\tau_\mathrm{e}}\dd\tau\,\frac{\sigma}{\Lambda}\bar z^2\left(\bar{\chi}_k^{\prime *2}-k^2\bar{\chi}_k^{*2}\right)\right)\nonumber\\
    &\simeq|\bar{\chi}_k(\tau_\mathrm{e})|^2\left(1+\mathrm{Im}\int_{-\infty}^{\tau_\mathrm{e}}\dd\tau\,\frac{\sigma}{\Lambda}\bar z^2\left(\bar{\chi}_k^{\prime *2}-k^2\bar{\chi}_k^{*2}\right)\right)\,,\label{eq:2ptapprox}
\end{align}
where the double angular brackets indicate a correlator without the delta function, i.e., $\langle\hat{\chi}_\mathbf{k}\hat{\chi}_{-\mathbf{k}}\rangle\equiv\llangle\hat{\chi}_\mathbf{k}\hat{\chi}_{-\mathbf{k}}\rrangle(2\pi)^3\delta^{(3)}(\mathbf{0})$. In the second line above, we used the late-time approximation $|\bar{\chi}_k(\tau_\mathrm{e})|^2\simeq-\bar{\chi}_k(\tau_\mathrm{e})^2$. If we define the $\chi$ dimensionless power spectrum according to
\begin{equation}
    \mathcal{P}_{s}(k)\equiv\frac{k^3}{2\pi^2}\llangle\hat\chi_\mathbf{k}\hat\chi_{-\mathbf{k}}\rrangle\,,
\end{equation}
the unperturbed power spectrum is then
\begin{equation}
    \mathcal{\bar P}_{s}(k)\simeq\frac{k^3}{2\pi^2}\left|\bar{\chi}_k(\tau_\mathrm{e})\right|^2\,.
\end{equation}
From \eqref{eq:2ptapprox}, we thus find the correction to the power spectrum due to the massive field,
$\Delta\mathcal{P}_{s}(k)\equiv\mathcal{P}_{s}(k)-\mathcal{\bar P}_{s}(k)$, satisfies
\begin{equation}\label{eq:correction_chi_power}
    \dfrac{\Delta\mathcal{P}_{s}}{\mathcal{\bar P}_{s}} \simeq \mathrm{Im}\int_{-\infty}^{\tau_\mathrm{e}}\dd\tau\,\frac{\sigma}{\Lambda}\bar z^2\left(\bar{\chi}_k^{\prime *2}-k^2\bar{\chi}_k^{*2}\right)\,.
\end{equation}
Let us already calculate the mode function structure of the integrand explicitly for future use:
\begin{equation}
    \bar z^2\left(\bar{\chi}_k^{\prime *2}-k^2\bar{\chi}_k^{*2}\right)=\frac{\pi}{4}k^2\tau\left(H_\nu^{(1)}(-k\tau)^2-H_{\nu-1}^{(1)}(-k\tau)^2\right)^*\stackrel{-k\tau\gg 1}{\simeq} -ik e^{i\pi\nu} e^{2ik\tau}\,.\label{eq:hankel_mode_funcs}
\end{equation}
In the second equality, we used the large-argument asymptotic form of the Hankel functions.

\subsubsection{Clock signal from oscillatory background}\label{sec:pscsignalanalytical}

In order to evaluate \eqref{eq:correction_chi_power}, we must specify a background solution for $\sigma(\tau)$. Motivated by our previous analytical approximations of the background, we could use \eqref{eq:sigmaAnalyticalFull}, though we recall the expression was fairly complicated with several combinations of special functions. Therefore, we will rather consider a simplified situation. Specifically, let us consider the case of a bump coupling function, which as explained in section \ref{sec:backAnalytical} approximately leads to only the homogeneous solutions $\bar\sigma_1,\bar\sigma_2$ in \eqref{eq:Besselapproxsigma}, which are Bessel functions. For further simplification, we will assume that for the purpose of evaluating \eqref{eq:correction_chi_power} it is justified to assume $-m_\sigma t\gg 1$, so that we may instead use \eqref{eq:sigmabarearlytime}. Thus, we may express the function we wish consider as
\begin{equation}
    \sigma(t)\simeq\Theta(t-t_0)\,\sigma_0\left(\frac{t_0}{t}\right)^{3p/2}\sin\left[-m_\sigma(t-t_0)\right]\,,\label{eq:sigmaClockParam}
\end{equation}
where $\sigma_0$ is an amplitude, which would generally have to be determined numerically. Note that since we expect $\sigma=0$ right until the critical time $t_0$ when the bump coupling occurs, we do not consider any additional phase in the above sine function.

The above function exhibits oscillations in $\sigma$ extending to arbitrarily late times, where the approximation $-m_\sigma t\gg 1$ would break down. However, the main contribution to the power spectrum signal in \eqref{eq:correction_chi_power} comes from a sub-horizon stationary point of the integrand, where the massive field $\sigma$ and the entropy field $\chi$ resonate \cite{Chen:2008wn,Chen:2011zf}. Indeed, since $\sigma\sim e^{-imt}$ in the sub-horizon regime, the integral of \eqref{eq:correction_chi_power} [also using the sub-horizon limit \eqref{eq:hankel_mode_funcs}] is schematically of the form
\begin{equation}
    \int\dd\tau\,e^{2ik\tau-imt}\,,
\end{equation}
which is dominated by the stationary point where $\frac{\dd}{\dd\tau}(2k\tau-mt)=0 \Leftrightarrow 2k/a=m$. Thus, resonance occurs at early times, and the integral is dominated at some instance in time where the approximations $-m_\sigma t\gg 1$ and $-k\tau\gg 1$ are expected to hold [the latter being the justification for using \eqref{eq:hankel_mode_funcs} in \eqref{eq:correction_chi_power}]. We shall soon explicitly check under what conditions this is indeed true, i.e., that resonance occurs before the horizon-exit time. Still, for the time being it shall be reasonable to consider \eqref{eq:sigmaClockParam} on the whole interval $-\infty<\tau<\tau_\mathrm{e}$ for the purpose of evaluating \eqref{eq:correction_chi_power}.

It is practical to recast the trigonometric function \eqref{eq:sigmaClockParam} in terms of exponentials. Further converting to conformal time [recall $-t\propto(-\tau)^{1/(1-p)}$], we have
\begin{equation}
    \sigma(\tau)\simeq\Theta(\tau-\tau_0)\left(-\frac{i\sigma_0}{2}\left(\frac{\tau_0}{\tau}\right)^{\frac{3p}{2(1-p)}}\exp[i\mu p\left(\left(\frac{\tau}{\tau_0}\right)^{\frac{1}{1-p}}-1\right)]+\mathrm{c.c.}\right)\,,
\end{equation}
where we defined
\begin{equation}
    \mu\equiv-\frac{m_\sigma}{H_0}=-\frac{m_\sigma t_0}{p}\,,\label{eq:mudef}
\end{equation}
and c.c.~indicates complex conjugation.
Substituting the above and \eqref{eq:hankel_mode_funcs} in \eqref{eq:correction_chi_power}, we are left with evaluating
\begin{align}
    \dfrac{\Delta \mathcal{P}_{s}}{\mathcal{\bar P}_{s}} &\simeq\frac{\sigma_0}{2\Lambda}\Im\left[-ie^{i\pi\nu}\int_{\tau_0}^{\tau_\mathrm{e}}\dd\tau\,ke^{2ik\tau}\left(-i\left(\frac{\tau_0}{\tau}\right)^{\frac{3p}{2(1-p)}}\exp[i\mu p\left(\left(\frac{\tau}{\tau_0}\right)^{\frac{1}{1-p}}-1\right)]+\mathrm{c.c.}\right)\right]\nonumber\\
    &=\frac{\sigma_0}{2\Lambda}\Im\left[e^{i\pi\nu}\int_{x_\mathrm{e}}^{x_0}\dd x\,\mathcal{A}(x)\left(e^{if_+(x)}-e^{if_-(x)}\right)\right]\,,\label{eq:DeltaPchioPevalInterStep}
\end{align}
where we performed a change of variable to $x\equiv -k\tau$ in going to the second line and defined
\begin{equation}
    \mathcal{A}(x)\equiv\left(\frac{x_0}{x}\right)^{\frac{3p}{2(1-p)}}\,,\qquad f_\pm(x)\equiv\mp \mu p\left(\left(\frac{x}{x_0}\right)^{\frac{1}{1-p}}-1\right)-2x\,.
\end{equation}
Naturally $x_0\equiv -k\tau_0$ and $x_\mathrm{e}\equiv -k\tau_\mathrm{e}$.

We perform the $x$ integrals by using the stationary phase (or saddle-point) approximation, which generally states that
\begin{equation}
    \int_{x_1}^{x_2}\dd x\,\mathcal{A}(x)e^{if(x)}\simeq\mathcal{A}(x_\mathrm{res})\sqrtb{\frac{2\pi}{\pm f_{,xx}(x_\mathrm{res})}}\exp[if(x_\mathrm{res})\pm i\frac{\pi}{4}]\,,
\end{equation}
for some functions $\mathcal{A}(x)$ and $f(x)$ such that $f_{,x}(x_\mathrm{res})=0$ and as long as $x_1<x_\mathrm{res}<x_2$ (the integral is approximately zero otherwise). In the above, the $+$ sign (the $-$ sign) applies when $f_{,xx}(x_\mathrm{res})>0$ [when $f_{,xx}(x_\mathrm{res})<0$]. The saddle-point approximation is most precise when $x_1\to-\infty$ and $x_2\to\infty$, or at least when $x_1\ll x_\mathrm{res}\ll x_2$. If we assumed $\sigma(t)\sim |t|^{-3p/2}\sin(m_\sigma|t|)$ for all $t\in(-\infty,\infty)$, then it would be perfectly justified to use the saddle-point approximation as stated, and this is how one derives the usual standard clock signal \cite{Chen:2011zf}. In our situation, we shall make a similar assumption computationally, but we shall ensure the resonance point lies within the finite integral limits through a Heaviside step function.

We find that the integral involving $f_+(x)$ in \eqref{eq:DeltaPchioPevalInterStep} has no real resonance point and thus approximately vanishes, while the integral involving $f_-(x)$ admits one resonance point at
\begin{equation}
    x_\mathrm{res}=x_0\left(\frac{2(1-p)x_0}{\mu p}\right)^{\frac{1-p}{p}}\,,
\end{equation}
as long as $x_\mathrm{res}<x_0$, i.e., if
\begin{equation}
    x_0<\frac{\mu p}{2(1-p)}\,.\label{eq:condRes}
\end{equation}
We may define $k_0\equiv-1/\tau_0$ as the mode which exits the horizon at the critical time $\tau_0$, so that $x_0=k/k_0$. We may further define
\begin{equation}
    K_\mathrm{res}\equiv\frac{\mu p}{1-p}k_0=\frac{m_\sigma}{m_{\mathrm{h}0}}k_0=a_0m_\sigma\,,\label{eq:Kresdef}
\end{equation}
$m_{\mathrm{h}0}\equiv m_\mathrm{h}(t_0)$ [recalling \eqref{eq:mh2}], and $a_0\equiv a(t_0)$,
such that the condition \eqref{eq:condRes} for resonance becomes
\begin{equation}\label{eq:krescond}
	2k<K_\mathrm{res}\,,
\end{equation}
or $k<k_\mathrm{res}$ with $k_\mathrm{res}\equiv K_\mathrm{res}/2$.
We then find
\begin{equation}
	\frac{x_\mathrm{res}}{x_0}=\frac{\tau_\mathrm{res}}{\tau_0}=\left(\frac{2k}{K_\mathrm{res}}\right)^{\frac{1-p}{p}}\,,\label{eq:tauvskres}
\end{equation}
implying that larger $k$-modes resonate earlier in the contracting phase. The condition \eqref{eq:krescond} thus ensures resonance indeed occurs after the massive field starts oscillating, $\tau_\mathrm{res}>\tau_0$.

The condition \eqref{eq:krescond} is not the only one to guarantee resonance. Indeed, we also need to make sure that resonance occurs before the $k$-mode exists the horizon, i.e., $\tau_\mathrm{res}<\tau_\star\equiv-1/k \Leftrightarrow -k\tau_\mathrm{res}>1$, as well as before the massive field stops oscillating, i.e., $\tau_\mathrm{res}<\tau_{\textrm{h-e}}$. Thus, requiring $\tau_\mathrm{res}(k)<\mathrm{min}\{\tau_{\textrm{h-e}},\tau_\star(k)\}$ sets a lower bound on $k$. One can check that $\tau_\mathrm{res}(k)<\tau_{\textrm{h-e}}$ is equivalent to $k>k_\textrm{h-e}/2$, where
\begin{equation}
    k_\textrm{h-e}\equiv a_\textrm{h-e}m_\sigma=\left(\frac{\mu p}{1-p}\right)^{1-p}k_0=\left(\frac{m_\sigma}{m_{\mathrm{h}0}}\right)^{1-p}k_0
\end{equation}
and $a_\textrm{h-e}\equiv a(\tau_\textrm{h-e})$. Similarly, $\tau_\mathrm{res}(k)<\tau_\star(k)$ is equivalent to $k>k_\textrm{h-e}/2^{1-p}$, and $\tau_\star(k)\gtrless\tau_\textrm{h-e}$ is equivalent to $k\gtrless k_\textrm{h-e}$. Also, for there to be massive field oscillations in the first place, we had better have $\tau_0<\tau_\textrm{h-e}$, which is equivalent to demanding [combining \eqref{eq:mudef} and \eqref{eq:th}]
\begin{equation}
    \mu>\frac{1-p}{p}\qquad\Leftrightarrow\qquad m_\sigma>m_{\mathrm{h}0}\,.\label{eq:mulb}
\end{equation}
Putting everything together, we arrive at different possible sufficient conditions for there to be resonance:
\begin{enumerate}
    \item[(1)] $k_\textrm{h-e}<k<k_\mathrm{res}$, i.e.,
    \begin{equation}
        \left(\frac{\mu p}{1-p}\right)^{1-p}k_0<k<\frac{\mu p}{2(1-p)}k_0\,,\label{eq:krangeres}
    \end{equation}
    provided
    \begin{equation}
        \mu>2^{1/p}\frac{1-p}{p}\,;\label{eq:murange1}
    \end{equation}
    \item[(2)] $k_\textrm{h-e}/2^{1-p}<k<k_\textrm{h-e}<k_\mathrm{res}$, i.e.,
    \begin{equation}
        \left(\frac{\mu p}{2(1-p)}\right)^{1-p}k_0<k<\left(\frac{\mu p}{1-p}\right)^{1-p}k_0<\frac{\mu p}{2(1-p)}k_0\,,
    \end{equation}
    provided $\mu$ is again respecting \eqref{eq:murange1};
    \item[(3)] or $k_\textrm{h-e}/2^{1-p}<k<k_\mathrm{res}<k_\textrm{h-e}$, i.e.,
    \begin{equation}
        \left(\frac{\mu p}{2(1-p)}\right)^{1-p}k_0<k<\frac{\mu p}{2(1-p)}k_0<\left(\frac{\mu p}{1-p}\right)^{1-p}k_0\,,
    \end{equation}
    provided
    \begin{equation}
        2\frac{1-p}{p}<\mu<2^{1/p}\frac{1-p}{p}\,.
    \end{equation}
\end{enumerate}
Case (1) implies $\tau_0<\tau_\mathrm{res}<\tau_\textrm{h-e}<\tau_\star$, while cases (2) and (3) imply $\tau_0<\tau_\mathrm{res}<\tau_\star<\tau_\textrm{h-e}$.
A necessary and sufficient condition may then simply be written as $k_\textrm{h-e}/2^{1-p}<k<k_\mathrm{res}$, i.e.,
\begin{equation}
    \left(\frac{\mu p}{2(1-p)}\right)^{1-p}k_0<k<\frac{\mu p}{2(1-p)}k_0\,,\label{eq:kcondiff}
\end{equation}
provided
\begin{equation}
    \mu>2\frac{1-p}{p}\qquad\Leftrightarrow\qquad m_\sigma>2m_{\mathrm{h}0}\,.
\end{equation}

Considering $k$-modes that fit in the range \eqref{eq:kcondiff},
we can put everything together and obtain
\begin{subequations}
\begin{align}
    \mathcal{A}(x_\mathrm{res})&=\left(\frac{\mu p}{2(1-p)x_0}\right)^{3/2}=\left(\frac{K_\mathrm{res}}{2k}\right)^{3/2}\,,\\
    f_-(x_\mathrm{res})&=-\mu p\left(\frac{p}{1-p}\left(\frac{2(1-p)x_0}{\mu p}\right)^{1/p}+1\right)=-\mu p\left(\frac{p}{1-p}\left(\frac{2k}{K_\mathrm{res}}\right)^{1/p}+1\right)\,,\\
    f_{-,xx}(x_\mathrm{res})&=\frac{4}{\mu}\left(\frac{\mu p}{2(1-p)x_0}\right)^{1/p}=\frac{4}{\mu}\left(\frac{K_\mathrm{res}}{2k}\right)^{1/p}\,,
\end{align}
\end{subequations}
hence \eqref{eq:DeltaPchioPevalInterStep} becomes
\begin{align}
    \dfrac{\Delta \mathcal{P}_{s}}{\mathcal{\bar P}_{s}}\simeq &-\frac{\sigma_0}{2\Lambda}\Im\left(e^{i\pi\nu}\Theta(x_0-x_\mathrm{res})\mathcal{A}(x_\mathrm{res})\sqrtb{\frac{2\pi}{f_{-,xx}(x_\mathrm{res})}}\exp[if_-(x_\mathrm{res})+i\frac{\pi}{4}]\right)\nonumber\\
    =&~\Theta(K_\mathrm{res}-2k)\frac{\sigma_0}{2\Lambda}\sqrtb{\frac{\pi\mu}{2}}\left(\frac{2k}{K_\mathrm{res}}\right)^{-\frac{3}{2}+\frac{1}{2p}}\sin\left(\mu\frac{p^2}{1-p}\left(\frac{2k}{K_\mathrm{res}}\right)^{1/p}+\mu p-\pi\left(\nu+\frac{1}{4}\right)\right)\,.
\end{align}
Note that the Heaviside function appears from the saddle-point approximation of the integral and ensures $2k<K_\mathrm{res}$, but the lower bound on $k$ that we derived in \eqref{eq:kcondiff} came from physical grounds other than the integral itself (making sure resonance occurs in the sub-horizon regime), hence it does not appear in the expression above. We shall thus include another Heaviside function by hand\footnote{Note that the approximation would not be so bad even without this Heaviside function, because $\Delta \mathcal{P}_{s}/\mathcal{\bar P}_{s}$ is approaching zero very fast for $k<k_\mathrm{res}$ as the amplitude scales as $(k/k_\mathrm{res})^{1/(2p)-3/2}$ --- recall the power is a large number when $p\ll 1$.} in the final expression,
\begin{align}
	\dfrac{\Delta \mathcal{P}_{s}}{\mathcal{\bar P}_{s}}
        &\simeq\Theta\big(K_\mathrm{res}-2k\big)\Theta\Big(k-\frac{k_\textrm{h-e}}{2^{1-p}}\Big)\frac{\sigma_0}{2\Lambda}\sqrtb{\frac{\pi m_\sigma}{2|H_0|}}\left(\frac{2k}{K_\mathrm{res}}\right)^{\frac{1-3p}{2p}}\sin\left(\frac{p^2}{1-p}\frac{m_\sigma}{|H_0|}\left(\frac{2k}{K_\mathrm{res}}\right)^{1/p}+\varphi\right)\, \nonumber
        \\
        &=\Theta\big(k_\mathrm{res}-k\big)\Theta\Big(k-\frac{k_\textrm{h-e}}{2^{1-p}}\Big)\frac{\sigma_0}{2\Lambda}\sqrtb{\frac{\pi m_\sigma}{2|H_0|}}
        \left(\frac{k}{k_\mathrm{res}}\right)^{\frac{1-3p}{2p}}
        \sin\left(p\frac{m_\sigma}{m_{\mathrm{h}0}}\left(\frac{k}{k_\mathrm{res}}\right)^{1/p}+\varphi\right)\,,\label{eq:clockSignalAnalytic}
\end{align}
where the phase has been gathered in
$\varphi\simeq pm_\sigma/|H_0|+\pi/4+(n_s-1)\pi/2$
(recall $\nu\simeq 2-n_s/2\approx 3/2$). This result is in agreement with the model-independent derivation of \cite{Chen:2011zf,Chen:2011tu,Chen:2014joa,Chen:2014cwa}, and the $k$ dependence of the envelope of the clock signal corresponds to the case of a direct coupling \cite{Chen:2014joa,Chen:2014cwa}.
An important property here is that the $k$ dependence in the phase of the clock signal is given by the inverse function of $a(t)\propto (-t)^p$. This is how primordial standard clocks work essentially.

\begin{figure}
	\centering
	\includegraphics[width=0.9\textwidth]{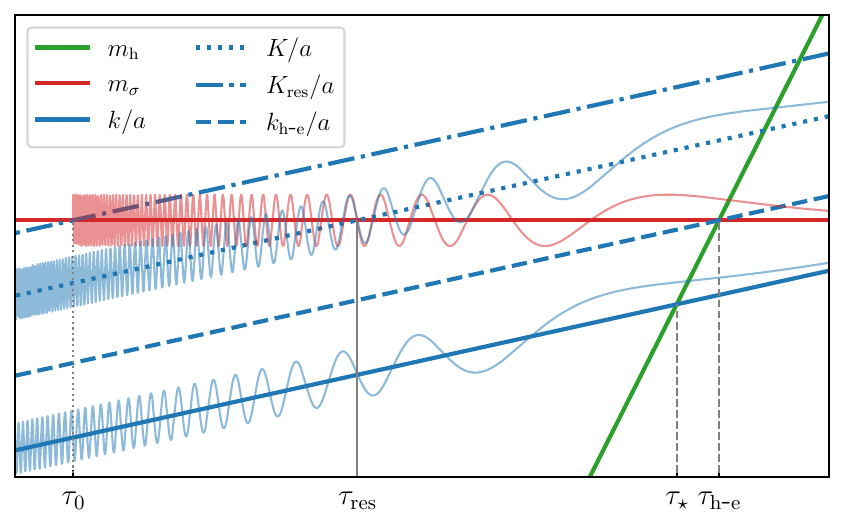}
	\caption{{\footnotesize{Depiction of the resonance between the quantum fluctuations in $\chi_k$ of physical wavenumber $k/a$ (in blue) and the massive field oscillations (in red), which occurs at a time $\tau_\mathrm{res}$ when $K/a=m_\sigma$, where $K\equiv 2k$. The fluctuations in $\chi_k$ freeze at horizon-exit time $\tau_\star$, while the massive field oscillations freeze at horizon-exit time $\tau_{\textrm{h-e}}$. The horizon scale $m_\mathrm{h}$ is shown in green, and $k_\textrm{h-e}$ and $K_\mathrm{res}$ represent modes that set lower and upper bounds on $K$-modes that can undergo resonance. The horizontal axis should be thought of as the logarithm of the conformal time (in absolute value), while the vertical axis represents energy scales on a logarithmic scale.}}}
	\label{fig:sketch}
\end{figure}

A sketch of an example of the relevant scales in this calculation is shown in Fig.~\ref{fig:sketch} to provide more intuition about the resonance process. A mode $\chi_k$ has a physical wavenumber $k/a$ that is depicted in thick solid blue. This $k$-mode is approximately oscillating as $\exp(-ik\tau)$ on sub-horizon scales and freezes on super-horizon scales after $\tau_\star=-1/k$. This is schematically represented by the thin solid blue curve oscillating on top of the $k/a$ line. The horizon scale $m_\mathrm{h}$ is depicted in green. In comparison to this $k$-mode, a massive field of mass $m_\sigma$ (this scale is shown by the thick red line) oscillates from the critical time $\tau_0$ until its horizon exit, at $\tau_{\textrm{h-e}}$, approximately as $\exp[\pm im_\sigma(-t)]\sim\exp[\pm im_\sigma(-\tau)^{1/(1-p)}]$ --- this is schematically represented by the thin red curve. The resonance between the two oscillations occurs at $\tau_\mathrm{res}$, that is when $2k/a=m_\sigma$, and induces a clock signal in the correlations of $\chi_k$. We use the notation $K\equiv 2k$, and we represent this mode as a dotted line and its corresponding oscillatory behavior, $e^{-iK\tau}$, as an overlaid thin solid curve again. We can see that the frequency of these oscillations indeed matches that of the massive field at $\tau_\mathrm{res}$, hence the resonance when $K/a=m_\sigma$. The range of $k$-modes that can undergo resonance may be expressed as $2^pk_{\textrm{h-e}}<K<K_\mathrm{res}$ (note $2^pk_{\textrm{h-e}}\approx k_{\textrm{h-e}}$ when $p\ll 1$), hence for visual purposes $K_\mathrm{res}$ is depicted as a dash-dotted line and $k_{\textrm{h-e}}$ as a dashed line in Fig.~\ref{fig:sketch}. The former crosses $m_\sigma$ at the time $\tau_0$, while the latter crosses both $m_\sigma$ and $m_\mathrm{h}$ at the time $\tau_\textrm{h-e}$.

\subsubsection{Sharp feature signal from step background}\label{sec:sharpsignalanalytical}

In this subsection, we analytically study the signature of a sharp feature alone, trying to isolate the effect of the massive field being suddenly excited. An illustrative example is a step in the massive field's vacuum expectation value, which can be parametrized as
\begin{equation}
    \sigma(\tau) = \sigma_0 \Theta(\tau-\tau_0)\,.\label{eq:sigmastep}
\end{equation}
In fact, if one has a bump coupling function at the background level, which triggers massive field oscillations without a continued driving force, then equation \eqref{eq:sigmabarlatetime} showed that $\sigma$ tends to a generally non-zero constant at late times, i.e., after horizon exit. Thus, \eqref{eq:sigmastep} captures this late-time dynamics as well, though the late-time value of $\sigma(\tau)$ might not be exactly the same as the sudden shift, $\sigma_0$, received at the trigger time $\tau_0$.

Similar to the calculation of the clock signal above, we use \eqref{eq:correction_chi_power} to calculate the correction to the power spectrum. Thereby, we wish to evaluate
\begin{equation}
    \dfrac{\Delta \mathcal{P}_{s}}{\mathcal{\bar P}_{s}} \simeq \frac{\sigma_0}{\Lambda}\,\mathrm{Im}\int_{-\infty}^{\tau_\mathrm{e}}\dd\tau\,\Theta(\tau-\tau_0)\bar z^2\left(\bar{\chi}_k^{\prime *2}-k^2\bar{\chi}_k^{*2}\right)\,.
\end{equation}
Using \eqref{eq:hankel_mode_funcs}, this becomes
\begin{equation}
    \dfrac{\Delta \mathcal{P}_{s}}{\mathcal{\bar P}_{s}} \simeq-\frac{\pi\sigma_0}{4\Lambda}k^2\,\mathrm{Im}\int_{\tau_0}^{\tau_\mathrm{e}}\dd\tau\,\tau\left(H_\nu^{(1)}(-k\tau)^2-H_{\nu-1}^{(1)}(-k\tau)^2\right)\,,
\end{equation}
and using the closed form of the above integral\footnote{The closed form of the relevant integral is given by
\begin{align*}
    \int\dd\tau\,\tau\left(H_{\nu}^{(1)}(-k\tau)^2-H_{\nu-1}^{(1)}(-k\tau)^2\right)=&-\frac{i}{\pi k^2}\Big(2-\pi k\tau\big[(\cot(\pi\nu)-i)J_{\nu-1}(-k\tau)\\
    &\qquad\qquad\qquad\quad\ +\csc(\pi\nu)J_{1-\nu}(-k\tau)\big]H_\nu^{(1)}(-k\tau)\Big)\,.
\end{align*}
} we obtain
\begin{equation}
    \dfrac{\Delta \mathcal{P}_{s}}{\mathcal{\bar P}_{s}} \simeq-\frac{\pi\sigma_0}{4\Lambda}\,\mathrm{Im}\Big[ik\tau\big[(\cot(\pi\nu)-i)J_{\nu-1}(-k\tau)+\csc(\pi\nu)J_{1-\nu}(-k\tau)\big]H_\nu^{(1)}(-k\tau)\Big]_{\tau_0}^{\tau_\mathrm{e}}\,.\label{eq:sharpFull}
\end{equation}
Keeping leading-order terms in $-k\tau_0\gg 1$, $-k\tau_\mathrm{e}\ll 1$, and $\nu\approx 3/2$, we finally find
\begin{equation}
    \dfrac{\Delta \mathcal{P}_{s}}{\mathcal{\bar P}_{s}} \simeq-\frac{\sigma_0}{2\Lambda}\left(1+\cos(-2k\tau_0)\right)=-\frac{\sigma_0}{2\Lambda}\left(1+\cos\left(\frac{2k}{k_0}\right)\right)\,.\label{eq:sharpFeature}
\end{equation}
This matches what has been derived for the sharp feature in \cite{Chen:2011zf}, up to a phase --- the expression goes as $1-\cos(2k/k_0)$ in \cite{Chen:2011zf}.
Note that both approximated results assume $-k\tau_0\gg 1$, so they do not approximate well the corrections to the super-horizon modes $-k\tau_0\ll 1$, even though the result in \cite{Chen:2011zf} vanishes in the limit $k\tau_0 \to 0^-$ coincidentally. Nevertheless, one can check that the full expression \eqref{eq:sharpFull} goes to zero in the deep infrared. 

Sharp features in inflationary models also induce oscillatory signals with the same sinusoidal running as in \eqref{eq:sharpFeature}, though with model-dependent phases and envelops \cite{Starobinsky:1992ts}. As shown in \cite{Chen:2011zf}, unlike the clock signals, such running behavior remains qualitatively the same for both inflationary and non-inflationary scenarios.

The full signal in $\Delta \mathcal{P}_{s}/\mathcal{\bar P}_{s}$ is then expected to be some combination of a sharp feature [e.g., \eqref{eq:sharpFeature}] and a clock signal [e.g., \eqref{eq:clockSignalAnalytic}], but a proper understanding of the superposition of signals requires numerical techniques, which is the subject of the next subsection. Nevertheless, there is a universal relation between the sharp feature signal and the clock signal: recalling \eqref{eq:Kresdef}, we have \cite{Chen:2011zf,Chen:2011tu,Chen:2014joa,Chen:2014cwa}
\begin{equation}
    \frac{2k_\mathrm{res}}{k_0}=\frac{p}{1-p}\mu=\frac{m_\sigma}{m_{\mathrm{h}0}}\label{eq:Kresokc}
\end{equation}
as the ratio of the clock signal resonance wavenumber to the sharp feature horizon-exit wavenumber.

\subsection{Numerical results}

\subsubsection{Methodology}

The equation we wish to solve numerically is $v_k''+(k^2-\tilde z''/\tilde z)v_k=0$, with the usual Bunch-Davies initial conditions, and where we recall $\tilde z=a\Omega(\phi)\sqrt{1+\sigma/\Lambda}$. Numerically, it is useful to rewrite the differential equation as a function of the $e$-folding number $\mathcal{N}$ rather than the conformal time $\tau$. Using the identities $\mathcal{N}'=aH(1-\epsilon)$ and $\mathcal{N}''=(aH)^2(1-\epsilon)(1-\epsilon-\epsilon_{,\mathcal{N}})$, we find
\begin{equation}
    v_{k,\mathcal{N}\mathcal{N}}+\frac{1-\epsilon-\epsilon_{,\mathcal{N}}}{1-\epsilon}v_{k,\mathcal{N}}+\left(\frac{k^2}{(aH)^2(1-\epsilon)^2}-\frac{\tilde z_{,\mathcal{N}\mathcal{N}}}{\tilde z}-\frac{1-\epsilon-\epsilon_{,\mathcal{N}}}{1-\epsilon}\frac{\tilde z_{,\mathcal{N}}}{\tilde z}\right)v_k=0\,.\label{eq:vsofNeom}
\end{equation}
Since we need $a(\mathcal{N})$, another equation must be solved numerically beforehand, namely $a_{,\mathcal{N}}=a/(1-\epsilon)$. [With no massive field, the solution is $\bar a(\mathcal{N})=a_\mathrm{ini}\exp(\mathcal{N}/(1-\bar\epsilon))$, where $a_\mathrm{ini}\equiv a(\mathcal{N}=0)$.] The equation of motion for $v_k$ can alternatively be written in terms of $\chi_k=v_k/\tilde z$ as
\begin{equation}
    \chi_{k,\mathcal{N}\mathcal{N}}+\left(\frac{1-\epsilon-\epsilon_{,\mathcal{N}}}{1-\epsilon}+2\frac{\tilde z_{,\mathcal{N}}}{\tilde z}\right)\chi_{k,\mathcal{N}}+\frac{k^2}{(aH)^2(1-\epsilon)^2}\chi_k=0\,,\label{eq:chiofNeom}
\end{equation}
where we can re-express the coefficient of the first-derivative term, $\chi_{k,\mathcal{N}}$, as
\begin{equation}
    \frac{1-\epsilon-\epsilon_{,\mathcal{N}}}{1-\epsilon}+2\frac{\tilde z_{,\mathcal{N}}}{\tilde z}=1+\frac{2-\epsilon_{,\mathcal{N}}}{1-\epsilon}-\sqrtb{\frac{2}{b}}\phi_{,\mathcal{N}}+\frac{\sigma_{,\mathcal{N}}}{\Lambda+\sigma}\,.
\end{equation}
Since \eqref{eq:chiofNeom} involves fewer derivatives on $\tilde z$ than \eqref{eq:vsofNeom}, and since the effects of the massive field oscillations are mainly captured by the time dependence of $\tilde z$, we choose to solve the $\chi_k$ equation directly rather than solving the equation for $v_k$ and then transforming to $\chi_k$. This has shown to be more robust in terms of numerical accuracy.\footnote{Since $\chi_k$ freezes while $|v_k|$ grows on super-horizon scales, it is often preferable anyway, numerically speaking, to solve the $\chi_k$ equation rather than the $v_k$ equation after Hubble crossing. This is done, e.g., in \cite{Price:2014xpa} in the case of inflation. Formally, though, it is completely equivalent to solve either the $v_k$ or $\chi_k$ equation.}

As the perturbations emerge from sub-horizon scales, the initial conditions at $\mathcal{N}_\mathrm{BD}$ are set by the Bunch-Davies (BD) state as
\begin{equation}
    v_k(\mathcal{N}_\mathrm{BD})=\frac{1}{\sqrtb{2k}}e^{-ik\bar\tau(\mathcal{N}_\mathrm{BD})}\,,\qquad \left.v_{k,\mathcal{N}}\right|_{\mathcal{N}_\mathrm{BD}}=\left.-\frac{i}{\bar a\bar H(1-\bar\epsilon)}\sqrtb{\frac{k}{2}}e^{-ik\bar\tau}\right|_{\mathcal{N}_\mathrm{BD}}\,,\label{eq:vICs}
\end{equation}
together with the relations
\begin{equation}
    \chi_k(\mathcal{N}_\mathrm{BD})=\frac{v_k(\mathcal{N}_\mathrm{BD})}{\tilde z(\mathcal{N}_\mathrm{BD})}\,,\qquad\left.\chi_{k,\mathcal{N}}\right|_{\mathcal{N}_\mathrm{BD}}=\left.\frac{v_k}{\tilde z}\left(\frac{v_{k,\mathcal{N}}}{v_k}-\frac{\tilde z_{,\mathcal{N}}}{\tilde z}\right)\right|_{\mathcal{N}_\mathrm{BD}}\,.
\end{equation}
Note that we make sure to always set the initial Bunch-Davies vacuum before the massive field contribution kicks in, i.e., $\mathcal{N}_\mathrm{BD}<\mathcal{N}_0$, so that the scaling solution is a good approximation for $a$, $H$, and $\epsilon$ initially.
We also need $\bar\tau(\mathcal{N})$, which we can get from the scaling solution as $\bar\tau(\mathcal{N})=\tau_\mathrm{ini}e^{-\mathcal{N}}$, with $\tau_\mathrm{ini}\equiv\tau(\mathcal{N}=0)=p/((1-p)a_\mathrm{ini}H_\mathrm{ini})$.
Moreover, we always choose the time $\mathcal{N}_\mathrm{BD}$ such that the mode is sufficiently sub-horizon, i.e., such that $k/(a(\mathcal{N}_\mathrm{BD})|H(\mathcal{N}_\mathrm{BD})|)$ is some constant factor, orders of magnitude larger than unity.
Then, fluctuations exit the horizon at a time $\mathcal{N}_\star$ when $k=a(\mathcal{N}_\star)H(\mathcal{N}_\star)$; we expect $\mathcal{N}_\star\approx\ln[k/(a_\mathrm{ini}|H_\mathrm{ini}|)]$ according to the scaling solution.
The final time $\mathcal{N}_\textrm{final}$ at which the power spectrum is evaluated --- $\mathcal{P}_s(k)=k^3\left|\chi_k(\mathcal{N}_\mathrm{final})\right|^2/(2\pi^2)$ --- is determined such that the mode is sufficiently super-horizon, i.e., such that $k/(a(\mathcal{N}_\textrm{final})|H(\mathcal{N}_\textrm{final})|)$ is some constant factor, orders of magnitude smaller than unity. Finally, the correction to the power spectrum is evaluated following
\begin{equation}
    \frac{\Delta\mathcal{P}_s}{\bar{\mathcal{P}}_s}(k)=\frac{\mathcal{P}_s(k)-\bar{\mathcal{P}}_s(k)}{\bar{\mathcal{P}}_s(k)}\,,\label{eq:DeltaPnum}
\end{equation}
where the unperturbed power spectrum is taken to be the standard power-law function, $\bar{\mathcal{P}}_s(k)=A_s(k/k_\mathrm{pivot})^{n_s-1}$. The tilt is set as explained below \eqref{eq:ns}, and the amplitude is read from the full numerical power spectrum $\mathcal{P}_s$ at some sufficiently large pivot scale (small $k_\mathrm{pivot}$) where we do not expect any deviation from a power-law.\footnote{This is well justified: smaller $k$-modes exit the horizon at earlier times, so $k_\mathrm{pivot}$ is chosen such that it exits the horizon before $\tau_0$ when the massive field gets excited. In other words, for $k\ll -1/\tau_0=k_0$, we expect $\Delta\mathcal{P}_s(k)\simeq 0$.}

\subsubsection{Sharp bump}

\begin{figure}
	\centering
	\includegraphics[width=0.9\textwidth]{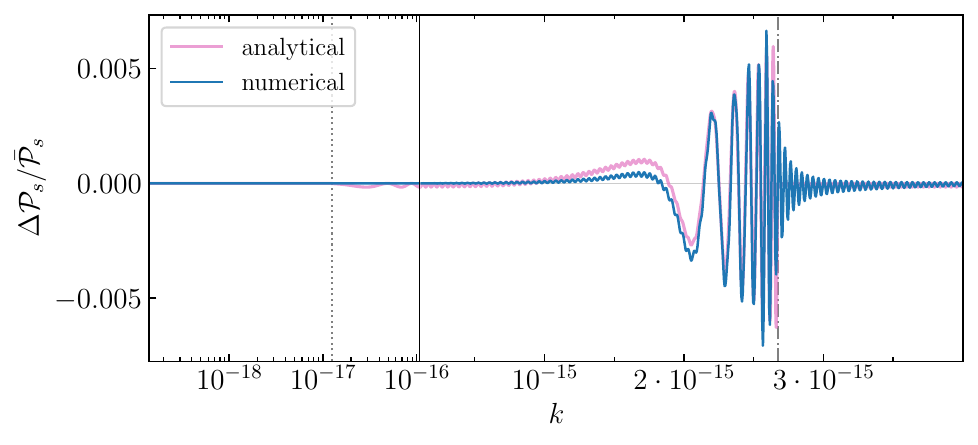}
	\caption{{\footnotesize{Plot of the relative deviation from an unperturbed power-law power spectrum as a function of comoving wavenumber $k$ in arbitrary units (see text for explanation). The blue curve is the result of a numerical computation, while the pink curve, mostly hidden behind the blue curve, is an analytically motivated fit to the numerical result (see text for further details). The model parameters are $p=1/10$, $\phi_\mathrm{ini}=19.5$, $\phi_0=13.2$, $m_\sigma=5\times 10^9|H_\mathrm{ini}|$, $\varrho=1$ (bump coupling function), and $\Lambda=0.01$, which translates into the following: $m_\sigma\approx 2.19\times 10^{-14}$, $|H_0|\approx 5.74\times 10^{-18}$, $\sigma_0\approx 8.7\times 10^{-7}$, $\mu\approx 3.81\times 10^3$, $k_\mathrm{res}\approx 2.67\times 10^{-15}$, and $k_\mathrm{res}/k_0\approx 211.7$. Horizontally, note that a logarithmic scale is used for $k\lesssim 10^{-16}$, while a linear scale is used for larger $k$-values. The separation is delineated by the vertical thin black solid line. The vertical gray dotted line denotes the value of $k_0\approx 1.26\times 10^{-17}$, and the vertical gray dash-dotted line denotes the value of $k_\mathrm{res}$.}}}
	\label{fig:pert_base2_full}
\end{figure}

Let us present a first example of correction to the power spectrum \eqref{eq:DeltaPnum} numerically computed following the above methodology. In Fig.~\ref{fig:pert_base2_full}, the model parameters are $p=1/10$, $\phi_\mathrm{ini}=19.5$, $\phi_0=13.2$, $m_\sigma=5\times 10^9|H_\mathrm{ini}|$, and $\varrho=1$, for the bump-like $\Xi$ coupling. The background evolution corresponding to this case has not been shown in the previous section, but closely resembles the blue curves in the left- and right-hand panels of Fig.~\ref{fig:1phirhom}, qualitatively speaking, in the sense that $\sigma$ undergoes many high-frequency oscillations. The small value of $\phi_0$ implies the oscillations in $\sigma$ are triggered when the background Hubble scale is quite high in absolute value (recall the $\phi$ vs $|H|$ relation that can be inferred from the left-hand panel of Fig.~\ref{fig:1p}). In turn, this implies that the $k$-modes that undergo resonance and lead to oscillatory features in the power spectrum are on very small scales. In Fig.~\ref{fig:pert_base2_full}, this corresponds to modes $k\gtrsim 10^{-18}$ in Planck units. Note, however, that this is a purely arbitrary choice, which was made for the purpose of improving numerical stability. Yet, in principle, the exact same signal could be shifted to arbitrarily smaller $k$-values, simply by exciting the massive field earlier in the evolution when $|H|$ was smaller (i.e., taking a larger value of $\phi_0$). Therefore, at fixed $m_\sigma/|H_\mathrm{ini}|$, the units of the $k$ values in the power spectra shown hereafter should be viewed as arbitrary.

Figure \ref{fig:pert_base2_full} presents the result of a fully numerical computation (in blue): for a range of $k$-modes, the full mode equation \eqref{eq:chiofNeom} is solved numerically given a numerically solved background, as described earlier. In pink, we contrast the numerical result with the expected result from the analytical estimates derived in section \ref{sec:analytical}. Specifically, this curve represents the sum of the expected standard clock signal \eqref{eq:clockSignalAnalytic} and sharp feature signal \eqref{eq:sharpFeature}. We use the same model parameters as in the numerical computations, but we allow for two parameters to be free: the overall amplitude, set by $\sigma_0$, and the phase of the oscillatory signals. While $\sigma_0$ can be approximately read off numerically from the amplitude of the first $\sigma$ oscillation after it has been triggered, this will lead to an analytical amplitude in $\Delta\mathcal{P}_s/\mathcal{\bar P}_s$ that is in the correct ballpark, but it will not be a precise fit. By allowing $\sigma_0$ to be free and fitting it to the full numerical $\Delta\mathcal{P}_s/\mathcal{\bar P}_s$, $\sigma_0$ is modified by an $\mathcal{O}(1)$ factor, and one gets a much better overall fit. Likewise for the phases: we start with the analytically predicted phases, but correct them by a small factor that we determine by fitting the analytical functions to the numerical result.

The two types of signals that are expected, the clock and the sharp feature signals, can be easily distinguished in Fig.~\ref{fig:pert_base2_full}. First, the clock signal is the largest oscillatory feature, which we expect to be within the interval $k_{\textrm{h-e}}/2^{1-p}\approx 1.56\times 10^{-15}\lesssim k \lesssim k_\mathrm{res}\approx 2.67\times 10^{-15}$. This is indeed where the clock signal manifests itself the most, but upon closer inspection of the plot, the clock signal seems to fade below $10^{-15}$. Also, something that is not captured by the leading-order analytical estimate is the damping of the signal for $k>k_\mathrm{res}$, to the right of the vertical gray dash-dotted line, damping that appears to be exponential in $k$. Second, overlaid on top of the clock signal are the sharp feature oscillations of constant (and much higher) frequency, but of smaller amplitude. The analytical sharp feature signal was derived assuming $k\gg k_0$, but the numerical result shows that not much signal appears until $k\gtrsim 5\times 10^{-16}$, more than an order of magnitude larger than $k_0\approx 1.26\times 10^{-17}$. One could ask how much the mixing between the clock and sharp feature signals could be altered if we changed the different coupling parameters. We now turn to this question.

\begin{figure}
	\centering
	\includegraphics[width=0.85\textwidth]{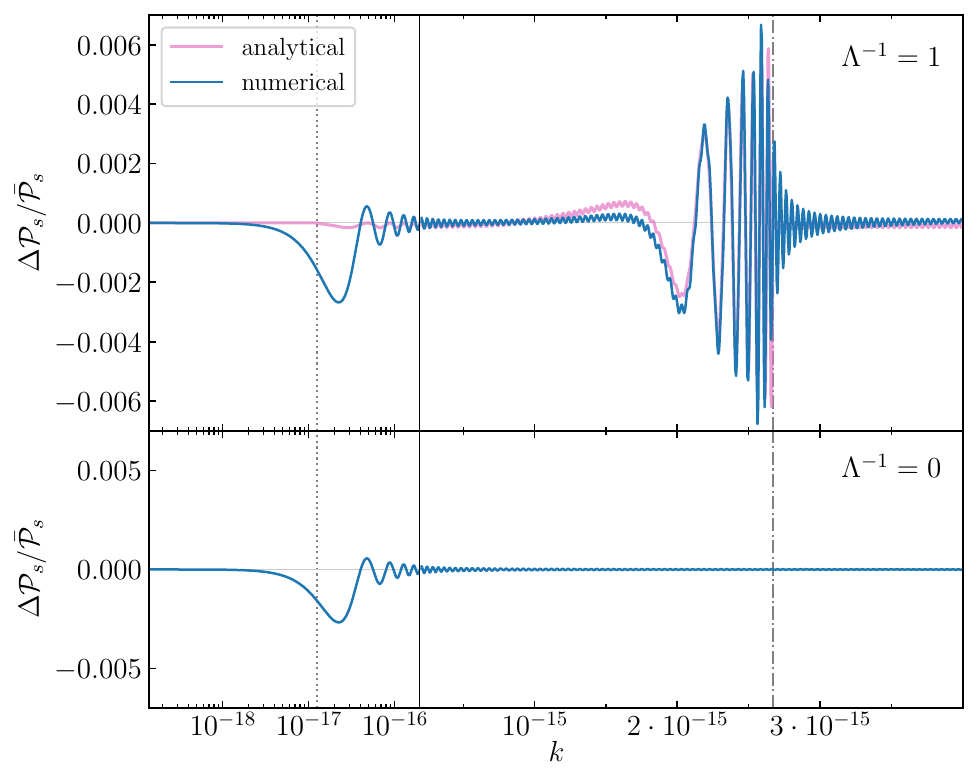}
	\caption{{\footnotesize{Same model parameters as in Fig.~\ref{fig:pert_base2_full}, except $\varrho=0.01$ for the background coupling (correspondingly $\sigma_0\approx 8.7\times 10^{-5}$) and for the perturbation coupling we have $\Lambda=1$ in the top plot and $\Lambda\to\infty$ (decoupling limit) in the bottom plot.}}}
	\label{fig:pert_base1_full}
\end{figure}

In Fig.~\ref{fig:pert_base1_full}, we use the same model parameters as in Fig.~\ref{fig:pert_base2_full}, except for the couplings: we reduce the turning radius $\varrho$ in $\phi$-$\sigma$ space by two orders of magnitude, hence the background coupling parameter that controls the amplitude of the $\sigma$ oscillations is enhanced by two orders of magnitude; and we reduce the $\chi$-$\sigma$ coupling $\Lambda^{-1}$ at the perturbation level by two orders of magnitude (top plot) and turn it off completely (bottom plot). We note that there is very little difference between $\varrho=0.01$ and $\Lambda=1$ (top plot of Fig.~\ref{fig:pert_base1_full}) versus $\varrho=1$ and $\Lambda=0.01$ (Fig.~\ref{fig:pert_base2_full}). Indeed, according to our analytical estimates \eqref{eq:clockSignalAnalytic} and \eqref{eq:sharpFeature}, the amplitude of the oscillatory signals in $\Delta\mathcal{P}_s/\mathcal{\bar P}_s$ is controlled by the ratio $\sigma_0/\Lambda\sim(\varrho\Lambda)^{-1}$, where $\sigma_0$ is more or less proportional to $1/\varrho$. Since the product $\varrho\Lambda$ is the same in the two plots, the signals are very similar, and the relative amplitude of the clock and sharp feature signals is thus expected to always be the same, i.e., both amplitudes are expected to be proportional to $(\varrho\Lambda)^{-1}$. Naturally, the overall amplitude of the signal can be modified by changing $(\varrho\Lambda)^{-1}$ as long as one remains under perturbative control, which is to say that $|\sigma|\ll\min\{\Lambda,\varrho\}$.

There is one major difference between the top plot of Fig.~\ref{fig:pert_base1_full} and Fig.~\ref{fig:pert_base2_full}, though: for $k\lesssim 1.5\times 10^{-15}$, the sharp feature oscillations are not damped, and in fact, they grow in amplitude as $k$ approaches $k_0$ (the vertical dotted gray line) with a decreasing frequency, until the signal goes to $0$ for smaller $k$-values (as expected). This effect is not captured by our analytical estimates. Indeed, as can be seen in the lower plot of Fig.~\ref{fig:pert_base1_full}, turning off the $\chi$-$\sigma$ coupling by sending $\Lambda\to\infty$ results in no clock signal and no sharp feature signal due to $\chi$-$\sigma$ interactions. However, what appears is a different type of sharp feature signal, solely due to gravitational interactions, i.e., through the time dependence of $\epsilon$ in the $\chi_k$ equation of motion.\footnote{This could also be seen from the in-in formalism following the same methodology as in section \ref{sec:analytical}, separating the equation of state as $\epsilon=\bar\epsilon+\Delta\epsilon$ in the expression for $z$ and finding the resulting interaction Hamiltonian. One would then need to analytically find $\Delta\epsilon$ to compute $\Delta\mathcal{P}_s/\mathcal{\bar P}_s$. We leave this for future work, but for some model-independent estimates, see \cite{Chen:2011zf,Chen:2011tu}.} We recall in the middle panel of Fig.~\ref{fig:1phirhom} that increasing the background coupling $1/\varrho$ increases the amplitude of the oscillatory corrections to the background equation of state $\epsilon$. Therefore, it is the oscillations in $\epsilon$ in that case that resonate with $\chi_k$ and lead to oscillations in the power spectrum as seen near $k_0$ in Fig.~\ref{fig:pert_base1_full}. We checked that, in all the examples shown in this section, turning off the direct perturbation coupling by sending $\Lambda\to\infty$ always leads to some oscillatory signal, i.e., the gravitational interaction is always present, though its effect is subdominant in most instances (for the choice of parameters explored, i.e., when $1/\varrho$ is small enough that $|\Delta\epsilon|/\bar\epsilon$ remains very small).

\begin{figure}
	\centering
    \includegraphics[width=0.85\textwidth]{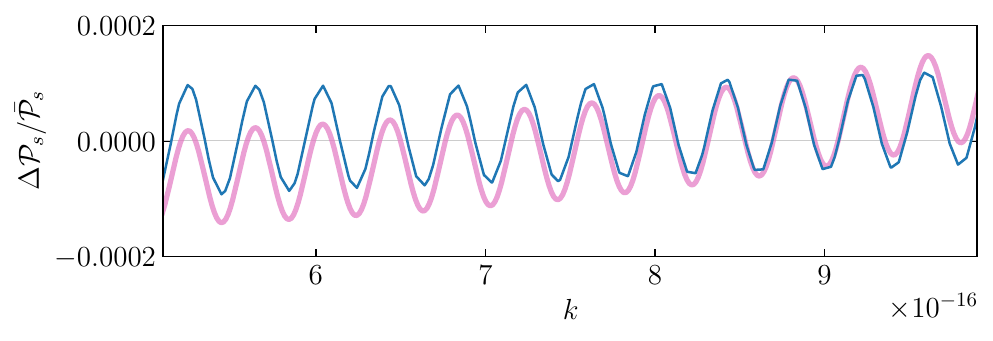}
    \includegraphics[width=0.85\textwidth]{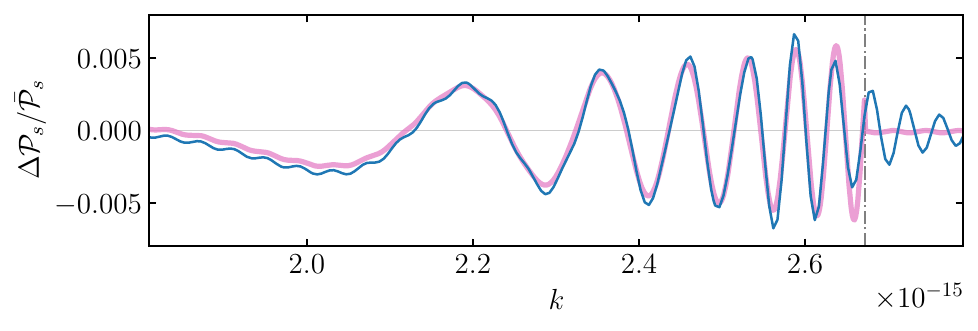}
	\caption{{\footnotesize{Same as the top plot of Fig.~\ref{fig:pert_base1_full}, but emphasizing two separate $k$-ranges: the top plot highlights the sharp feature signal, while the bottom plot highlights the clock signal. As before, the blue and pink curves are the numerical and analytical results, respectively.}}}
	\label{fig:pert_base1_full2}
\end{figure}

When the direct $\chi$-$\sigma$ coupling is turned on, say with $\Lambda=1$ as in the top plot of Fig.~\ref{fig:pert_base1_full}, we must emphasize that the fit between the analytical estimates and the numerical results is quite good when zooming on the respective regimes of the sharp feature and clock signals. We demonstrate this in Fig.~\ref{fig:pert_base1_full2}, which shows the same results as in the top plot of Fig.~\ref{fig:pert_base1_full}, but where specific $k$-ranges are selected to highlight the sharp feature signal (top plot of Fig.~\ref{fig:pert_base1_full2}) and the clock signal (bottom plot of Fig.~\ref{fig:pert_base1_full2}). The frequency, amplitude, and phase of both signals are quite well fit, modulo the slight scale dependence in the top plot due to the mixture of sharp feature and clock signals, and modulo the damping of the clock signal for $k>k_\mathrm{res}$ (to the right of the vertical dash-dotted gray line), which is not captured by our leading-order analytic estimate.

\begin{figure}
	\centering
	\includegraphics[width=0.9\textwidth]{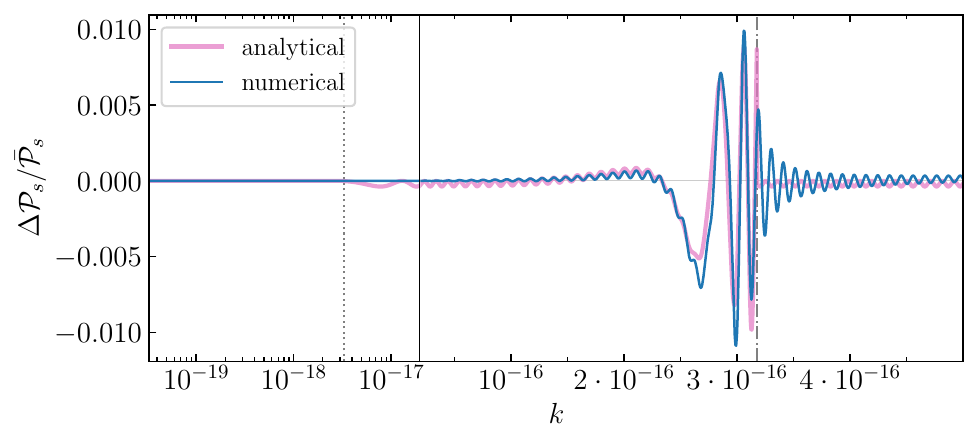}
	\caption{{\footnotesize{Same parameters as in Fig.~\ref{fig:pert_base2_full}, except with $p=1/11$, hence $m_\sigma\approx 2.43\times 10^{-15}$, $|H_0|\approx 1.27\times 10^{-18}$, $\sigma_0\approx 2.1\times 10^{-6}$, $\mu\approx 1.92\times 10^3$, $k_\mathrm{res}\approx 3.18\times 10^{-16}$, and $k_\mathrm{res}/k_0\approx 96$.}}}
	\label{fig:pert_base4_full}
\end{figure}

Putting the emphasis on the clock signal and the sharp feature signal due to a direct $\chi$-$\sigma$ coupling, let us show more examples similar to Fig.~\ref{fig:pert_base2_full} with $\varrho=1$ and $\Lambda=0.01$. In Fig.~\ref{fig:pert_base4_full}, we use the same value for $m_\sigma/|H_\mathrm{ini}|$ and $\phi_0$, but we set $p=1/11$, which results in $m_\sigma\approx 2.43\times 10^{-15}$, $|H_0|\approx 1.27\times 10^{-18}$, $\sigma_0\approx 2.1\times 10^{-6}$, $\mu\approx 1.92\times 10^3$, $k_\mathrm{res}\approx 3.18\times 10^{-16}$, and $k_\mathrm{res}/k_0\approx 96$. Thus, this case has in effect a slightly smaller value of $m_\sigma$ and $\mu$, and the ratio $k_\mathrm{res}/k_0$ is smaller as per \eqref{eq:Kresokc}. Therefore, Fig.~\ref{fig:pert_base4_full} resembles Fig.~\ref{fig:pert_base2_full} (recall the $k$ units should be viewed as arbitrary), except for a few main differences: the smaller $\mu$ value in Fig.~\ref{fig:pert_base4_full} implies a smaller clock signal frequency at any given $k$, and as such, there are fewer clock signal oscillations in the power spectrum; and the ratio $k_\mathrm{res}/k_0$ being smaller implies a smaller spacing between the dotted ($k_0$) and dash-dotted ($k_\mathrm{res}$) vertical gray lines, and as such, there is more of a mix between the clock and sharp feature signals, though they are still quite distinct from one another.

\begin{figure}
	\centering
	\includegraphics[width=0.9\textwidth]{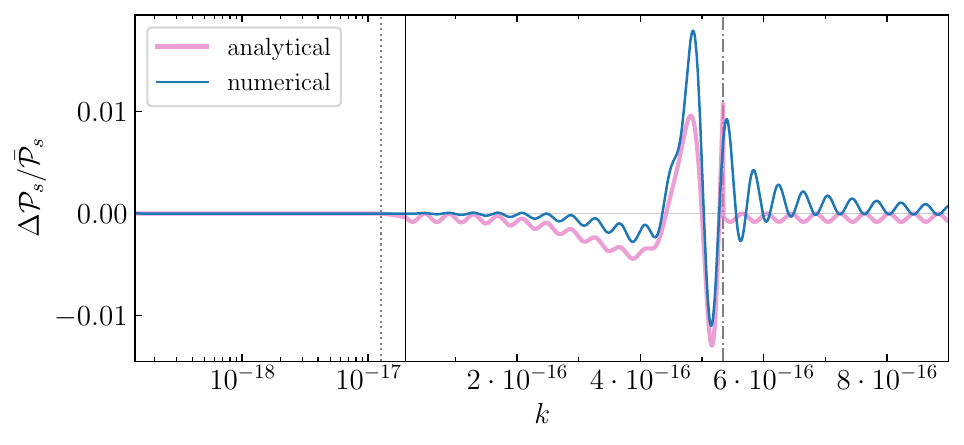}
	\caption{{\footnotesize{Same parameters as in Fig.~\ref{fig:pert_base2_full}, except for $m_\sigma=10^9|H_\mathrm{ini}|$, hence $m_\sigma\approx 4.37\times 10^{-15}$, $\sigma_0\approx 5.2\times 10^{-6}$, $\mu\approx 7.62\times 10^2$, $k_\mathrm{res}\approx 5.34\times 10^{-16}$, and $k_\mathrm{res}/k_0\approx 42$.}}}
	\label{fig:pert_base3_full}
\end{figure}

\begin{figure}
	\centering
	\includegraphics[width=0.9\textwidth]{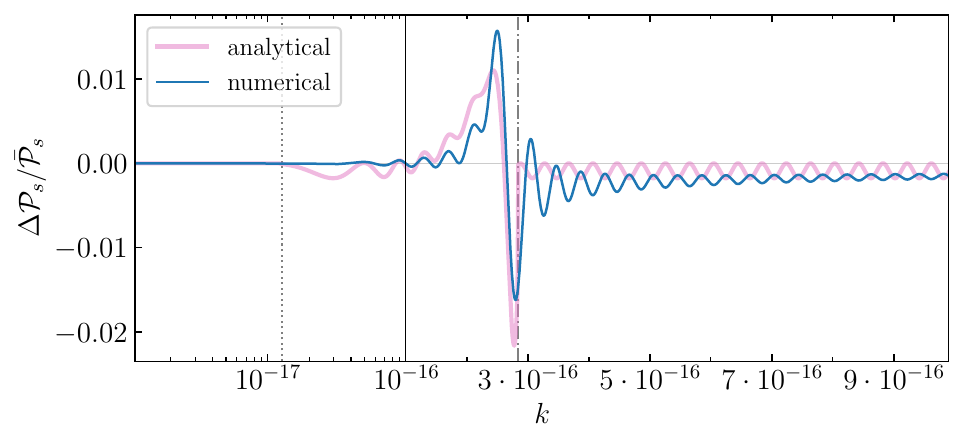}
	\caption{{\footnotesize{Same parameters as in Fig.~\ref{fig:pert_base2_full}, except for $m_\sigma=5.3\times 10^8|H_\mathrm{ini}|$, hence $m_\sigma\approx 2.32\times 10^{-17}$, $\sigma_0\approx 9.85\times 10^{-6}$, $\mu\approx 403.9$, $k_\mathrm{res}\approx 2.83\times 10^{-16}$, and $k_\mathrm{res}/k_0\approx 22.44$.}}}
	\label{fig:pert_base5_2_full}
\end{figure}

\begin{figure}
	\centering
	\includegraphics[width=0.9\textwidth]{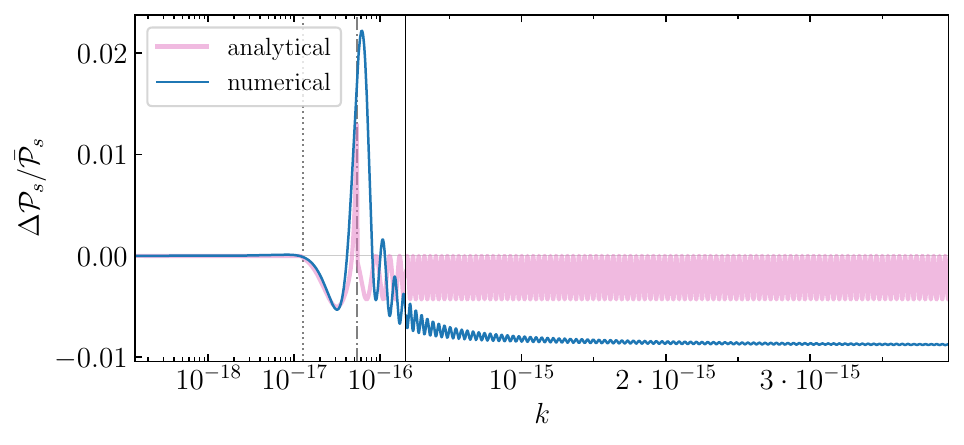}
	\caption{{\footnotesize{Same parameters as in Fig.~\ref{fig:pert_base2_full}, except for $m_\sigma=10^8|H_\mathrm{ini}|$, hence $m_\sigma\approx 4.37\times 10^{-16}$, $\sigma_0\approx 5.4\times 10^{-5}$, $\mu\approx 76.2$, $k_\mathrm{res}\approx 5.34\times 10^{-17}$, and $k_\mathrm{res}/k_0\approx 4.2$.}}}
	\label{fig:pert_base5_full}
\end{figure}

These effects can be isolated and amplified by using the same parameters as in Fig.~\ref{fig:pert_base2_full} and just lowering the mass parameter, set by $m_\sigma/|H_\mathrm{ini}|$. Figures \ref{fig:pert_base3_full}, \ref{fig:pert_base5_2_full}, and \ref{fig:pert_base5_full} show such examples, where one sees the effect of having successively lower values of $m_\sigma$, $\mu$, and $k_\mathrm{res}/k_0$ (in fact, one can compare Figs.~\ref{fig:pert_base2_full}, \ref{fig:pert_base4_full}, \ref{fig:pert_base3_full}, \ref{fig:pert_base5_2_full}, and \ref{fig:pert_base5_full}, which all have successively lower values of $m_\sigma$, $\mu$, and $k_\mathrm{res}/k_0$). As those are lowered, we see fewer clock signal oscillations, a smaller relative difference between the amplitude of the clock and sharp feature signals, and a tightened window between the dotted and dash-dotted vertical gray lines.

Some related observations: in Fig.~\ref{fig:pert_base3_full}, it is interesting to note that the damping of the clock signal to the right of $k_\mathrm{res}$ is less severe than in the examples with larger values of $\mu$, precisely because of the smaller relative difference between the amplitude of the clock and sharp feature signals. Note, once again, that, besides knowing it is sinusoidal oscillation, we cannot quite capture the phase of the oscillation analytically since, in particular, the analytically predicted sharp feature signal is out of phase with the numerical result to the right of the clock signal. We further observe that in Figs.~\ref{fig:pert_base5_2_full} and \ref{fig:pert_base5_full},
$k_\mathrm{res}/k_0$ is so small there is barely a single clock signal oscillation, and in Fig.~\ref{fig:pert_base5_full} this is to the extent that the sharp feature signal only manifests itself for $k>k_\mathrm{res}$. Also, note that in these cases even in the analytical approximations there are very few oscillations in the clock signals.

There is one additional noticeable effect the analytical function cannot capture, which is the fact that we are left with a (negative) offset in $\Delta\mathcal{P}_s/\mathcal{\bar P}_s$ that approaches a constant toward large $k$-values in Figs.~\ref{fig:pert_base5_2_full} and \ref{fig:pert_base5_full}. The reason for this offset is due to the background freezing of $\sigma$ when $m_\sigma$ becomes significantly smaller than $m_\mathrm{h}$, the horizon scale. (In section \ref{sec:pscsignalanalytical}, we assumed $\sigma$ to be purely sinusoidal, which did not take into account its late-time freezing; a refined analytical treatment could use Bessel functions.) As we saw in Figs.~\ref{fig:1p} and \ref{fig:1Comp}, as well as equation \eqref{eq:sigmabarlatetime}, $\sigma$ approaches a constant at late times, and the value of this constant --- let us call it $\sigma_\mathrm{late}$ --- sensitively depends on the phase of the oscillation at horizon crossing. Thus, $\sigma_\mathrm{late}$ depends on $t_0$, $t_\textrm{h-e}$, as well as $m_\sigma$, $\varrho$, and all other model parameters [recall \eqref{eq:Nc} and \eqref{eq:Nh}] in a non-trivial way. In some cases, $|\sigma_\mathrm{late}|$ can be somewhat larger than $\sigma_0$; see, e.g., the red curve in Fig.~\ref{fig:1p}. It turns out this offset was rather small in comparison to the dominant clock signal in the cases of Figs.~\ref{fig:pert_base2_full}, \ref{fig:pert_base1_full}, \ref{fig:pert_base4_full}, \ref{fig:pert_base3_full}, though still discernible upon closer inspection (compare the heights of the blue and pink curves in the high-$k$ limit). In Figs.~\ref{fig:pert_base5_2_full} and \ref{fig:pert_base5_full}, the offset is definitely noticeable (especially in Fig.~\ref{fig:pert_base5_full}), indicating that $\sigma_\mathrm{late}$ has a relatively large effect in comparison to the clock signal. We generally expect $\mathcal{P}_s(k)\propto k^3|v_k|^2/\tilde z^2$, with $\tilde{z}^2\approx(1+\sigma/\Lambda)\bar{z}^2$ when the direct coupling dominates over the gravitational coupling, so a constant deviation from zero, $\sigma_\mathrm{late}\neq 0$, at late times (assuming $|\sigma_\mathrm{late}|\ll\Lambda$) implies $\mathcal{P}_s(k)\simeq(1-\sigma_\mathrm{late}/\Lambda)\mathcal{\bar P}_s(k)$ at large $k$, hence
\begin{equation}
    \frac{\Delta\mathcal{P}_s}{\mathcal{\bar P}_s}\stackrel{k\gg k_\mathrm{res}}{\loooongrightarrow}-\frac{\sigma_\mathrm{late}}{\Lambda}\,.\label{eq:offset}
\end{equation}
We have numerically checked that \eqref{eq:offset} indeed holds from reading off the approximate value of $\sigma_\mathrm{late}$ in the background numerical solutions and comparing $-\sigma_\mathrm{late}/\Lambda$ to the offset seen in the above figures. We note that, while the offset is always present, it is most discernible when $m_\sigma$ is small (and correspondingly when $\mu$ and $k_\mathrm{res}/k_0$ are also small), because then there is little $k$-space for the clock signal to comparatively grow in amplitude. However, irrespective of the clock signal, the significance of the offset \eqref{eq:offset} is solely controlled by $\sigma_\mathrm{late}/\Lambda$, which does not depend on $m_\sigma$ in any obvious and direct way.

\subsubsection{Mild bump}

In the previous subsection, we saw examples where the clock signal was generally the most dominant signal (e.g., Fig.~\ref{fig:pert_base2_full}). As we will further discuss in the next section, detecting such a clock signal would have profound implications since it would be a very strong indication that primordial perturbations were generated in an ekpyrotic phase of contraction. The region of parameter space where the clock signal is the most `isolated' is when $m_\sigma$, $\mu$, and $k_\mathrm{res}/k_0$ are sufficiently large. So far we have studied the case where $\Xi$ is a bump-like (Gaussian) coupling function, i.e., at the background level, the coupling between $\sigma$ and $\phi$ is quickly turned on and off; recall \eqref{eq:Xibumpexp}. As mentioned at the end of section \ref{sec:backNumMethod}, the sharpness parameter $\delta$ has always been fixed to $10^{-3}$. In this subsection, we will explore what happens if the transition is made less sharp, i.e., if it is milder, by looking at larger values of $\delta$.

\begin{figure}
	\centering
	\includegraphics[width=0.9\textwidth]{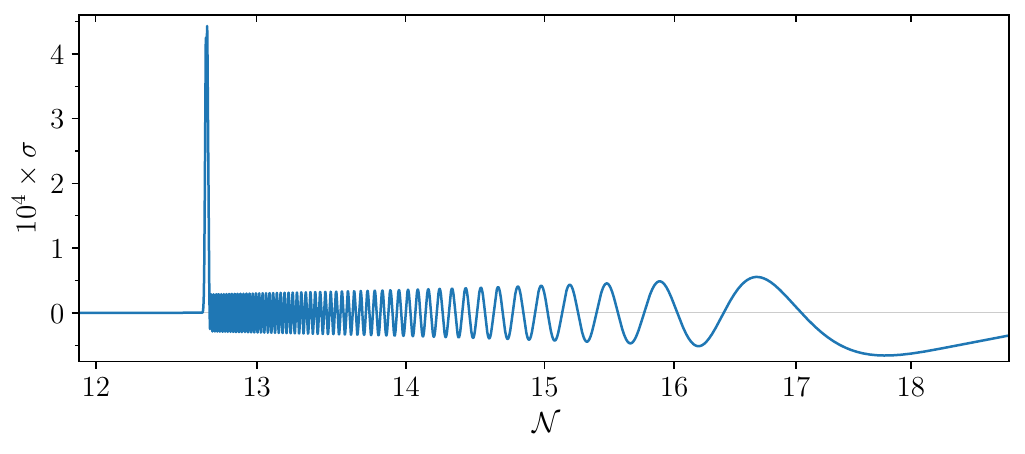}
	\caption{{\footnotesize{Background solution for the massive field $\sigma$ as a function of the $e$-folding number $\mathcal{N}$ in the case of a mild bump, i.e., the coupling function $\Xi$ is taken according to \eqref{eq:Xibumpexp} as in Figs.~\ref{fig:1p}, \ref{fig:1Comp}, and \ref{fig:1phirhom}, but the sharpness parameter $\delta$ is multiplied by a factor of $5$. The other parameters are the same as in Fig.~\ref{fig:pert_base2_full}, i.e., $p=1/10$, $\phi_\mathrm{ini}=19.5$, $\phi_0=13.2$, $m_\sigma/|H_\mathrm{ini}|=5\times 10^{9}$, except for the turning radius, which we take to be $\varrho=10^{-3}$ here.}}}
	\label{fig:sigmaOfNmildBump}
\end{figure}

Let us first understand the effect of a milder bump at the background level. An example is shown in Fig.~\ref{fig:sigmaOfNmildBump}, where $\delta=5\times 10^{-3}$ (five times larger than before). Already, one can see the effect of a wider, longer bump: when the heavy field is excited, it does not immediately start oscillating as a massive clock according to \eqref{eq:sigmabarearlytime} with initial amplitude $\sigma_0$; instead, at first it `slowly' goes up and down its potential while the trajectory turns in field space. This is the larger and wider peak at around $\mathcal{N}\sim 12.6$ in Fig.~\ref{fig:sigmaOfNmildBump}. Once the `bending' is done, the massive field settles into its standard oscillatory behavior at the bottom of its potential, and this yields the subsequent oscillations that match \eqref{eq:sigmabarearlytime}. Note that, in general, a milder transition results in a smaller initial amplitude $\sigma_0$ for the oscillations, hence in Fig.~\ref{fig:sigmaOfNmildBump} we purposely amplified the coupling constant to $1/\varrho=10^3$ (note that $|\sigma|/\varrho$ remains $<1$ at all times). In this example, one reads off $\sigma_0\approx 2.88\times 10^{-5}$.

\begin{figure}
	\centering
	\includegraphics[width=0.9\textwidth]{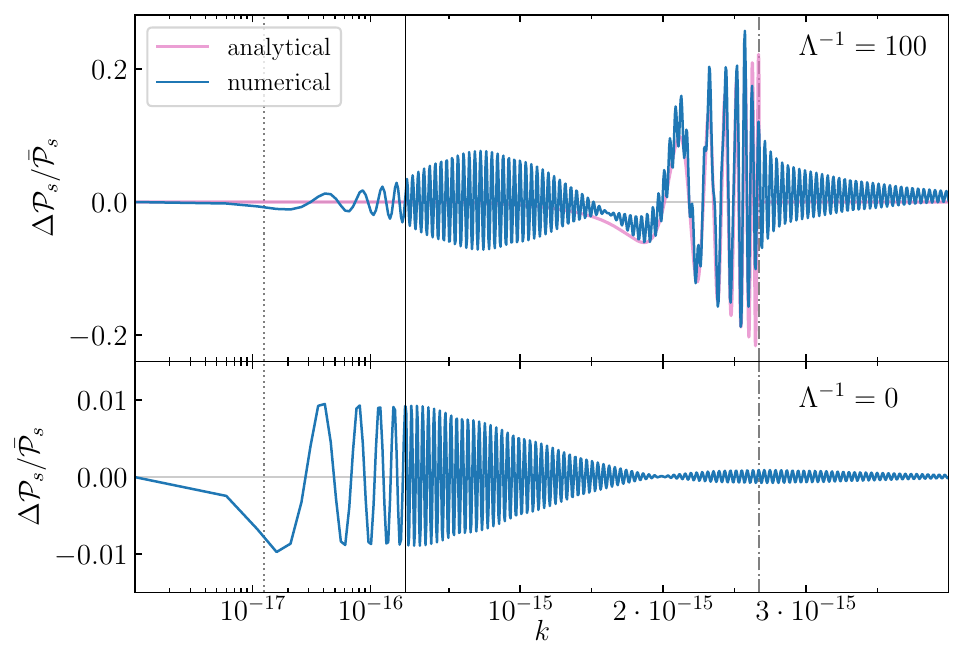}
	\caption{{\footnotesize{Perturbations for the background shown in Fig.~\ref{fig:sigmaOfNmildBump} --- see the caption of that figure for the parameter values. The analytical curve in pink only accounts for the clock signal.}}}
	\label{fig:pert_base7_2}
\end{figure}

Let us then explore the resulting perturbations and how the power spectrum is modified. In Fig.~\ref{fig:pert_base7_2}, we solve for the perturbations with a $\sigma$-$\chi$ coupling constant $1/\Lambda=0.01$ (top) and with no direct coupling (only gravitational; bottom). In the former case, we notice that the clock signal is still present in the range $10^{-15}\lesssim k\lesssim 2.7\times 10^{-15}$ as in Fig.~\ref{fig:pert_base2_full}, but the sharp feature signal is somewhat different. Indeed, it now has an envelope that is itself oscillatory; for example, it peaks at around $k\sim 0.75\times 10^{-15}$ with an amplitude that is about half of the clock signal's amplitude and has a trough at around $k\sim 1.6\times 10^{-15}$. The signal from the purely gravitational coupling (bottom plot in Fig.~\ref{fig:pert_base7_2}) is subdominant, but also has an interesting sharp feature pattern (with no apparent clock signal). Note that the analytical fit shown in the figure in pink only considers the estimated clock signal \eqref{eq:clockSignalAnalytic}, i.e., we did not include the estimated sharp feature signal \eqref{eq:sharpFeature} since it cannot capture the envelope evolution. This is not surprising since the step function \eqref{eq:sigmastep} becomes less appropriate to model a smooth transition. We expect the reason for the modulated envelop of the sharp feature signal to be an interference between two sharp feature signals, introduced at the start and at the end of the bump, respectively. As such, the oscillating wavelength of the envelop decreases as the width of the bump increases.

\begin{figure}
	\centering
	\includegraphics[width=0.9\textwidth]{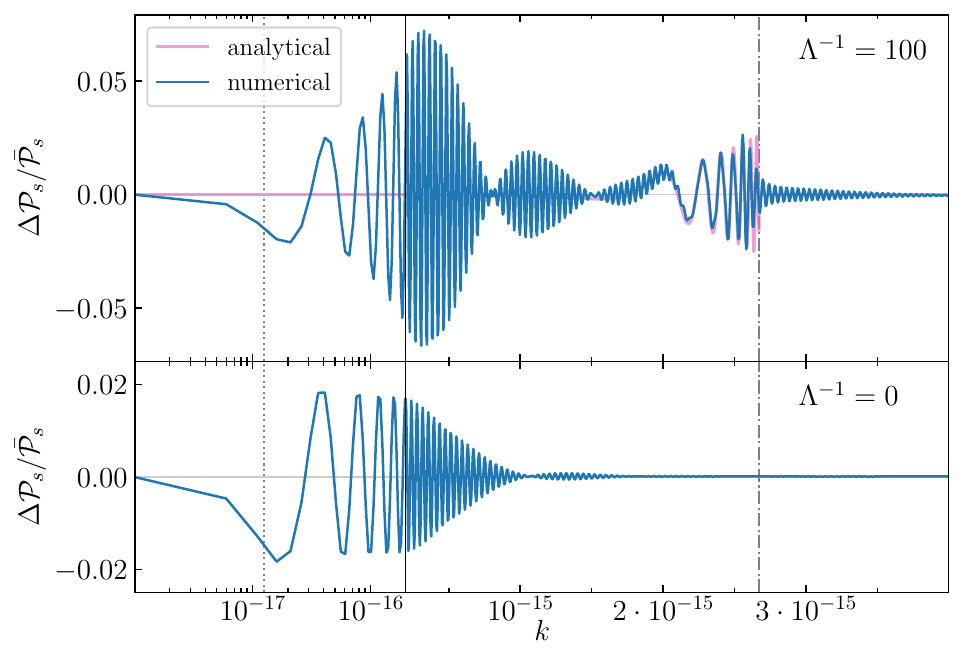}
	\caption{{\footnotesize{Same model parameters as in Fig.~\ref{fig:pert_base7_2}, except $\delta$ is twice as large, i.e., $\delta=10^{-2}$.}}}
	\label{fig:pert_base7_1}
\end{figure}

The sharp feature signal can be even more important if we further increase the $\delta$ parameter, i.e., if we further smooth the $\Xi$ coupling, as shown by Fig.~\ref{fig:pert_base7_1}. The sharp feature signal has an even more complex envelope evolution, and it dominates in amplitude over the clock signal. Note that a significant fraction of the sharp feature signal here comes from the gravitational coupling (as can be seen in the bottom plot). This is due to the fact that we kept a very tight turning radius ($\varrho=10^{-3}$), despite the transition being smoother. (The background evolution for $\sigma$ is very similar to that of Fig.~\ref{fig:sigmaOfNmildBump}, except the first peak is even wider.) Interestingly, despite the large amplitude of the sharp feature, its signal and the clock signal are still very distinct from one another: the sharp feature signal seems to decay just fast enough for the clock signal to stand out in the interval $2\times 10^{-15}\lesssim k\lesssim k_\mathrm{res}$. This is in stark contrast with the full model examples studied in the inflationary scenario, where the sharp feature and clock signals are typically much more entangled \cite{Chen:2014cwa}.

Let us comment more on this last observation. In inflation, the amplitudes of the sharp feature and clock signals are similar in the transition region. This means that, in data analyses, if one only uses the analytic clock signal template, one would get significant misrepresentation. In the ekpyrotic models here, we seem to see the opposite trend. In nearly all examples, the sharp feature signals always already decay away before they reach the clock signals. In fact, one could have thought that the transition would have been more complicated than in the inflation case, because in the ekpyrotic models, the clock signals lie in the $k$-region that has $k<k_\mathrm{res}$ and so are closer to the sharp feature signals. (In inflation, the clock signals lie in the $k$-region that has $k>k_\mathrm{res}$.) But despite this, they decay before they reach the clock signals. This is good news for the data analyses, because we can use the analytic template of the clock signals more reliably than in inflation.

Starting from $k=k_\mathrm{res}$, an inflationary clock signal runs to larger $k$, while an ekpyrotic clock signal runs to smaller $k$. Although the clock signal in ekpyrosis decays fast [towards smaller $k$ as $k^{(1-3p)/(2p)}$ according to \eqref{eq:clockSignalAnalytic}], it runs to smaller values of $k$ (i.e., towards the sharp feature signal) in the first place. This is why the separation of the signals has more to do with the fact that, in ekpyrosis, the sharp feature signal somehow decays faster (towards larger $k$) or has a smaller amplitude than in inflation.

Two things might be at play here that offer an explanation: the size of $\delta$ and $\mu$. For the former, we saw that the transition sharpness controls the heights of the first $\sigma$ peak and thus the amplitude of the sharp feature signal compared to the clock signal. For the latter, we saw that the mass-to-Hubble ratio $\mu$ controls the ratio $k_\mathrm{res}/k_0$, i.e., the relative separation between the two signals. Therefore, let us explore a little more how varying these parameters affects our conclusions.

\begin{figure}
	\centering
	\includegraphics[width=0.9\textwidth]{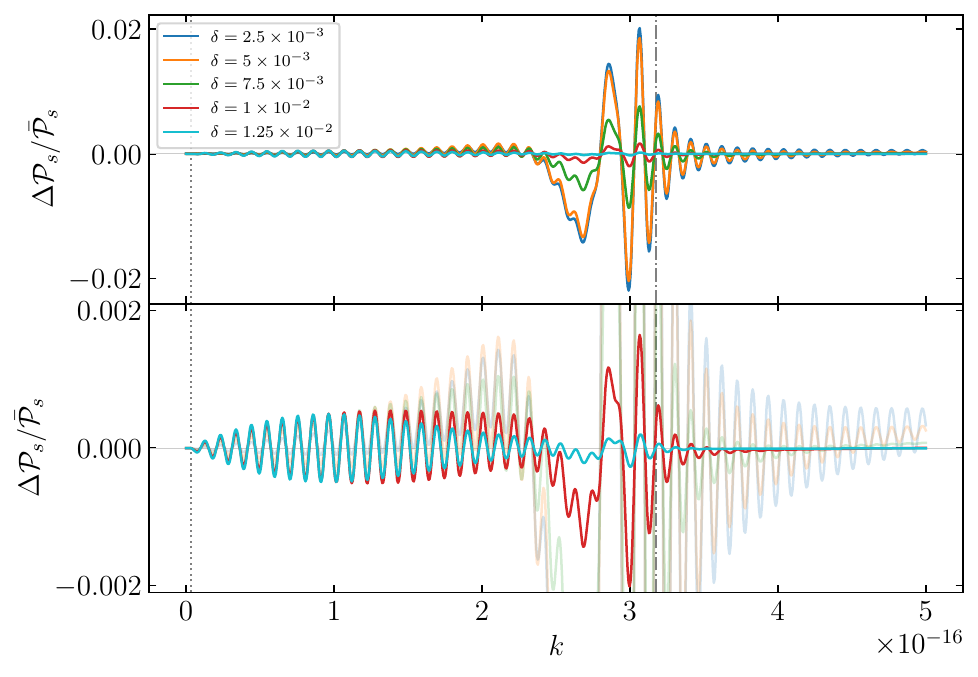}
	\caption{{\footnotesize{The model parameters are the same as in Fig.~\ref{fig:pert_base4_full}, i.e., $p=1/11$, $\phi_\mathrm{ini}=19.5$, $\phi_0=13.2$, $m_\sigma/|H_\mathrm{ini}|=5\times 10^9$, $\varrho=1$, and $\Lambda=0.01$, except $\delta$ is varied. Curves of different color represent different values of $\delta$ as indicated by the legend. The bottom plot is a zoom-in of the top plot --- the range of the vertical axis is ten times smaller --- for the sake of better visualizing the larger $\delta$ values (we further make the cyan and red curves stand out by reducing the opacity of the other curves).}}}
	\label{fig:pert_base8}
\end{figure}

\begin{figure}
	\centering
	\includegraphics[width=0.9\textwidth]{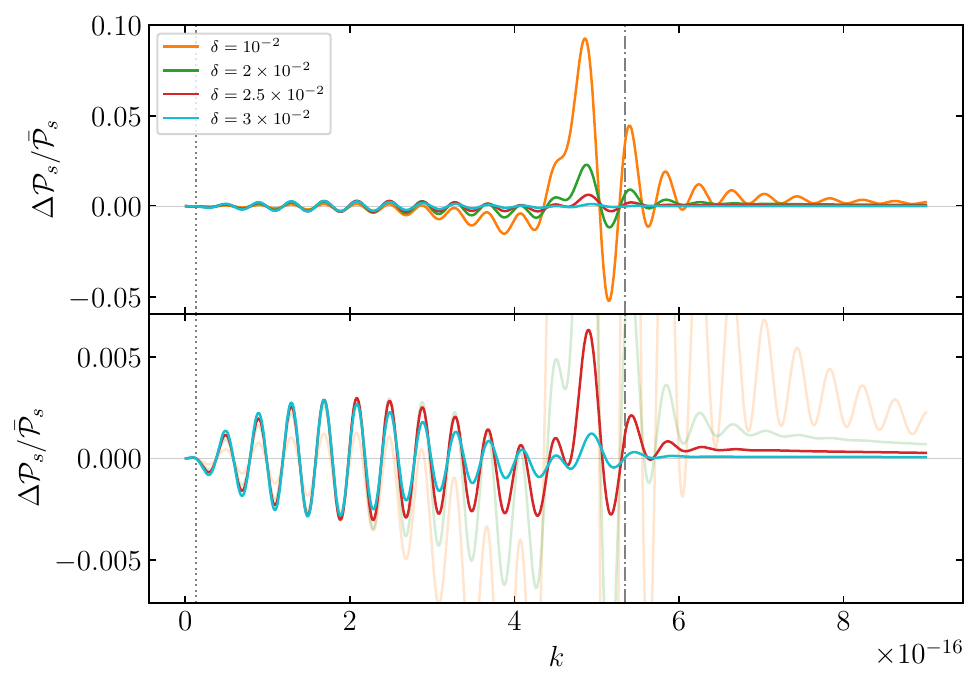}
	\caption{{\footnotesize{The model parameters are the same as in Fig.~\ref{fig:pert_base3_full}, i.e., $p=1/10$, $\phi_\mathrm{ini}=19.5$, $\phi_0=13.2$, $m_\sigma/|H_\mathrm{ini}|=10^9$, $\varrho=1$, and $\Lambda=0.01$, except $\delta$ is varied as in Fig.~\ref{fig:pert_base8}. The bottom plot is a zoom-in of the top plot also as in Fig.~\ref{fig:pert_base8}.}}}
	\label{fig:pert_base9}
\end{figure}

In Fig.~\ref{fig:pert_base8}, we use parameter values that yield $\mu\approx 1.92\times 10^3$ and $k_\mathrm{res}/k_0\approx 96$, while in Fig.~\ref{fig:pert_base9} we have $\mu\approx 7.62\times 10^2$ and $k_\mathrm{res}/k_0\approx 42$ (see their respective captions), and in both figures curves of different color show the results for different values of $\delta$. The trend we observe is that the larger $\delta$ is (so the smoother the transition is), the smaller the clock signal is, especially relative to the sharp feature signal. In Fig.~\ref{fig:pert_base8}, we see that when $\delta>10^{-2}$, we get to a point where the sharp feature and clock signals start mixing with one another a little more. This is even more the case in Fig.~\ref{fig:pert_base9} since we recall a smaller $k_\mathrm{res}/k_0$ ratio implies fewer clock oscillations and a clock signal that gets closer to the sharp feature signal in $k$-space. Thus, for $\delta>2.5\times 10^{-2}$ in Fig.~\ref{fig:pert_base9}, the clock signal is barely distinguishable from the sharp feature signal.

From these examples, a conclusion is that determining how much the two signals mix cannot be answered in a model-independent way, emphasizing the importance of studying full models.
On the other hand, the full model examples studied so far in inflation all have relatively small $k_\mathrm{res}/k_0$, of order ${\cal O}(10)$; so it is worth examining higher values of $k_\mathrm{res}/k_0$ for the inflationary scenario, like what we did here, to see if the two types of signals would become less mixed or remain the same.

\subsubsection{Plateau and step}

Let us end our survey with plateau-like and step-like coupling functions. As we saw in section \ref{sec:plateauStepBack}, the behavior of the massive field $\sigma$ at the background level can be quite different when the $\sigma$-$\phi$ coupling remains `on', as opposed to being quickly switched `on and off' as with a bump-like coupling function. Indeed, the coupling resulted in some sort of centrifugal force pushing the massive field up its potential, modulated by some small oscillations along the way. At the background level, such oscillations are very different from `standard clocks' oscillating as $\sim\sin(-m_\sigma t)$. In fact, recall from section \ref{sec:backAnalytical} that the sum of the homogeneous solutions [$\sim\sin(-m_\sigma t)$ at early times] and particular solutions involved some complicated cancellations among special functions, which were hard to track analytically. Correspondingly, we do not have analytical insight into the expected signal at the perturbation level. Nevertheless, we can expect the following: the figures of section \ref{sec:plateauStepBack} showed that the longer the coupling remained `on', the higher up its potential $\sigma$ was pushed, and the more the ekpyrotic background scaling solution was disrupted. In fact, for a step-like coupling function, for which the coupling always remains `on' after the transition at $\phi_0$, we expect the effective background equation of state $\epsilon$ to asymptote $3$, and for such an equation of state we do not expect a near scale-invariant scalar power spectrum for $\chi$ [unless the coupling function $\Omega(\phi)$ is adjusted, but we do not wish to do that]. Therefore, we can expect the modes that exit the horizon at late times (large-$k$, small-scale modes) to receive a large $\Delta\mathcal{P}_s$ correction. We will confirm this numerically below.

\begin{figure}
	\centering
	\includegraphics[width=0.9\textwidth]{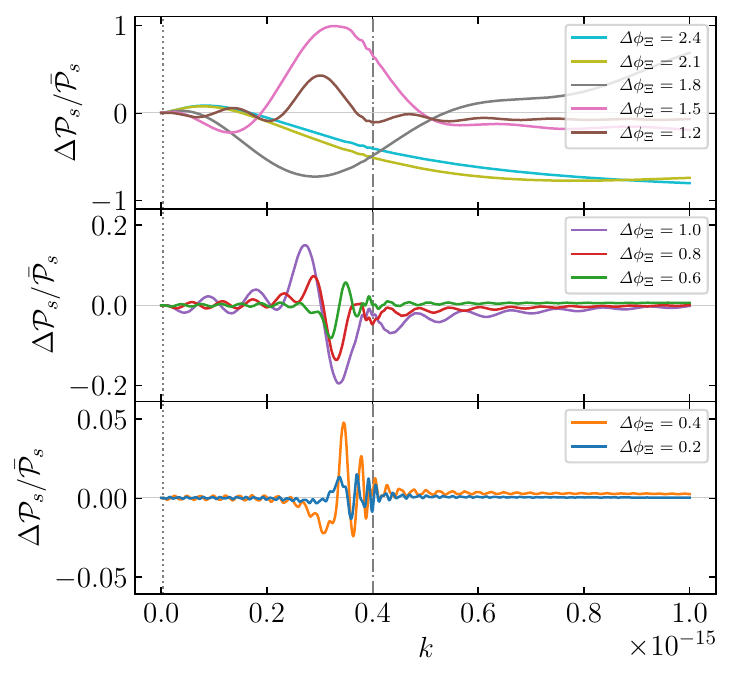}
	\caption{{\footnotesize{Plots of the relative deviation from an unperturbed power-law power spectrum as a function of comoving wavenumber $k$ in arbitrary units when the background coupling function $\Xi$ is a plateau-like function \eqref{eq:Xiplateautanh} of various widths $\mathit{\Delta}\phi_\Xi\equiv\phi_0-\phi_\mathrm{e}$. The corresponding background solutions for $\sigma$ have been shown in Fig.~\ref{fig:plateau_more} --- the color coding is the same. We recall the model parameters are $\phi_\mathrm{ini}=19.5$, $\phi_0=13.2$, $p=1/11$, $\varrho=1$ and $m_\sigma=10^{9.8}|H_\mathrm{ini}|$, and we further set $\Lambda=0.01$ here. Note that the three plots have three different vertical ranges that cover different orders of magnitude as $\mathit{\Delta}\phi_\Xi$ is varied.}}}
	\label{fig:pert_base10}
\end{figure}

Figure \ref{fig:pert_base10} shows the numerical results for $\Delta\mathcal{P}_s/\mathcal{\bar P}_s$ given the background solutions already shown in Fig.~\ref{fig:plateau_more}, namely for a plateau-like coupling function $\Xi(\phi)$ of increasing field-space width $\mathit{\Delta}\phi_\Xi$. All other parameters are kept fixed (see the caption of Fig.~\ref{fig:pert_base10} for their values). We recall from Fig.~\ref{fig:plateau_more} that the larger $\mathit{\Delta}\phi_\Xi$ is, the higher up the potential $\sigma$ gets before oscillating at the bottom of its potential, and the higher the amplitude of these oscillations. This explains why the larger $\mathit{\Delta}\phi_\Xi$ is in Fig.~\ref{fig:pert_base10}, the larger the overall amplitude of $\Delta\mathcal{P}_s/\mathcal{\bar P}_s$.

An important thing we notice from Fig.~\ref{fig:pert_base10} is that, despite the background massive field eventually undergoing `standard clock oscillations' once the coupling turns off (as long as this happens before freezing, so if $\mathit{\Delta}\phi_\Xi\lesssim 2$; again, recall Fig.~\ref{fig:plateau_more} and the discussion around it), the perturbations do not exhibit a clear-cut standard clock signal as for the bump-like coupling we have seen in the previous subsections. Indeed, even when the plateau is somewhat narrow (say the blue curve with $\mathit{\Delta}\phi_\Xi=0.2$), the standard clock signal that we recognize just to the left of the vertical dash-dotted gray line (depicting $k_\mathrm{res}$) is somewhat altered --- both its frequency dependence and its envelope behavior. One particular reason is the following: recalling Fig.~\ref{fig:plateau_more}, $\sigma$ in all the examples undergoes the same small oscillations after the critical time while being pushed up to higher values. These (`non-standard') oscillations lead to a feature that is different from the standard clock signal, but which is very much in the same $k$-window. We note that this signal is always there --- even in the top plot showing the larger values of $\mathit{\Delta}\phi_\Xi$, one can see some small wiggle just to the left of $k_\mathrm{res}$ indicative of these oscillations --- but it is engulfed by some larger oscillatory feature, which we discuss next.

For smaller $\mathit{\Delta}\phi_\Xi$ values in the bottom plot, we notice the presence of some kind of secondary sharp feature signal. However, as $\mathit{\Delta}\phi_\Xi$ is increased, this signal is washed out, and a more important oscillatory signal appears, peaking in amplitude at around $k\sim 0.33\times 10^{-15}$. As $\mathit{\Delta}\phi_\Xi$ is increased, the frequency of these oscillations shrinks, while their amplitude grows.

The reason the sharp feature signals are more complicated in the plateau case is that the plateau introduces two sharp features. One at the start of the plateau and another at the end. We have already seen some effects of this in the bump case, but in the plateau case, the two sharp features are more separated in scales and strength, and the net effect becomes very different. From Fig.~\ref{fig:plateau_more}, we can see that the end of plateau introduces a sharp edge in the evolution of the $\sigma$ field, and this sharp feature is more pronounced than the one at the start of the plateau. Moreover, the oscillation amplitude of the $\sigma$ field is also more pronounced after the end of the plateau than the beginning of the plateau. Therefore, in the plateau case, we should really treat the mode that exits the horizon at the time when the plateau ends as $k_0$.

In the top plot of Fig.~\ref{fig:pert_base10}, we see that the correction to the power spectrum can become $\mathcal{O}(1)$ the larger $\mathit{\Delta}\phi_\Xi$ is. In fact, increasing the plateau width even more, and certainly with a step-like coupling function, we obtain a very large correction to the power spectrum, eventually deviating completely from being just red-tilted. This was to be expected given the backreaction at the background level we alluded to when the coupling remains `on' for long enough. While we do not show the small-$k$ regime on a logarithmic scale in Fig.~\ref{fig:pert_base10}, we can nevertheless mention that $\Delta\mathcal{P}_s/\mathcal{\bar P}_s$ still goes to $0$ below $k_0$ (the vertical dotted gray line).

Let us end by mentioning that while the signals found in Fig.~\ref{fig:pert_base10} are not necessarily desired for the sake of finding standard clocks and model-independent evidence of ekpyrosis, they still represent interesting phenomenology, which could deserve further attention in follow-up work.

\section{Discussion and conclusions}\label{sec:conclusions}

In this paper, we constructed the first full models of classical primordial standard clocks in an alternative to inflation, specifically in the slowly contracting ekpyrotic scenario.
Our setup consisted in introducing a coupling between the ekpyrotic field $\phi$ and a massive field $\sigma$ though an operator of the form $\sigma(\partial\phi)^2$. Thinking of the adiabatic trajectory as the bottom of an inclined valley in field space, the massive field is initially at rest, but a turning of the valley gives some initial momentum for $\sigma$ to start oscillating on the slopes perpendicular to the adiabatic trajectory.
We investigated the observational signatures of this entire process, motivated by the significant implication arising from the signature of an oscillating massive field. The setup and questions we asked here are very similar to quasi-single field inflation \cite{Chen:2009zp} (see \cite{Chen:2010xka,Wang:2013zva} for reviews) and many constructions of classical primordial standard clocks in inflation \cite{Chen:2014cwa,Braglia:2021ckn,Braglia:2021sun,Braglia:2021rej}.

When the turning occurs over a short time scale, which is to say that the operator $\sigma(\partial\phi)^2$ is quickly turned on and off through a field-dependent coupling $\Xi(\phi)$, and if the massive field is sufficiently heavy at that moment in time (compared to the background horizon mass scale), then it starts oscillating in the standard way to lead to a clock signal in the power spectrum. Such a signal would be a telling sign of an ekpyrotic phase of contraction. However, when the coupling term $\Xi\sigma(\partial\phi)^2$ remains present for a sufficiently long time with constant $\Xi$, the standard clock oscillations are highly modulated during their evolution. Correspondingly, the clock signals get more mixed with other signals, and as such, they become more difficult to be disentangled. In such cases, we find a vast phenomenology of features in the power spectrum besides the clock signal, none of which would be model-independent evidence of ekpyrosis to the same extent as a clear-cut clock signal. We leave for future study the question of whether or how any potential clock signals could be clearly separated in this case.

At the background level, there is already a clear difference between inflation and a contraction alternative when it comes to implementing a classical primordial standard clock. Indeed, in inflation the horizon scale $m_{\mathrm{h}}$ is nearly constant, hence if $m_\sigma\gg m_{\mathrm{h}}$ initially, $\sigma$ remains heavy at all times. In a contracting scenario, any massive field would be `heavy' as long as $m_{\mathrm{h}}$ is sufficiently small initially. However, this does not hold forever, and as the universe contracts and $m_{\mathrm{h}}$ becomes large, massive fields eventually become `light', with $m_\sigma\ll m_{\mathrm{h}}$. For these reasons, inflation probes massive fields with a nearly constant energy scale, as a `cosmological collider' \cite{Chen:2009zp,Arkani-Hamed:2015bza}, whereas a contracting alternative scans massive fields over a vast range of energy scales, as a `particle scanner' \cite{Chen:2018cgg}.

The fact that heavy fields eventually become light in a contracting cosmology comes at a price, though.
For example, there is an issue of stability for non-attractor models such as the matter bounce scenario (see, e.g., \cite{Cai:2013vm,Levy:2016xcl,Lin:2017fec,Ganguly:2021pke} and references therein for a discussion of the issue and possible resolutions, as well as \cite{Baumann:2011dt,Quintin:2015rta,Li:2016xjb} regarding further challenges).
Slow contraction (ekpyrosis), though, has the advantage of being an attractor like inflation, hence
one may not have to worry about massive fields potentially disrupting the background evolution and the leading-order nearly scale-invariant power spectrum of scalar perturbations. In that sense, any massive field (whether effectively heavy or light) is just a spectator. We have shown this to be indeed true, although only when the interaction term $\sigma(\partial\phi)^2$ does not remain present for too long, because otherwise $\sigma$ drifts (as opposed to in inflation, where it would damp out and settle down) and eventually backreacts on $\phi$ to the extent that $\phi$ no longer acts as an ekpyrotic field (instead, all fields become effectively massless).

We should stress that this last conclusion is quite specific to our classical primordial standard clock implementation, i.e., having an operator $\sigma(\partial\phi)^2$, which may or may not remain present. This is in no way exhaustive and unique. One could certainly think of alternative models, which would not involve such an operator and which would thus not have the same issue\footnote{Let us emphasize that this is not a critical issue of the ekpyrotic scenario in itself; it is simply a challenge for implementing classical primordial standard clocks. In fact, more complete and realistic models of ekpyrosis have a potential $V(\phi)$ that goes to $0$ as the bounce is approached at late times (i.e., ekpyrosis is followed by a phase of kinetion; see, e.g., \cite{Steinhardt:2001st,Lehners:2008vx,Cai:2012va}), thus in that sense having an evolution from $\epsilon\gg 3$ to $\epsilon\to 3$ is not problematic at all; it is rather desired and expected.} --- for instance, one could explore the massive field as coupling to the background through operators of the form $\partial\sigma\partial\phi$, $\sigma^2R$, etc. We have explored a few alternative phenomenological implementations of massive scalar fields classically excited in a phase of ekpyrotic contraction, but we left them out for follow-up work, which could also explore the phenomenology of massive particles of different spin (e.g., fermions, vector fields, etc.), as well as perhaps more fundamental constructions (e.g., string theoretic).

A key conclusion of our work is that a sharp bump coupling leads to clear-cut, easily distinguishable sharp feature and clock signals. A further important conclusion is that whenever we find a clock signal, this signal has a frequency dependence and a characteristic envelop,
\begin{equation}
    \frac{\Delta\mathcal{P}_s}{\mathcal{\bar P}_s}\sim
    \left(\frac{k}{k_\mathrm{res}}\right)^{\frac{1-3p}{2p}}
    \sin\left[p\frac{m_\sigma}{m_{\mathrm{h}0}}\left(\frac{k}{k_\mathrm{res}}\right)^{1/p}\right]\,,
\end{equation}
which robustly correspond to what is expected from a slowly contracting cosmology with $0<p\ll 1$ (and which is in some ways exactly opposite of what one has in inflation with $p\gg 1$). In fact, the analytical estimates we derived are quite accurate in describing the sharp feature and clock signals for a sharp bump coupling function as long as
$m_\sigma/m_{\mathrm{h}0}=K_\mathrm{res}/k_0$
is large. Thus in that regime, the analytical estimates may be used to construct robust templates for the sake of searching for such signals in data. At small $m_\sigma/m_{\mathrm{h}0}$, though, the analytical estimates become less accurate, and one needs to solve the full model to compute the correct signal (which further motivates this work). We note that while having a very large mass compared to the horizon scale at the start of the oscillations leads to more oscillations in the clock signal, it would also make the signal harder to detect, because it increases its frequency, hence one would require more angular resolution in the observations. On the other hand, low frequency signals are easier to detect, but small-$m_\sigma/m_{\mathrm{h}0}$ signals have very few oscillations or even just a spike. Yet, this might still be distinctive evidence of ekpyrosis as we do not know of any other models (including other than ekpyrotic and alternative features such as sharp ones) that would produce such a signal. It will be interesting to determine given some observational forecast what ranges of $m_\sigma/m_{\mathrm{h}0}$ values we may be able to probe in the future.

The conclusion that the sharp feature and clock signals are distinct from one another is somewhat dependent on the sharpness of the transition: we found that the smoother the transition, the more the two signals can mix. More work could be done to better understand the model dependence of this aspect of the results in both inflation and ekpyrosis. Let us stress that, in this work, the clock analytical template is found to be good for both sharp and smooth bumps as long as $m_\sigma/m_{\mathrm{h}0}$ is large, but the sharp feature signal is definitely more complex for smooth bumps (regardless of the value of $m_\sigma/m_{\mathrm{h}0}$). And regardless of the transition sharpness, a generic prediction of the bump coupling is a constant shift in the amplitude of the power spectrum at large $k$. This is due to the freezing of the massive field to a non-zero constant value at late times, hence the presence of the operator $\sigma\Omega^2(\phi)(\partial\chi)^2$ in the perturbation sector leads to a constant offset in the amplitude of the power spectrum on small length scales. Such a prediction could be easily tested with data, though this would be specific to this particular model. Similarly to the previous discussion, one could further explore different ways the massive field can couple to the cosmological perturbations. Most of our focus has been on a direct coupling between $\sigma$ and the isocurvature perturbation $\chi$, but other avenues could be further explored. A gravitational coupling always appears to be present, though in this work we find its effect to be less important than a direct coupling, and the gravitational coupling seems more efficient in producing sharp feature signals than clock signals. Certainly, in the future it would be interesting to have a better analytical understanding of the signals due to a pure gravitational coupling in the present model and to better understand how this compares to the situation in inflation.

Let us recall that primordial standard clocks have been proposed to provide model-independent evidence of the evolution of the scale factor during the primordial phase, but this proposal first came from parametrized estimates of the behavior of massive fields in some given background. Actual models of primordial standard clocks in inflation supported the proposal, but it is the first time we can address how this fares in comparison to an actual model in a non-inflationary scenario. In general, we confirm that a clear-cut clock signal would be proper evidence of a given scenario over another. More specifically, we showed that if a massive field is properly classically excited, it leads to a signal in the two-point function that robustly discriminates between ekpyrosis and inflation. This is crucial since, as we stressed in our introduction, very few predictions have this power.

There are some caveats since we made some simplifying assumptions in our computations. For instance, we ignored any $\chi$ potential term at reheating and did not model the conversion of isocurvature perturbations ($\chi$) into curvature perturbations ($\mathcal{R}$). This process is known to produce non-Gaussianities, and it would be interesting to determine how the presence of features affects the process. In fact, it would be important to support the conclusion of \cite{Raveendran:2018yyh} that features in $\mathcal{P}_s(k)$ are imprinted in $\mathcal{P}_\mathcal{R}(k)$ after conversion.
Relatedly, a critical issue in alternatives to inflationary is how the density perturbations survive through the bounce. Explicitly evolving the perturbations through a non-singular bouncing model and checking the assumption that the predicted $\Delta\mathcal{P}_s/\mathcal{\bar P}_s$ at the end of ekpyrosis translates into $\Delta\mathcal{P}_\mathcal{R}/\mathcal{\bar P}_\mathcal{R}$ at the onset of radiation-dominated expansion should be done in the future. Nevertheless, we believe that standard clock signals should be quite robust against the details of the bounce, as the key information is in the scale dependence of the oscillatory phases.

Something else that we ignored in this work is the effect of other perturbations, such as $\delta\sigma$ and $\delta\phi$. The latter is typically subdominant compared to $\chi$ in standard ekpyrotic cosmology. We expect this to still hold in the presence of massive fields and the same to be true for $\delta\sigma$. Nevertheless, it would be interesting to study the \emph{full} effect of $\sigma$ perturbations in future work [i.e., $\sigma(t)+\delta\sigma(t,\mathbf{x})$] and in particular looking at purely quantum effects, namely quantum standard clocks (e.g., \cite{Chen:2015lza}). Indeed, quantum standard clocks have only been explored in alternatives to inflation in \cite{Chen:2015lza,Chen:2018cgg,Domenech:2020qay}, and once more it would be relevant to have a better understanding of the robustness of their predictions in actual models to see how things compare to inflation. There has been a lot of work on quantum standard clocks or cosmological collider physics in the context of inflation, but again with the prospect of finding new ways of discriminating whole scenarios of the very early universe, more work should be done on non-inflationary theories. In particular, for both classical and quantum standard clocks, a lot of information is encoded in higher correlation functions beyond the power spectrum, e.g., in the bi- and trispectra. In general, studying non-Gaussian signals due to the presence of massive fields in alternatives to inflation is largely unexplored.

We end with a list of follow-up questions that could further deserve attention:
\begin{itemize}
    \item What if we put in the standard model or some particles beyond the standard model --- how can we couple things consistently, are some massive fields excited, do we get specific clock signals, and how does it compare to inflation such as in \cite{Chen:2016uwp,Hook:2019vcn}?
    In particular, it would be interesting to have a relation between the mass and energy ranges of the theories and the observable scales (e.g., in terms of the CMB angular scale $\ell$ or large-scale structure wavenumber $k$) where we would expect to find these signals.
    It would also be interesting to explore the signatures of massive fields beyond the power spectrum, in primordial non-Gaussianities; and in cases with or without sharp features, with contributions from either classical or quantum oscillations of massive fields.
    Overall, these questions are similar to those asked in the program of cosmological collider physics for inflation \cite{Chen:2009we,Chen:2009zp,Baumann:2011nk,Assassi:2012zq,Chen:2012ge,Sefusatti:2012ye,Norena:2012yi,Pi:2012gf,Noumi:2012vr,Gong:2013sma,Arkani-Hamed:2015bza,Chen:2015lza,Dimastrogiovanni:2015pla,Kehagias:2015jha,Chen:2016nrs,Lee:2016vti,Meerburg:2016zdz,Chen:2016uwp,Chen:2016hrz,Chen:2017ryl,Kehagias:2017cym,An:2017hlx,An:2017rwo,Iyer:2017qzw,Kumar:2017ecc,Chen:2018sce,Chen:2018xck,Chua:2018dqh,Kumar:2018jxz,Wu:2018lmx,MoradinezhadDizgah:2018ssw,Saito:2018omt,Tong:2018tqf,Fan:2019udt,Alexander:2019vtb,Lu:2019tjj,Hook:2019zxa,Hook:2019vcn,Kumar:2019ebj,Wang:2019gbi,Liu:2019fag,Wang:2019gok,Wang:2020uic,Li:2020xwr,Wang:2020ioa,Fan:2020xgh,Kogai:2020vzz,Bodas:2020yho,Aoki:2020zbj,Arkani-Hamed:2018kmz,Baumann:2019oyu,Baumann:2020dch,Maru:2021ezc,Lu:2021gso,Lu:2021wxu,Wang:2021qez,Tong:2021wai,Pinol:2021aun,Cui:2021iie,Tong:2022cdz,Reece:2022soh,Pimentel:2022fsc,Qin:2022lva,Chen:2022vzh,Jazayeri:2022kjy,Ghosh:2022cny,Qin:2022fbv,Xianyu:2022jwk,Niu:2022quw,Niu:2022fki,Wang:2022eop,Aoki:2023tjm,Werth:2023pfl,Chen:2023txq,Tong:2023krn,Qin:2023bjk,Jazayeri:2023xcj,Xianyu:2023ytd,Ema:2023dxm,Chakraborty:2023qbp,Green:2023ids,Green:2023uyz,Chakraborty:2023eoq,Pinol:2023oux,Craig:2024qgy,McCulloch:2024hiz,Fan:2024iek,Cabass:2024wob,Melville:2024ove,Wu:2024wti,Sohn:2024xzd,Aoki:2024uyi}.
    
    \item It would be interesting to explore if primordial features in ekpyrosis, such as the ones studied here or other types of features, may be used to explain certain feature candidates in the CMB, such as the low-$\ell$ dip and high-$\ell$ wiggle (such as explored in \cite{WMAP:2003syu,Adams:2001vc,Bean:2008na,Mortonson:2009qv,Hazra:2010ve,Hazra:2014goa,Miranda:2014fwa,Chen:2014joa,Chen:2014cwa,Cai:2015xla,Planck:2018jri,Canas-Herrera:2020mme,Braglia:2021sun,Braglia:2021rej}) or various anomalies (as explored in \cite{Domenech:2020qay,Hazra:2022rdl}).
    
    \item How can we improve templates for the sake of data fitting (as in, e.g., \cite{Chen:2012ja,Chen:2014joa,Chen:2014cwa,Domenech:2020qay,Braglia:2021ckn,Braglia:2021sun,Braglia:2021rej}) to better account for the phenomenology of features found in this work?
    \item What is the prospect of detecting clock signals in the future (building on, e.g., \cite{Chen:2016zuu,Chen:2016vvw,Slosar:2019gvt,Beutler:2019ojk,Biagetti:2019bnp,Braglia:2022ftm,Euclid:2023shr,Antony:2024vrx}) and is a certain kind of clock signal (inflationary, ekpyrotic, etc.) more easily detectable due to the specific frequency dependence?
    
    \item What about non-standard clocks, i.e., massive fields with non-constant (time-dependent) mass? Is it always possible to engineer a model (see \cite{Chen:2011zf,Huang:2016quc,Domenech:2018bnf,Wang:2020aqc}) that would mimic the standard clock signal of another scenario, or is this unrealistic/unrealizable? If possible, it would be important to construct the full models; and if not, to what extent could this be done?
    
    \item What about different models of ekpyrosis altogether, e.g., models that do not have any entropic field $\chi$ and where it is the adiabatic field $\phi$ that acquires near scale invariance (see, e.g., the model of \cite{Brandenberger:2020tcr,Brandenberger:2020eyf,Brandenberger:2020wha}) --- can one consistently implement standard clocks in such models and do they predict the same clock signals?
    \item What about other alternatives to inflation, like slowly expanding ones (e.g., \cite{Brandenberger:1988aj,Nayeri:2005ck,Brandenberger:2006pr,Creminelli:2010ba,Joyce:2011ta,Geshnizjani:2011dk}) or holographic ones (e.g., \cite{Hinterbichler:2011qk,Hinterbichler:2014tka,Carrillo-Gonzalez:2020ejs})? Can there be heavy spectator fields in such theories and what are their effects and predicted signals?
\end{itemize}
The future will be bright if progress in these directions can be achieved, and this will be beneficial for both inflationary cosmology and its alternatives since the common goal is to find robust evidence for the actual origin of our universe.

\vskip23pt
\subsection*{Acknowledgments}
{\small
J.Q.~and X.C.~thank Robert Brandenberger for initial collaboration and discussions many years ago. J.Q.~further thanks Jean-Luc Lehners for encouragements and the Institute for Theory and Computation at Harvard for hospitality in the early stages of this work. X.C.~further thanks Mohammad Hossein Namjoo and Amin Nassiri-Rad for discussions on a related subject.
J.Q.~acknowledges financial support from the University of Waterloo's Faculty of Mathematics William T.~Tutte Postdoctoral Fellowship, and his research at the Perimeter Institute for Theoretical Physics is supported by the Government of Canada through the Department of Innovation, Science and Economic Development and by the Province of Ontario through the Ministry of Colleges and Universities.
J.Q.~further acknowledges financial support over the years from the Natural Sciences and Engineering Research Council of Canada and the Fonds de recherche du Qu\'ebec --- Nature et technologie.
R.E.~is supported by the Quantum Technology Center and Grant \#63034 from the John Templeton Foundation. The opinions expressed in this publication are those of the authors and do not necessarily reflect the views of the John Templeton Foundation.
This research made use of \texttt{Jupyter} \cite{soton403913}, \texttt{NumPy} \cite{2020Natur.585..357H}, \texttt{SciPy} \cite{2020NatMe..17..261V}, \texttt{matplotlib} \cite{2007CSE.....9...90H}, and \texttt{MathGR} \cite{Wang:2013mea}.
The codes underlying this article may be shared upon reasonable request to the corresponding authors.}
\vskip23pt


\appendix

\section{Types of horizons}\label{app:mass-horizon}

Let us review the different notions of horizon that are relevant in this work. First, there is the scale at which a massive field freezes and stops oscillating.
In a flat FLRW background with scale factor $a(t)$, the equation of motion of a minimally coupled homogeneous massive scalar field $\sigma(t)$ of mass $m_\sigma$ is
\begin{equation}
    \ddot\sigma+3\frac{\dot a}{a}\dot\sigma+m_\sigma^2\sigma=0\,.
\end{equation}
Here, $t$ is the physical time, and $\dot{}\equiv\partial_t{}$. In terms of conformal time $\tau$ (where $\dd\tau\equiv a^{-1}\dd t$, ${}'\equiv\partial_\tau{}$), the equation of motion is
\begin{equation}
    \sigma''+2\frac{a'}{a}\sigma'+m_\sigma^2a^2\sigma=0\,.
\end{equation}
Letting $\varsigma\equiv a\sigma$, the equation becomes
\begin{equation}
    \varsigma''+\left(m_\sigma^2a^2-\frac{a''}{a}\right)\varsigma=0\,.
\end{equation}
This is the equation of a simple harmonic oscillator with time-dependent frequency $\omega(\tau)$ satisfying
\begin{equation}
    \omega^2=m_\sigma^2a^2-\frac{a''}{a}\,.
\end{equation}
The transition time when $\varsigma$ (and equivalently $\sigma$) goes from oscillating (when $\omega^2$ is dominated by $m_\sigma^2a^2$) to being frozen (when $\omega^2$ is dominated by $a''/a$) is the moment when $m_\sigma^2a^2=|a''|/a$. We thus use this to define the `mass' (reciprocal length or frequency) of the corresponding horizon scale,
\begin{equation}
    m_\mathrm{h}(\tau)\equiv\sqrtb{\frac{|a''|}{a^3}}\,.
\end{equation}
Using
$a(\tau)=(-\tau)^{\frac{p}{1-p}}$ (assuming a contracting cosmology here with $\tau<0$ and $0<p<1$),
one has
\begin{equation}
    \frac{a''}{a^3}=-\frac{p(1-2p)}{(1-p)^2}(-\tau)^{-\frac{2}{1-p}}\,,
\end{equation}
and one can verify that the Hubble parameter, $H\equiv\dot a/a=a'/a^2$, satisfies
\begin{equation}
    H(t)=-\frac{p}{(1-p)}\left(-\tau(t)\right)^{-\frac{1}{1-p}}\,.
\end{equation}
Therefore, we can write
\begin{equation}
    m_\mathrm{h}(t)=\sqrtb{\frac{\left|1-2p\right|}{p}}\left|H(t)\right|=\frac{\sqrtb{p\left|1-2p\right|}}{|t|}\stackrel{p\ll 1}{\simeq}\frac{\sqrtb{p}}{|t|}\label{eq:mh1}
\end{equation}
as the mass of the horizon.

Note that the same horizon scale determines whether adiabatic fluctuations in a constant equation of state background (contracting with $0<p<1$) are oscillating or frozen. Indeed, they follow
\begin{equation}
    u_k''+\left(k^2-\frac{a''}{a}\right)u_k=0\,,
\end{equation}
where $u_k$ denotes the normalized mode function of an adiabatic fluctuation with comoving wavenumber $k$. Accordingly, the freezing of adiabatic modes occurs when $k^2=|a''|/a$, so when their physical wavenumber $q\equiv k/a$ equates $m_\mathrm{h}$ as defined above. If we define the corresponding comoving scale as $k_\mathrm{h}\equiv am_\mathrm{h}$, then we notice that
\begin{equation}
    k_\mathrm{h}(\tau)=\frac{\sqrtb{p|1-2p|}}{(1-p)|\tau|}\stackrel{p\ll 1}{\simeq}\frac{\sqrtb{p}}{|\tau|}\,.\label{eq:kh1}
\end{equation}
A mode that acquires scale invariance (whether adiabatic or entropic) satisfies $v_k''+(k^2-2/\tau^2)v_k=0$, in which case the horizon scale associated with freezing is simply $k_\mathrm{h}=\sqrtb{2}/|\tau|$.

An alternative definition of horizon, used in \cite{Chen:2011zf,Chen:2014cwa}, is the horizon scale defined by
\begin{equation}
    m_\mathrm{h}^{-1}(t)\equiv a(t)\int_t^0\frac{\dd\tilde t}{a(\tilde t)}=-a(t)\tau(t)\,,\label{eq:defEventHor}
\end{equation}
in which case one can deduce
\begin{equation}
    m_\mathrm{h}(t)=\frac{1-p}{p}\left|H(t)\right|=\frac{1-p}{|t|}\stackrel{p\ll 1}{\simeq}\frac{1}{|t|}\,,\qquad k_\mathrm{h}\equiv am_\mathrm{h}=\frac{1}{|\tau|}\,.\label{eq:mh2}
\end{equation}
The integral in \eqref{eq:defEventHor} should be taken assuming $a(t)\propto(-t)^p$ in a contracting cosmology ($t<0$, $0<p<1$) that ends in a big crunch at $t=0$. In this case, \eqref{eq:defEventHor} is the event horizon \cite{Mukhanov:2005sc}. In reality, a bounce would occur before a crunch is reached for the model to be realistic, in which case there may be no such concept as an event horizon. Nevertheless, as long as one is far away from the bouncing time, i.e., deep in the contracting phase when modes of observational relevance would freeze, the above captures a relevant notion of horizon. (It is the same story as in quasi-de Sitter inflation, where the event horizon of de Sitter satisfies $m_\mathrm{h}=H$, $k_\mathrm{h}=1/|\tau|$; recall $a(\tau)=-1/(H\tau)$ for de Sitter.)

For concreteness, we follow the convention of \cite{Chen:2011zf,Chen:2014cwa} and use \eqref{eq:mh2} as the definition of the horizon scale throughout this work. As such, the horizon-exit time of a massive field is
\begin{equation}\label{eq:thA}
    t_{\textrm{h-e}}=-\frac{1-p}{m_\sigma}\stackrel{p\ll 1}{\simeq}-\frac{1}{m_\sigma}\,,
\end{equation}
which is the solution to when $m_\mathrm{h}(t)=m_\sigma \Leftrightarrow -m_\sigma t\simeq 1$. Likewise, the horizon-exit time of a perturbation is when $k_\mathrm{h}(\tau)=k \Leftrightarrow -k\tau=1$; we shall often denote this time as $\tau_\star=-1/k$. Note, though, that these are not quite the same as the freezing times according to \eqref{eq:mh1} and \eqref{eq:kh1}; the scales differ by a factor of about $1/\sqrtb{p}$. Naturally, in the limit $p\to 0^+$ one recovers Minkowski, hence there is no freezing time, only oscillations. However, for some reasonable values of $p$ (e.g., $1/20\lesssim p\lesssim 1/3$), the difference may be $\mathcal{O}(1)$, akin to the $\sqrtb{2}$ difference between \eqref{eq:mh2} and the expected freezing scale of a scale-invariant mode. In any case, the concept of freezing time is somewhat imprecise, in the sense that it is never a sharp transition. It is rather more of a smooth transition for which it is difficult to pinpoint exactly where oscillations stop. This can be seen explicitly in analytic solutions that involve the Bessel or Hankel functions.

\section{Second-order action}\label{app:2ndaction}

Let us briefly review how one derives \eqref{eq:S21}. Starting from the full action \eqref{eq:actionGenModels} and expanding the fields according to \eqref{eq:perturbedFields} in the spatially flat gauge ($\Psi\equiv 0$) up to second order in perturbations yields
\begin{align}
    S^{(2)}=&~\frac{1}{2}\int\dd^3x\,\dd t\,a^3\Bigg(\left(-6H^2+(1+\Xi\sigma)\dot\phi^2+\dot\sigma^2\right)\Phi^2+2\left(-2H\Phi+(1+\Xi\sigma)\dot\phi\,\delta\phi+\dot\sigma\,\delta\sigma\right)\frac{\partial^2B}{a^2}\nonumber\\
    &\qquad+(1+\Xi\sigma)\left(\dot{\delta\phi}^2-\frac{(\partial_i\delta\phi)^2}{a^2}\right)+(1+\Upsilon\sigma)\Omega^2\left(\dot\chi^2-\frac{(\partial_i\chi)^2}{a^2}\right)+\dot{\delta\sigma}^2-\frac{(\partial_i\delta\sigma)^2}{a^2}\nonumber\\
    &\qquad-\left(V_{,\phi\phi}-\frac{1}{2}\Xi_{,\phi\phi}\sigma\dot\phi^2\right)\delta\phi^2-m_\sigma^2\,\delta\sigma^2+\Xi_{,\phi}\dot\phi^2\,\delta\sigma\,\delta\phi+2\Xi\dot\phi\,\delta\sigma\,\dot{\delta\phi}+2\Xi_{,\phi}\sigma\dot\phi\,\delta\phi\,\dot{\delta\phi}\nonumber\\
    &\qquad-2\Phi\bigg(\dot\sigma\,\dot{\delta\sigma}+\Big(m_\sigma^2\sigma+\frac{1}{2}\Xi\dot\phi^2\Big)\delta\sigma+(1+\Xi\sigma)\dot\phi\,\dot{\delta\phi}+\Big(V_{,\phi}+\frac{1}{2}\Xi_{,\phi}\dot\phi^2\sigma\Big)\delta\phi\bigg)\Bigg)\,.
\end{align}
The expanded action also has the following contribution,
\[
    \frac{1}{2}\int\dd^3x\,\dd t\,a^3\bigg(\frac{(\partial_i\partial_jB)^2}{a^4}-\frac{(\partial^2B)^2}{a^4}\bigg)\,,
\]
but this combination is a boundary term, hence we omit it in the expression.
We also performed integration by parts of the form $\int\partial_i\delta\phi\,\partial^iB=-\int\delta\phi\,\partial^2B$ (and similarly with $\delta\phi$ replaced by $\delta\sigma$).
In the above, we notice that the perturbations of the lapse ($\Phi$) and of the shift ($B$) appear as Lagrange multipliers (no time derivatives act on them), hence we can solve for the corresponding constraints (the Hamiltonian and momentum constraints, respectively).
The solutions to the constraints read
\begin{subequations}
\begin{align}
    \Phi=&~\frac{1}{2H}\left((1+\Xi\sigma)\dot\phi\,\delta\phi+\dot\sigma\,\delta\sigma\right)\,,\\
    \frac{\partial^2B}{a^2}=&-\frac{1}{2H}\left((1+\Xi\sigma)\dot\phi\,\dot{\delta\phi}+\dot\sigma\,\dot{\delta\sigma}\right)-\frac{1}{4H^2}\Big((1+\Xi\sigma)\dot\phi(6H^2-\dot\sigma^2)-(1+\Xi\sigma)^2\dot\phi^3\nonumber\\
    &+2HV_{,\phi}+\Xi_{,\phi}\sigma\dot\phi^2H\Big)\delta\phi-\frac{1}{4H^2}\Big((6H^2-(1+\Xi\sigma)\dot\phi^2)\dot\sigma-\dot\sigma^3+2Hm_\sigma^2\sigma+\Xi\dot\phi^2H\Big)\delta\sigma\,.
\end{align}
\end{subequations}
Upon substituting these solutions back into the perturbed action to eliminate the lapse and shift perturbations, we obtain an action that solely depends on the dynamical degrees of freedom $\chi$, $\delta\phi$, and $\delta\sigma$. Doing some more integration by parts to eliminate terms like $\delta\phi\,\dot{\delta\phi}$ and using the background equations of motion \eqref{eq:backEOMsFull} to simplify the expression, we finally arrive at equation \eqref{eq:S21}. We expressed it in Fourier space for later convenience, but the expression in real space is just as straightforwardly deduced.


\addcontentsline{toc}{section}{References}

\let\oldbibliography\thebibliography
\renewcommand{\thebibliography}[1]{
  \oldbibliography{#1}
  \setlength{\itemsep}{0pt}
  \footnotesize 
}

\bibliographystyle{JHEP2}
\bibliography{refs}

\end{document}